\documentclass[10pt, aps, prd, superscriptaddress, nofootinbib, 
               amsmath, amssymb, twocolumn,
               preprintnumbers, showpacs,
               raggedbottom,
               floatfix]{revtex4-1}

\usepackage[T1]{fontenc}
\usepackage{textcomp}
\usepackage{etoolbox}
\usepackage{etextools}
\usepackage{comment}
\usepackage[bottom]{footmisc}

\usepackage[acronym]{glossaries}
\glsdisablehyper
\makeglossaries
\usepackage[utf8]{inputenc}

\usepackage{hyperref}
\usepackage{graphicx}
\usepackage{dcolumn}
\usepackage{color}

\newcommand{\dthree}[1]{\frac{d^3#1}{(2\pi)^3}}
\newcommand{\deltafour}{(2\pi)^4\delta(\sum p^\mu)}

\newcommand{\sdot}{{\dot{s}}}
\newcommand{\dota}{{\dot{a}}}
\newcommand{\dotb}{{\dot{b}}}

\newcommand{\calP}{{\cal{P}}}
\newcommand{\calM}{{\cal{M}}}
\newcommand{\calL}{{\cal{L}}}

\newcommand{\slsh}[1]{#1\!\!\!/}
\newcommand{\tr}{{\rm{tr}}}

\newcommand{\bfr}{{\bf{r}}}

\newcommand{\bfp}{{\bf{p}}}
\newcommand{\bfq}{{\bf{q}}}
\newcommand{\bfl}{{\bf{l}}}
\newcommand{\bfb}{{\bf{b}}}

\newcommand{\ph}{{\rm{ph}}}
\newcommand{\hatb}{{\hat{b}}}
\newcommand{\hatf}{{\hat{f}}}
\newcommand{\bfv}{{\bf{v}}}
\newcommand{\f}{{\rm{f}}}
\newcommand{\dmu}{{\delta\mu}}
\newcommand{\bsigma}{{\bar{\sigma}}}

\newcommand{\SU}{{\rm{SU}}}
\newcommand{\U}{{\rm{U}}}
\newcommand{\CFL}{{\rm{CFL}}}
\newcommand{\SC}{{\rm{SC}}}
\newcommand{\un}{{\rm{un}}}

\newcommand{\diag}{{\rm{diag}}}

\begin{document}

\title{The shear viscosity of two-flavor crystalline color superconducting
quark matter}
\author{Sreemoyee Sarkar}
\email{sreemoyee.sinp@gmail.com}
\affiliation{TIFR, Homi Bhabha Road, Navy Nagar, Mumbai 400005, India}
\affiliation{UM-DAE Centre for Excellence in Basic Sciences
Health Centre, University of Mumbai,
Vidyanagari Campus, Kalina, Santacruz (East), Mumbai 400098, India.}
\author{Rishi Sharma}
\email{rishi@theory.tifr.res.in}
\affiliation{TIFR, Homi Bhabha Road, Navy Nagar, Mumbai 400005, India}

\date{\today}

\begin{abstract}
\noindent
We present the first calculation of the shear viscosity for two-flavor plane wave
(FF) color superconducting quark matter. This is a member of the family of
crystalline color superconducting phases of dense quark matter that may be
present in the cores of neutron stars. The paired quarks in the FF phase
feature gapless excitations on surfaces of crescent shaped blocking regions in
momentum space and participate in transport. We
calculate their contribution to the shear viscosity. We note that they also
lead to dynamic screening of transverse $t^1$, $t^2$, $t^3$ gluons which are
undamped in the $2\SC$ phase. The exchange of these gluons is the most
important mechanism of the scattering of the paired quarks. We find that the
shear viscosity of the paired quarks is roughly a factor of $100$ smaller
compared to the shear viscosity of unpaired quark matter. Our results may have
implications for the damping of $r-$modes in rapidly rotating, cold neutron
stars.
\end{abstract}
\preprint{TIFR/TH/16-48}
\pacs{21.65.Qr,67.10.Jn,97.60.Jd,67.85.-d}

\maketitle
\tableofcontents
\section{Introduction}
~\label{sec:introduction}

The discovery of neutron stars with masses close to
$2M_\odot$~\cite{Demorest:2010bx,Antoniadis:2013pzd}, (see
Ref.~\cite{Ozel:2016oaf} for a recent review), has provided a strong constraint
on the equation of state of matter in neutron
stars~\cite{page2006dense,Ozel:2010bz,Potekhin:2010,Lattimer:2012nd,Prakash:2014tva}
ruling out large parameter spaces in various models of dense matter.  (For
quark matter see Ref.~\cite{Ranea-Sandoval:2015ldr}.) Refinements in the
measurements on the radii of neutron stars provide additional constraints on
the equation of state~\cite{Steiner:2012xt,Lattimer:2013hma}. 

In addition to analyzing constraints on the equation of state, characterising
the nature of the phases of matter in neutron stars will require
observationally constraining the transport properties of neutron stars. These
observations can help eliminate models of dense matter inconsistent with the
data~\cite{page2006dense}. Transport properties are
sensitive to the spectrum of excitations above the equilibrium state (which is
essentially the ground state because the temperatures of neutron stars are much
smaller than the other relevant energy scales). These excitations can
differ substantially for phases with similar equations of state.

For example, the short time (time scales of many days) thermal evolution
already constrains the thermal conductivity and the specific heat of matter in
the neutron star crust (\cite{Chamel:2008ca,Page:2012zt} and references
therein). Neutrino cooling on much longer time scales ($10^5$ years) depends on
the phase of matter inside the cores (see
Ref.~\cite{Pethick:1992,Yakovlev:2000jp,Yakovlev:2003} for a review). A neutron
star of mass around $1.4M_\odot$, with a core of only protons, electrons and
neutrons cools ``slowly''. The presence of condensates, strange particles, or
unpaired quark matter in the cores, leads to ``fast'' cooling.  One hopes that
observations of the temperatures and the ages of neutron stars will be able to
tell us whether neutron star cores feature such exotic phases.

A set of observables sensitive to the viscosity of matter in the cores of neutron
stars are the spin frequencies, temperatures, and the
spin-down rates of fast rotating neutron stars~\cite{Alford:2013pma}. In the
absence of viscous damping, the fluid in rotating neutron stars
is~\cite{Andersson:1997xt,Andersson:1997rn} unstable to a mode which
couples to gravity which radiates away the angular momentum of the
star. If the mode grows the neutron stars are expected to spin down rapidly.
This is the famous $r$-mode instability.

The connection between the viscosity and the spin observables is subtle because
it depends on the amplitude~\cite{Lindblom:2000az,Alford:2011pi} (determined by
non-linear physics) at which the $r$-mode saturates if the star winds up in the
regime where the $r$-mode is unstable in the linear approximation.  But the
essence of the connection can be understood
easily~\cite{Andersson:1997xt,Andersson:1997rn,Alford:2010fd} by ignoring the
saturation dynamics.  In this regime the amplitude of the mode changes with
time as,
\begin{equation}
{\cal{A}}e^{-(\Gamma_{\rm{GW}}+\Gamma_{\rm{Bulk}}+\Gamma_{\rm{Shear}})t}
\end{equation}
where $\Gamma_{\rm{GW}}<0$. $\Gamma_{\rm{Bulk}}$ and $\Gamma_{\rm{Shear}}$ are
both positive and depend on the bulk and shear viscosities throughout the star
and therefore depend on the phase of matter in the core and the temperature of
the star. The magnitude of $\Gamma_{\rm{GW}}$ grows with the rotational
frequency, $\Omega$, of the star~\cite{Andersson:1997xt}.

If $\Omega$ is large enough such that 
$\Gamma_{\rm{GW}}+\Gamma_{\rm{Bulk}}+\Gamma_{\rm{Shear}}<0$, the neutron star
can be expected to spin down rapidly. This will continue till $\Omega$ is small 
enough that the shear and bulk viscosities can damp the $r$-modes. This
argument implies that at any given temperature $T$, the neutron star frequency
should be below a maximum~\cite{Alford:2013pma} determined by the shear and bulk viscosities at that
temperature. The shear moduli dominate at smaller $T$ and the bulk moduli at
larger temperatures, and the crossover point depends on the phase. 

Assuming there are no other damping mechanisms and that the $r$-modes do not
saturate at unnaturally small amplitudes, fluids in neutron
stars~\cite{Flowers:1976,Flowers:1979} made up of only neutrons, protons and
electrons do not
have sufficient viscosity to damp $r$-modes in many rapidly rotating neutron
stars~\cite{Andersson:2000pt,Bildsten:1999zn,Jaikumar:2008kh,Alford:2013pma}.
Large damping at the crust-core
interface~\cite{Bildsten:1999zn,Levin:2000vq,Lindblom:2000gu} could stabilize
the $r$-mode in such stars, but would require unnaturally large shear 
moduli for hadronic matter~\cite{Rupak:2012wk} and may not be sufficient even for extremely
favourable assumptions about this contribution~\cite{Alford:2013pma}.
Appearance of various condensates and strange particles like hyperons could 
enhance the viscosity of the hadronic phase. This is a very active field of 
research~\cite{Lindblom:1999wi,Lindblom:2002,Haensel:2000vz,Haensel:2001,Haensel:2002,
Shternin:2008,Haskell:2010,Manuel:2012rd,Colucci:2013sra}. 

At some high enough density we expect that a description based on deconfined 
quarks (and gluons) is a better description for dense matter than a description 
in terms of hadrons (though it is hard to say {\it{how}} high with existing
techniques)~\cite{Jaikumar:2006rh} and hence it is worthwhile if the transport
properties of quark matter are consistent with observations. 

Viscosities of unpaired quark matter have been extensively analyzed in
the literature~\cite{Madsen:1992,Heiselberg:1993}. They are dominated by excitations of
quarks near their Fermi surfaces and are efficient due to the large density of
states of low energy excitations. Models of neutron stars featuring a core of
unpaired quarks~\cite{Jaikumar:2008kh,Alford:2013pma} are consistent with the
observations of their rotation frequencies. (Interactions between quarks might
play an important role~\cite{Schwenzer:2012ga} in this agreement.) Similarly,
neutrino emission in unpaired quark matter would lead to ``fast'' cooling of
neutron stars~\cite{Iwamoto:1980,Iwamoto:1982}. 

However, quarks in the cores of neutron stars are likely to be in a paired
phase (see Refs.~\cite{Rajagopal:2000wf,Alford:2000sx,Alford:2007xm} for reviews).
Pairing affects the spectrum of quasi-particles and can change the transport
properties qualitatively. For example, at asymptotically high density, quark
matter exists in the Color Flavor Locked ($\CFL$) phase~\cite{Alford:1998mk}.
All the fermionic excitations in this phase are gapped and transport is carried
out by Goldstone modes. The shear viscosity of the Goldstone mode associated
with ${\rm{U}}(1)_B$ breaking was calculated
in~\cite{Manuel:2004iv,Mannarelli:2008je}. A star made only of CFL matter is
not consistent with the observed rotational
frequencies~\cite{Manuel:2004iv,Rupak:2010qg}, but in a star featuring a core
of CFL surrounded by hadronic matter (hybrid neutron star) some mechanism
involving dynamics at the interface (analogous to the one discussed in
Ref.~\cite{Alford:2015gna}) might be able to saturate $r$-mode amplitudes at a
level consistent with observations.

At intermediate densities, the nature of the pairing pattern of quark matter is not
known~\cite{Alford:2007xm}. We review some of the candidate phases below
(Sec.~\ref{sec:review}). One
exciting possibility is that quarks form a crystalline
color superconductor~\cite{Alford:2000ze}. (See Ref.~\cite{Anglani:2013gfu} for a recent
review.) These phases are well motivated ground states for quark matter at
intermediate densities~\cite{Rajagopal:2006ig,Ippolito:2007uz} although their
analysis is challenging because the condensate is position 
dependent~\cite{Cao:2015rea}. Unlike the $\CFL$ phase, crystalline color superconductors feature gapless
fermionic excitations. Therefore, we expect transport properties of these phases to
resemble unpaired quark matter.

Neutrino emission in the crystalline color superconducting phases for the
simplest three-flavor condensate was computed in Ref.~\cite{Anglani:2006br}.
Stars featuring these phases in the core do indeed cool
rapidly~\cite{Anglani:2006br,Hess:2011} and this rules out the presence of
these phases in several neutron stars which have been observed to cool
slowly~\cite{Yakovlev:2000jp,Yakovlev:2003}. It is possible that these stars have a
smaller central density (because they are lighter) than the fast rotating
neutron stars for which observations are consistent with a phase with a large 
viscosity to damp $r$-modes. Such interesting questions can be answered by more observations and
microscopic calculations of the transport properties of various phases of quark
matter.

In this paper we present the first calculation of the shear viscosity in the
simplest member in the family of the crystalline color superconducting phases:
the two-flavor Fulde-Ferrel (FF)~\cite{fulde1964superconductivity} phase. The
shear viscosity depends on the spectrum of the low energy modes as well
as their strong interactions. Hence it is different from the neutrino
emissivity where the strong interactions between quasi-particles do not play 
a role.  

In the two-flavor FF phase, (just like the isotropic $2\SC$
phase~\cite{Alford:1997zt,Rapp:1997zu}, reviewed
below) the ``blue'' ($b$) colored up ($u$) and down ($d$) quarks do not
participate in pairing. Their transport properties were analyzed in
Ref.~\cite{Alford:2014doa}. But because of the presence of gapless modes,
(unlike the $2\SC$ phase), the ``red'' ($r$) and ``green'' ($g$) colored $u$
and $d$ quarks also contribute to the viscosity.

We argue that $ur-dg-ug-dr$ quarks scatter dominantly via exchange of
transverse $t^1$, $t^2$, and $t^3$ gluons (for details see
Sec.~\ref{sec:interactions}). These gluons are Landau damped (in the $2\SC$
phase the longitudinal and transverse $t^1$, $t^2$, and $t^3$ gluons are 
neither screened nor Landau damped~\cite{Alford:1999pb,Rischke:2000qz,
Rischke:2000cn,Rischke:2002rz}). The polarization tensor of the $t^1$, $t^2$,
and $t^3$ gluons are anisotropic.

Therefore, both the quasi-particle dispersions and their interactions are
anisotropic, and the usual techniques to simplify the collision integral in the
Boltzmann equation~\cite{Heiselberg:1993} are not applicable, making its
evaluation challenging. Furthermore, the Boltzamann analysis needs to be
modified to accommodate the fact that the excitations are Bogoliubov
quasi-particles. To address this we find it convenient to separate the
modes in the two Bogoliubov branches (Eq.~\ref{eq:E1E2polar}) into modes
(Eq.~\ref{eq:fourspecies}) corresponding to momenta ($|\bfp|$) greater than the chemical
potential ($\mu$) (in the absence of pairing these are associated
with particle states) and $|\bfp|<\mu$ (in the absence of pairing these are
associated with hole states). 

Quasi-particle modes near the gapless surface dominate transport, but the shape of the surface
of gapless modes in the FF phase is non-trivial. In addition, the momentum
transferred between the quasi-particles can be large and a small momentum
expansion can not be always made. Therefore we evaluate the collision integral
(Eq.~\ref{eq:master}) numerically. 
 
The main result of the computation is given in Fig.~\ref{fig:Leta_vs_T_aniso} 
and Eq.~\ref{eq:Leta_vs_T_aniso}. The central conclusion is that the viscosity
of the $ur-dg-ug-dr$ quarks is reduced compared to their contribution in
unpaired quark matter by a factor of roughly $100$. The detailed analyses and
the dependence on the shear viscosity on $T$ and the splitting
between the Fermi surfaces $\delta\mu$, are shown in Sec.~\ref{sec:t123}.

The reduction of the viscosity by a large factor depends on the properties of
the mediators between the quasi-particles. For example, if we use Debye
screened longitudinal gluons (this is appropriate for one flavor FF pairing
and is also a good model for condensed matter systems like the FF phase in cold
atoms) then the viscosity of the paired fermions is the unchanged from its
value in the absence of pairing: the geometric factors associated with the
reduced area of the Fermi surface cancel out. (See
Sec.~\ref{sec:simple_anisotropic} for details.) We give an intuitive argument
to clarify the difference between long ranged and short ranged interactions.
These results (Sec.~\ref{sec:simple_anisotropic}), though not directly relevant for the two-flavor FF phase,
provide intuition for the three-flavor crystalline phases where both
longitudinal and transverse gluons are screened, and may also be relevant for
condensed matter systems where transverse gauge bosons don't play a role.   

To understand some aspects of the numerical results obtained for the FF phase
(Sec.~\ref{sec:anisotropic}) we use our formalism to calculate the viscosity in
isotropically paired systems with Fermi surface splitting in
Sec.~\ref{sec:simple}. For these systems it is possible to compare the
numerical results with simple analytic expressions in certain limits.  The
results of Sec.~\ref{sec:simple} --- the shear viscosity of fermions
participating in isotropic pairing and interacting via a simple model
interaction (the exchange of Debye screened longitudinal gluons) --- are not
novel, but clarify some physical aspects of the of the problem. For example we
study the role played by the scattering of paired fermions with phonons in
suppressing their transport that have not been highlighted before. While the
role played by phonon-fermion scattering is only of academic importance in the
extreme limits $\Delta\gg T$ and $\Delta\ll T$, it may be important in the
intermediate regime where $\Delta>T$ but not $\Delta\gg T$.

 
The plan of the paper is as follows. We quickly review the low energy
excitations in some relevant phases of quark matter in Sec.~\ref{sec:review} to
compare and contrast with the FF phase. In Sec.~\ref{formalism} we set up the
problem. The basic formalism is the multi-component Boltzmann transport
equation (Sec.~\ref{sec:Boltzmann}) which we solve in the relaxation time
approximation. We describe the low energy modes
(Secs.~\ref{sec:quarks},~\ref{sec:spectrum}) and their interactions
(Sec.~\ref{sec:interactions}). We also clarify the role played by phonons in
Sec.~\ref{sec:phonons}. In Sec.~\ref{sec:simple} we show results for isotropic
pairing. In Sec.~\ref{sec:anisotropic} we show results for the FF phase.  We
summarize the results and speculate about some implications for neutron star
phenomenology in Sec.~\ref{sec:conclusions}. A quick review of the gapless
fermionic modes in FF phases (Appendix~\ref{sec:blocking}) and the details
about the numerical implementation of the collision integrals
(Appendix~\ref{sec:numerics}) are given in the Appendix. 

\section{Review}
\label{sec:review}
We now review some proposed phases of quark matter in neutron stars.  We
discuss the excitation spectra and the interactions between the quasi-particles
in the phases and this will help us in identifying the ingredients required in
setting up the Boltzmann transport equation for the crystalline phase. Experts
in the field can skip to the end of the section and start from
Sec.~\ref{sec:Boltzmann}.

In the absence of attractive interactions, fermions at a finite chemical
potential $\mu$ and a temperature much smaller than $\mu$ are expected to
form a Fermi gas, filling up energy levels up to the Fermi sphere.

For massless weakly coupled quarks in the absence of pairing, the excitation spectrum is simply
\begin{equation}
E=|\xi|=||{\bf{p}}|-\mu|
\end{equation}
where $\xi=|{\bf{p}}|-\mu$ is the radial displacement of the momentum vector
from the Fermi surface. The excitations at the Fermi surface (defined by
$\xi=0$) are gapless, can be excited thermally, and therefore fermions near the
Fermi surface are very efficient at transporting momentum and charge. They
exhibit ``fast'' neutrino cooling and sufficiently large viscosities to damp
$r$-modes.

The interactions between the quarks are mediated by gluons (eight gluons
corresponding to the generators $t^1,\dots t^8$~\footnote{We use the standard
notation for the Gell-Mann matrices~\cite{Peskin:257493} as the generators of the color $\SU(3)$.}) and the photon. In
the absence of pairing, the longitudinal components of these mediators are
Debye screened~\cite{Madsen:1992}. The transverse components of the mediators (magnetic
components) are unscreened in the presence of static fluctuations of the
current, and are only dynamically screened (Landau damping). Consequently, they
have a longer range compared to the longitudinal gauge bosons and dominate 
scattering in relativistic systems~\cite{Heiselberg:1993}.

Pairing, induced by the attractive color interaction between the quarks,
qualitatively affects the transport properties of quark matter. 

At asymptotically high densities (corresponding to a quark number chemical
potential $\mu$ sufficiently larger than the strange quark mass), the
strange quark mass can be ignored, and the lagrangian is symmetric under
${\rm{SU}}(3)$ transformations between the up ($u$ or $1$), down ($d$ or $2$)
and ($s$ or $3$) quarks. They can all be treated as massless and form Cooper
pairs in a pattern that locks the color and flavor symmetries (CFL
phase)~\cite{Alford:1998mk} 
\begin{equation}
\begin{split}
\langle\psi_{cfsL}(r)\psi_{c'f's'L}(r)\rangle &= \sum_I \Theta 
\epsilon_{Icc'}\epsilon_{Iff'}\epsilon_{ss'}\\
\langle\psi_{cfR}^{\sdot}(r)\psi_{c'f'R}^{\sdot'}(r)\rangle &= -\sum_I \Theta
\epsilon_{Icc'}\epsilon_{Iff'}\epsilon^{\sdot\sdot'}\;.~\label{eq:CFLcondensate}
\end{split}
\end{equation}
$s, s'$ are the Weyl spinor indices, $f$ are flavor indices that run from $1$
to $3$. $c,c'$ are color labels that run over $1$ (colloquially red or $r$),
$2$ (green or $g$), and $3$ (blue or $b$). The left handed quarks ($L$) and the
right handed quarks ($R$) pair among themselves and can be treated
independently. The condensate is translationally invariant, which corresponds to
pairing between quarks of opposite momenta. 

The $\SU(3)$ color symmetries and the global $\SU(3)_L$ and $\SU(3)_R$ flavor
symmetries are broken by the condensate to a global subgroup consisting of
simultaneous color and flavor transformations,
\begin{equation}
\begin{split}
\SU(3)_c\times\SU(3)_L\times\SU(3)_R\times {\rm{U}}(1)_B
\rightarrow \SU(3)_{c+L+R}Z_2\label{eq:CFLsymmetry}\;.
\end{split}
\end{equation}
A diagonal subgroup of the $\SU(3)_L\times\SU(3)_R$ is weakly
gauged by the electric charge $Qe$, where $Qe$ is a diagonal matrix in the
flavor space with entries equalling the electric charges of the $u$, $d$
and $s$ quarks, and a linear combination of the $t^8$ and $Q$ (known as
$\tilde{Q}$) is unbroken~\cite{Alford:1998mk,Alford:1999pb}.

The fermionic excitations are $9$ Bogoliubov quasi-particles~\cite{Alford:1998mk} (for each
hand) which are all gapped. In the NJL model~\cite{nambu1961dynamical}, the condensate
$\Theta$ is related to the gap in the excitation spectrum,
$\Delta_{0{\rm{CFL}}}$ as follows~\cite{Alford:1998mk},
\begin{equation}
\begin{split}
\Delta_{0{\rm{CFL}}} = \lambda \Theta\;,
\end{split}
\end{equation}
where $\lambda$ is a measure of the interaction strength between quarks (the
condensate $\Theta$, as well as $\Delta_{0{\rm{CFL}}}$ depend on $\mu$, but we
are not explicitly writing the dependence here.) 

Using the BCS theory one can show that eight fermionic quasi-particles in
the CFL phase have excitation energies~\cite{Alford:1999pa,Alford:1999xc,Casalbuoni:2000na}
\begin{equation}
E=\sqrt{\xi^2+\Delta_{0{\rm{CFL}}}^2}
\end{equation}
and another branch of quasi-particles have (approximately) the spectrum of
excitation
\begin{equation}
E=\sqrt{\xi^2+4\Delta_{0{\rm{CFL}}}^2}\;.
\end{equation}
$\Delta_{0{\rm{CFL}}}$ is expected to be of the order of a few $10$s 
of MeV while the temperatures of the neutron stars of interest is at most a few
keV, and therefore the quarks do not participate in transport. 

Pairing also qualitatively modifies the propagation of the gauge fields. The
Debye screening of the longitudinal gauge bosons is proportional to the
susceptibility of the free energy to changes in the color gauge potential and
therefore is largely unaffected if $(\Delta_{0{\rm{CFL}}}/\mu)^2\ll1$ (as we
shall assume). But pairing generates a Meissner mass for the transverse gluons.
In the limit $e\rightarrow 0$ all the eight gluons have equal Meissner
masses. Turning on the weak electromagnetic interaction, ($e\ll g$ where $g$ is
the strong coupling)~\cite{Alford:1998mk,Rischke:2000ra,Casalbuoni:2000na}
leads to a mixing between the transverse gauge fields and a linear combination
of the gauge fields associated with the $\tilde{Q}$ charge does not develop a
Meissner mass while the orthogonal combination has a Meissner mass
approximately equal to that of the other gluons.

Since the fermions are all gapped, the low energy theory consists of the
Goldstone modes (``phonons'') associated with the broken global
symmetries~\cite{Alford:1998mk,Son:1999cm,Son:2000tu,Casalbuoni:1999zi,Rho:1999xf,Hong:1999ei,Manuel:2000wm,Rho:2000ww}.
While the phonon viscosity~\cite{Manuel:2004iv} formally diverges at small $T$,
what this really means is that the hydrodynamic approximation breaks down at a
temperature small enough that mean free path becomes equal to the size of the
neutron star (or vortex separation~\cite{Mannarelli:2008je}). Flow on smaller
length scales is dissipationless, and the $r$-modes can not be efficiently damped at very
small temperatures.  The conclusion from the discussion of the unpaired and the
CFL phase of quark matter is that the phenomenology of $r$-mode damping
suggests that phases featuring gapless fermionic excitations might be
consistent with the data.

Even at the highest densities expected in neutron stars~\cite{page2006dense},
the strange quark mass can not be ignored. The finite strange quark mass
stresses~\cite{Rajagopal:2005dg} the cross species pairing
(Eq.~\ref{eq:CFLcondensate}) in the CFL phase. 

To understand the origin of this stress, note that in the absence of pairing,
the Fermi surfaces of the quarks in neutral quark matter in weak equilibrium
Refs.~\cite{Alford:1999pa,Alford:2000ze} are given~\cite{Bowers:2003ye} by
\begin{equation}
\begin{split}
p_F^d = p_F^u + \frac{m_s^2}{4\mu},\;\;\;\;
p_F^s = p_F^u + \frac{m_s^2}{4\mu} ~\label{eq:pFs}
\end{split}
\end{equation}
implying in particular that the splitting between the $u-d$ and the $d-s$ Fermi 
surfaces 
\begin{equation}
2\delta p_F=\frac{m_s^2}{4\mu}\;.~\label{eq:dmu}
\end{equation}

On the other hand pairing between fermions of opposite momenta
(Eq.~\ref{eq:CFLcondensate}) is strongest if the pairing species have equal
Fermi momenta. This argument suggests that when $\delta p_F\sim\Delta_{0{\rm{CFL}}}$, the
symmetric pairing pattern in Eq.~\ref{eq:CFLcondensate} may get disrupted. A
detailed analysis~\cite{Alford:2005} bears out this intuition. For
${m_s^2}/{\mu}>2\Delta_{0{\rm{CFL}}}$, a condensate with unequal pairing
strengths between various species has a lower free energy than the condensate
in Eq.~\ref{eq:CFLcondensate}~\footnote{Other ways by which the $\CFL$ phase
can respond to the stress on pairing include the formation of $K^0$ condensates
($\CFL-K^0$)~\cite{Bedaque:2001je,Schafer:2002yy,Buballa:2004sx,Forbes:2005jya,Warringa:2006dk}
and $K^0$ condensates with a current
(curr$\CFL-K^0$)~\cite{Forbes:2005jya,Kryjevski:2004kt,Gerhold:2006np,Kryjevski:2008zz}. The bulk viscosity in
the $\CFL-K^0$ phase was calculated in Refs.~\cite{Alford:2007rw,Alford:2008pb}. In the absence of
additional damping mechanisms, the viscosity of $\CFL-K^0$ appears to be
insufficient to damp $r$-modes~\cite{Rupak:2010qg}.}.
\begin{equation}
\begin{split}
\langle\psi_{cfsL}(r)\psi_{c'f's'L}(r)\rangle = \sum_I \Theta_I 
\epsilon_{Icc'}\epsilon_{Iff'}\epsilon_{ss'}\;.~\label{eq:gCFLcondensate}
\end{split}
\end{equation}
The pairing between the $s$ and the $d$ quarks is the weakest because the
splitting between their Fermi surfaces is the largest
(Eq.\ref{eq:pFs}).  The $s$ and the $u$ pairing is also reduced, while the $u-d$
pairing is not significantly affected~\cite{Alford:2005}. 

The resultant phase has a remarkable property that certain fermionic
excitations are gapless~\cite{Alford:1999xc}. To see how this behavior arises,
note that if two fermions $i$ and $j$ with a chemical potential difference
$|\mu_i-\mu_j|=2\delta\mu$, form Cooper pairs with a gap parameter $\Delta$
($\Delta=\lambda\Theta$ is not the gap in the excitation spectrum for finite
$\delta\mu$), the Bogoliubov quasi-particles have eigen-energies
\begin{equation}
E_{\pm} = \delta\mu \pm \sqrt{\xi^2+\Delta^2}~\label{eq:SplitDispersion}\;.
\end{equation}
For $\Delta<\delta\mu$, the set of gapless fermions lie on the surface 
\begin{equation}
|\xi| = \pm \sqrt{\delta\mu^2-\Delta^2}~\label{eq:GaplessXi}\;.
\end{equation}

The gapless CFL phase was found to be
unstable in
Refs.~\cite{Casalbuoni:2004tb,Fukushima:2005cm,Alford:2005qw}. The Meissner mass squared of some of the gluons is
negative in this phase. (This {\it{chromomagnetic instability}} was found
earlier in the $2\SC$ phase~\cite{Huang:2004bg,Huang:2004am} that we discuss below.) This
instability can be seen as an instability towards the formation of a position
dependent condensate~\cite{Giannakis:2004pf,Fukushima:2006su}, which bear resemblance to the
LOFF (Larkin, Ovcinnikov, Fulde, Ferrell)
phases~\cite{fulde1964superconductivity,larkin1964nonuniform} previously
considered in condensed matter systems.  (It has been argued in
Refs.~\cite{Gorbar:2005rx,Kiriyama:2006ui} that the chromomagnetic instability
might instead lead to a condensation of gluons, a possibility we won't explore
further here.) We review the LOFF phases in
Sec.~\ref{loff}, and the possibility that LOFF phases are the ground state of
baryonic matter in the cores of neutron stars motivates the analysis of
transport in FF phases, which is the prime objective of present manuscript. 

Restricting, for the moment, to spatially homogeneous and isotropic condensates,
another possibility that has been considered in detail in the
literature~\cite{Alford:1997zt,Rapp:1997zu} is one where the stress due to the
$s$ quark mass lead to the $s$ quarks dropping out from pairing. The $u$ and
$d$ quarks form a two-flavor, two color condensate ($2{\rm{SC}}$ pairing)
\begin{equation}
\begin{split}
\langle\psi_{cfs} \psi_{c'f's'}\rangle = \Theta_3  
\epsilon_{3cc'}\epsilon_{3ff'}\epsilon_{ss'}~\label{eq:2SCcondensate}\;.
\end{split}
\end{equation} 
The $u-b$, the $d-b$ are also unpaired, while the $ur$ quarks pair with the
$dg$ quarks and the $ug$ with the $dr$.

Taking, for the moment, equal $u$ and $d$ Fermi surfaces, $2{\rm{SC}}$ pairing
(Eq.~\ref{eq:2SCcondensate}) leaves a ${\rm{SU}}(2)$ sub group of color
unbroken. The symmetry breaking pattern is 
\begin{equation}
\begin{split}
\SU(3)_c\times\SU(2)_L\times\SU(2)_R&\times {\rm{U}}(1)_B
\rightarrow\\ \SU(2)_{(r-g)}\times\SU(2)_{L}\times&\SU(2)_{R}
\times\U(1)_{\tilde{B}}\label{eq:2SCsymmetry}\;,
\end{split}
\end{equation}

Since the $\SU(2)$ transformations associated with $r-g$ quarks are unbroken,
the $t^1, t^2, t^3$ gluons do not pick up a Meissner
mass~\cite{Alford:1999pb,Rischke:2000qz}. As we shall see, because of this, the
$t^1, t^2, t^3$ gluons play a special role in the two-flavor FF phase that we
consider.

If the strange quark mass is large enough that their contribution to the
thermodynamics can be ignored, the longitudinal components of the $t^1, t^2,
t^3$ gluons remain un-screened~\cite{Rischke:2000qz,Rischke:2002rz}. This can
be intuitively understood as follows. Debye screening (at low $T$) requires the presence of
ungapped excitations (here ungapped fermions) that can couple with the relevant
gauge field. Here, the $r$ and $g$ quarks of both $u$ and $d$ quarks are gapped
(with a gap $\Delta_{\rm{2SC}0}$) due to pairing. Furthermore, the condensate is
also neutral under the $t^1, t^2, t^3$ gluons. Therefore, both longitudinal and
transverse gluons $t^1, t^2, t^3$ can mediate long range interactions between
quarks, and give rise to confinement on an energy scale much smaller than
$\Lambda_{\rm{QCD}}$~\cite{Rischke:2000cn}. The color transformations
corresponding to the $t^4\dots t^7$ generators and the associated transverse
gauge fields do develop a Meissner mass. The longitudinal components of the
$t^4\dots t^7$ gauge fields are Debye
screened~\cite{Rischke:2000qz,Rischke:2002rz}. Similarly, the $t^8$ gluons
feature Meissner and Debye screening~\cite{Rischke:2000qz,Rischke:2002rz}. As
in the case of the CFL phase, the transverse components of a linear combination
of $t^8$ and $Q$ gauge bosons ($\tilde{Q}$ photon) have $0$ Meissner mass.
Electrical neutrality is maintained by electrons. Finally, since no global
symmetries are broken by the condensate, there are no Goldstone modes. 

The low energy dynamics are therefore dominated by the unpaired $u-b$ and $d-b$
quarks interacting predominantly via the $\tilde{Q}$ photon and the electrons
interacting via the photons. Transport in this phase has been analyzed in
detail in Ref.~\cite{Alford:2014doa}. The bulk viscosity for the $2\SC$ phase
was computed in Ref.~\cite{Alford:2006gy}.  Since the $b$ quarks are
unpaired, the transport properties in this mode are similar to that in unpaired
quark matter and hence we expect that viscosities should be large enough to damp
$r$-mode instabilities if a large volume of $2\SC$ matter is present in the cores
of neutron stars.

For weak and intermediate coupling
strengths~\cite{Abuki:2005ms,Ruester:2005jc}, the $2\SC$ phase has a smaller
free energy compared to the CFL phase and unpaired quark matter, only for
temperatures larger than a few
MeV~\cite{Alford:2002kj,Steiner:2002gx,Abuki:2005ms,Ruester:2005jc}. For large
couplings~\cite{Abuki:2005ms,Ruester:2005jc} however, it is favoured over the
$\CFL$ and the unpaired phase over a range of chemical potentials expected to
be present in some region in the cores of neutron stars ($350$ to $400$MeV) and
the occurrence of a $2\SC$ phase may provide a plausible mechanism for the
damping of $r$-modes. It is however worth exploring other compelling
possibilities, viable in particular for intermediate and weak coupling.

As in the case of three-flavor pairing, the requirements of neutrality and weak
equilibrium tend to split the $u-d$ Fermi surfaces~\cite{Shovkovy:2003uu} and
impose a stress on pairing. For large enough
stress ($\delta\mu\sim\Delta_{\rm{2SC}0}$), the $u$ and $d$ quarks that participate in
pairing exhibit gapless excitations as suggested by
Eqs.~\ref{eq:GaplessXi} and~\ref{eq:2SCcondensate}~\cite{Shovkovy:2003uu,Huang:2003xd}. 

The low energy theory of the gapless $2\SC$ phase features the unpaired $b$
quarks near the Fermi surface, as well as Bogoliubov quasi-particles (linear
combinations of $ur-dg-ug-dr$ quarks and holes) near the gapless spheres
(Eq.~\ref{eq:GaplessXi}). The gapless quasi-particles interact via the $t^1,
t^2, t^3$, and $t^8$ gluons and the photon with each other. They can also
exchange $t^4\dots t^7$ gluons to change to $b$ quarks.  In terms of the
participants in the low energy theory, this phase resembles the two-flavor FF
phase that we shall study in detail in the paper.

The transverse $t^1, t^2, t^3$ gluons remain massless since the
$\SU(2)_{c(r-g)}$ is unbroken in the gapless phase. (The global
$\SU(2)_L\times\SU(2)_R$ in Eq.~\ref{eq:2SCsymmetry} are no longer relevant
because it is broken by $\delta\mu$.) The presence of gapless excitations
generates a Debye screening mass~\cite{Huang:2004bg,Huang:2004am} for the
longitudinal modes of all the gauge fields. 

However, like three-flavor pairing, the $2{\rm{SC}}$ phase with gapless Bogoliubov excitations is
unstable~\cite{Huang:2004bg} since the Meissner mass squared of a linear
combination of photon and the $t^8$ gluon (orthogonal to the $\tilde{Q}$
photon) becomes negative for $\delta\mu>\Delta_{2\SC0}$. In addition, the mass
squared of $t^4, t^5, t^6$, and the $t^7$ gluons become negative for
$\delta\mu>\Delta_{2\SC0}/\sqrt{2}$~\cite{Huang:2004bg}. As in three-flavor
case, this instability can be seen as an instability towards the formation for
a LOFF phase.

Finally, it is possible that the stress due to $\delta\mu$ disrupts the 
inter species pairing altogether and leads to the formation of Cooper pairs of a
single flavor~\cite{Alford:2002rz,Schafer:2000cj,Buballa:2002wy,Schmitt:2004et}.  
If the pairing is weaker than keV scale, then for hotter neutron stars it
will be irrelevant and results found for unpaired quark matter shall apply. For
stronger pairing, only  few transport properties of these single flavor states have
been studied (see Ref.~\cite{Schmitt:2005wg} for the calculation of neutrino
emissivity and Ref.~\cite{Schmitt:2003xq} for electronic properties.) 
and it will be interesting to calculate their viscosities. Some of
the phases feature gapless fermionic modes and would be expected to behave
similarly to unpaired quark matter, though more detailed analyses would be
interesting.

The two points we want to take away from this brief review are (a) the
analyses of $r$-mode damping suggest that if a quark matter core damps the
$r$-modes,  then it features gapless fermionic excitations (b) at neutron
star densities for a range of parameters, uniform and isotropic pairing phases are
unstable towards the formation of a position dependent condensate. We now 
review the salient features of phases with such pairing.

\subsection{LOFF phase}
\label{loff}
A natural candidate for a position dependent pairing condensate is the LOFF
phase which was proposed as the plausible ground state for stressed quark
matter~\cite{Alford:2000ze,Kundu:2001tt} before the discovery of the
chromomagnetic instabilities. The motivation for this proposal is that a
condensate of the form,
\begin{equation}
\begin{split}
\langle\psi_{i}(r)\psi_{j}(r)\rangle &=  \Theta
e^{2i\bfb\cdot\bfr}\;,~\label{eq:pwcondensate}
\end{split}
\end{equation}
allows pairing along rings on split Fermi surfaces for
$b=|\bfb|>\delta\mu$~\cite{Alford:2000ze}~\footnote{The real number $b$ refers
to $|\bfb|$ which is different from the ``blue'' colored quark. $b$ as an index
in the set $\{a$, $b$, $c$, $d\}$ refers to the branch of the dispersion as we
discuss below. We apologize for the degeneracy in notation but the contexts
are quite different and hence unlikely to cause confusion.} ($b,
\delta\mu$ and $\Delta$ are all taken
to be much smaller than $\mu$). $\bfb$ defines the wave-vector for the periodic
variation of the condensate.

In the NJL model, the phase with condensate Eq.~\ref{eq:pwcondensate} is preferred over
unpaired matter as well as the space independent condensate for
$\delta\mu\in[0.707\Delta_0,0.754\Delta_0]$~\cite{Alford:2000ze}. At the
upper end, the  transition from the crystalline phase to the normal phase is
second order as we increase $\delta\mu$, and $\Theta\rightarrow 0$ smoothly as
$\delta\mu\rightarrow0.754\Delta_0$ from the left. The
crystalline phase is favoured over normal matter for $\delta\mu<0.754\Delta_0$,
where $\Delta_0$ is the two-flavor gap for $\delta\mu=0$. The most favoured
momentum $b$ near $\delta\mu=0.754\Delta_0$ is
\begin{equation}
b=\zeta\delta\mu\;,~\label{eq:eta}
\end{equation}
with $\zeta\approx1.1996786...$~\cite{larkin1964nonuniform,fulde1964superconductivity,
Bowers:2002xr,Bowers:2003ye,Mannarelli:2006}.  (This number is
conventionally called $\eta$ in the literature but in this manuscript we give
it a different symbol to avoid confusion with the viscosity $\eta$.) The
homogeneous phase with pairing parameter $\Delta_0$ is favoured for
$\delta\mu<0.707$. (For single gluon exchange the window of favorability is
larger~\cite{Leibovich:2001xr}.) 

Intuitively one expects~\cite{Bowers:2002xr} that condensates featuring
multiple plane waves 
\begin{equation}
\begin{split}
\langle\psi_{i}(r)\psi_{j}(r)\rangle &=  \Theta \sum_{m}
e^{2i\bfb_m\cdot\bfr}\;,~\label{eq:LOcondensate}
\end{split}
\end{equation}
can pair quarks along multiple rings and give a stronger
Free energy benefit as long at the pairing rings do not overlap. The
set of plane waves $\{\bfb_m\}$ define a crystal structure. A detailed
calculation~\cite{Bowers:2002xr} till the $6$th order in the pairing parameter in
the Ginsburg-Landau approximation confirms this. A more recent sophisticated
numerical analyses reveals~\cite{Cao:2015rea} that higher order terms are 
important for determining the favoured crystal structure, and may predict
different favoured crystal structures than what the Ginsburg-Landau analysis
predicts.

For the three-flavor problem, the form of the LOFF
condensate~\cite{Casalbuoni:2005zp,Mannarelli:2006,Rajagopal:2006ig} is
\begin{equation}
\begin{split}
\langle\psi_{cfs}(r)\psi_{c'f's'}(r)\rangle = 
\sum_I \sum_{\{\bfq_m\}^I} \Theta_I
e^{2i\bfq_m\cdot\bfr} 
\epsilon_{Icc'}\epsilon_{Iff'}\epsilon_{ss'}.~\label{eq:3SCLOFFcondensate}
\end{split}
\end{equation}
Within the Ginzburg-Landau approximation~\cite{Rajagopal:2006ig},
condensates of the form Eq.~\ref{eq:3SCLOFFcondensate}
for two crystalline phases have a lower free energy than unpaired quark 
matter as well as homogeneous pairing phases over a wide range of parameters of
$\mu$, $\Delta$ and $m_s$ that are expected to exist in neutron star cores
~\cite{Ippolito:2007uz}.

Therefore it is natural to evaluate its transport properties and test whether
they are consistent with existing and future observations. As mentioned above,
neutrino emissivity for a three-flavor LOFF phase with the simplest three
flavor crystal structure were computed in Ref.~\cite{Anglani:2006br}. 

In this paper we take the first step in the calculation of the shear viscosity
in crystalline color superconductors. To simplify the calculations we ignore
the $s$ quarks completely and consider phases with a single plane wave
condensate,
\begin{equation}
\begin{split}
\langle\psi_{cfs}(r)\psi_{c'f's'}(r)\rangle = \Theta_3 e^{2i\bfb\cdot\bfr}  
\epsilon_{3cc'}\epsilon_{3ff'}\epsilon_{ss'}\;,~\label{eq:2SCFFcondensate}
\end{split}
\end{equation}
which corresponds to taking $\Theta_1=\Theta_2=0$ in
Eq.~\ref{eq:3SCLOFFcondensate}, as well as
limiting the set of momentum vectors $\{\bfb_{m}\}^3$ to just one vector $\bfb$. This 
is also known as the Fulde-Ferrel (FF) state.  

Eq.~\ref{eq:2SCFFcondensate} models FF pairing between $u$ and $d$ quarks with
Fermi surfaces split by $2\delta\mu=\mu_d-\mu_u$ [which can be thought of as
the measure of the strange quark mass $\delta\mu\sim
m_s^2/(4\mu)$~\cite{Alford:1999pa} (Eq.~\ref{eq:pFs}) in
$2\SC+s$~\cite{Alford:2007xm} or the electron chemical potential $\delta\mu\sim
\mu_e/2$~\cite{Shovkovy:2003uu} in the $2\SC$ phase without $s$ quarks]. This
simplifies the calculations significantly since the dispersions of the
fermions~\cite{Alford:2000ze} in the FF state have a compact analytic form
(Eq.~\ref{eq:E1E2}). We shall see that even with these approximations, the
calculation of the viscosity contributions of the $ur-ug-dr-dg$ quarks is
non-trivial because of pairing.

Eq.~\ref{eq:2SCFFcondensate} can be seen as denoting pairing between two Fermi
surfaces with radii $\mu\pm\delta\mu$ and centres displaced by $2\bfb$.
For $b>\delta\mu$, the two Fermi surfaces intersect. For $\Theta_3\rightarrow 0$
(true near the second order phase transition between the inhomogeneous
and unpaired phase), the pairing parameter is small and pairing can not occur when
either the $u$ or the $d$ momentum state is unoccupied~\cite{Alford:2000ze}. (See
Appendix~\ref{sec:blocking} for a quick reminder.) The boundary of these
``pairing regions'' feature gapless fermionic excitations. This suggests that
the contributions of the paired $ur-dg-ug-dr$ quarks is not very different from
their contributions in unpaired quark matter. 

However, the shapes of the gapless ``Fermi surfaces'' in LOFF pairing is quite
complicated, and their areas drop rapidly as $\Theta_3$ increases as we
decrease $\delta\mu$ from $0.754\Delta_{0}$. Therefore, it is not clear how
their contributions behave in the neutron star core. We answer this question in
this paper. In the following section we develop the formalism to calculate
shear viscosity coefficient in crystalline color superconducting phase.
 
\section{Formalism}
\label{formalism}
This section develops the theoretical aspect of calculation of transport
coefficients in the LOFF phase. We start our discussion with Boltzmann equation
in an anisotropic system.

\subsection{Boltzmann transport equation}
\label{sec:Boltzmann}
In a system of multiple species, the relaxation times $\tau_i$ for the species
$i$ can be found by solving a matrix equation,
\begin{equation}
L_i^{(n)}=\sum_j [R^{(n)}_{ij}]\tau^{(n)}_j\;,~\label{eq:matrix}
\end{equation}
where $L_i$ is related to the phase space of quasi-particles that participate
in transport, and $[R_{ij}]$ is the collision integral. We have labelled the
collisional integral with an additional index $(n)$ associated with the tensor
structure of the transport property we are considering.

To be concrete, consider a situation where transport is dominated by fermionic
particles and their interaction with each other provides the most important
scattering mechanism. Following the notation of Ref.~\cite{Alford:2014doa} the Boltzmann
transport equation for each species $i$ can be written as,
\begin{equation}
\begin{split}
L_i^{(n)} &= 
\frac{1}{\gamma^{(n)}} \int \frac{d^3p_i}{(2\pi)^3} \frac{df_0^i}{d\epsilon}
(\phi_i^{ab}\psi_i^{ab})\\
\sum_j [R^{(n)}_{ij}]\tau^{(n)}_j&=-\sum_{234} \frac{1}{\gamma^{(n)}}\frac{1}{T}\nu_2 \int\;
\dthree{p_i}\dthree{p_2}\dthree{p_3}\dthree{p_4}\\
&
|\calM(i2\rightarrow34)|^2\\
&
\deltafour
[f_if_2(1-f_3)(1-f_4)]\\
&3\phi_i.[\tau^{(n)}_i\psi_i^{(n)}
+\tau^{(n)}_2\psi_2^{(n)}
-\tau^{(n)}_3\psi_3^{(n)}
-\tau^{(n)}_4\psi_4^{(n)}]\\
\;,~\label{eq:master}
\end{split}
\end{equation}
where $f$ is the Fermi-Dirac distribution function. ${\cal{M}}(12\rightarrow
34)$ refers to the transition matrix element for the scattering of the initial
state featuring $1, 2$ (defined by momenta $p_1$, $p_2$ and additional quantum
numbers like spin, color and flavor) to the final state $2,4$. The sum over $2,
3, 4$ runs over all the species that interact with $i$.  

The form of the flows $\phi$ and $\psi$ in Eq.~\ref{eq:master}, relevant for
the calculation of the shear viscosity, are given by
\begin{equation}
\begin{split}
\phi_i^{ab}&=p^av^b\;\\
\psi_i^{ab}&=\Pi^{(n)ab\alpha\beta}\phi_i^{ab}, 
\end{split}
\end{equation}
where
\begin{equation}
v^a = \frac{d E}{dp^a}\;.
\end{equation}

$\Pi^{(n)ab\alpha\beta}$ are operators that project the shear
viscosity tensor into susbspaces, $(n)$, invariant under the rotational 
symmetries of the system. $\gamma^{n}$ defined
by
\begin{equation}
\begin{split}
\gamma^{(n)} =  \Pi^{(n)ab\alpha\beta}\delta^{a\alpha}\delta^{b\beta}
\end{split}
\end{equation}
are the dimensions of these subspaces.

For example, in an isotropic system, the shear viscosity tensor should be
invariant under all rotations, and the only projection operator is the
traceless symmetric tensor
\begin{equation}
\begin{split}
\Pi^{ab\alpha\beta} &= ( \frac{1}{2}\delta^{a\alpha}\delta^{b\beta} 
    + \frac{1}{2}\delta^{a\beta}\delta^{b\alpha}\\
    &- \frac{1}{3}\delta^{ab}\delta^{\alpha\beta})\;,
\end{split}
\end{equation}
with $\gamma=5$.

We will consider system where the condensate chooses a particular direction and
such systems have $5$ independent forms.  In particular, we will focus on $n=0$
for which
\begin{equation}
\begin{split}
\Pi^{(0)} &=(\frac{3}{2})(b_ab_b-\frac{1}{3}\delta_{ab})
(b_\alpha b_\beta-\frac{1}{3}\delta_{\alpha\beta})\\
\gamma^{(0)} &= 1\;.
\end{split}
\end{equation}

The contribution to the viscosity tensor for each species $i$ is given by
\begin{equation}
\eta_i^{ab\alpha\beta} = \sum_{(n)} \eta_i^{(n)} \Pi^{(n)ab\alpha\beta}\;,
\end{equation}
where
\begin{equation}
\begin{split}
\eta_i^{(n)} =
(-\frac{3}{2}\gamma^{(n)}\nu_i)[L_i^{(n)}]\tau_i^{(n)}~\label{eq:eta_ioftau_i}\;.
\end{split}
\end{equation}

To evaluate Eq.~\ref{eq:master} we need to identify the relevant species and
the interactions between them. We do these in turn in the next two sections.

\subsection{Quark species}
\label{sec:quarks}
We shall consider phases with a condensate of the form
\begin{equation}
\begin{split}
\langle\psi_{cfs}(r)\psi_{c'f's'}(r)\rangle = \Theta_3(r) 
\epsilon_{3cc'}\epsilon_{3ff'}\epsilon_{ss'}\;.~\label{eq:2SCLOFFcondensate2}
\end{split}
\end{equation}
We shall ignore the contribution of the $s$ quarks which, if present ($2\SC+s$
phase~\cite{Shovkovy:2003uu}), are unpaired. Only $ur-dg$ and $ug-dr$ quark pairs participate
in pairing. The $ub$ and $db$ ($b$ color) quarks as well as the electrons are
unpaired.  

Transport effected by the $ub$ and the $db$ quarks, as well as by the electrons
in the homogeneous and isotropic ${\rm{2SC}}$ phase has been studied in detail
in Ref.~\cite{Alford:2014doa}. Since they are unpaired, techniques from
condensed matter theory for calculating the transport in Fermi liquids can be
used to simplify the calculation, although there are new features associated
with the fact that the quarks are relativistic~\cite{Baym:1990} and due to the
non-trivial color and flavor structure of the
interaction~\cite{Alford:2014doa}. 

Here we want to focus on the effect of crystalline pairing on the quark
transport. In the full three-flavor theory with $\Theta_1, \Theta_2\neq0$, the
$ub$ and $db$ species as well as the strange quarks will participate in
crystalline pairing (Eq.~\ref{eq:3SCLOFFcondensate}). Therefore we need to
develop techniques to calculate fermionic transport properties in the presence
of a crystalline order parameter. In this paper, we shall limit ourselves to the
calculation of transport in the two-color two-flavor subsystem of $ur-dg-ug-dr$
quarks. Even in this two-color, two-flavor subspace, the theory of transport is
quite rich and we will learn valuable lessons that will help in future attempts
to extend the calculations to the three-flavor problem.

\begin{figure}[tbp]
\includegraphics[width=0.49\textwidth,clip=]{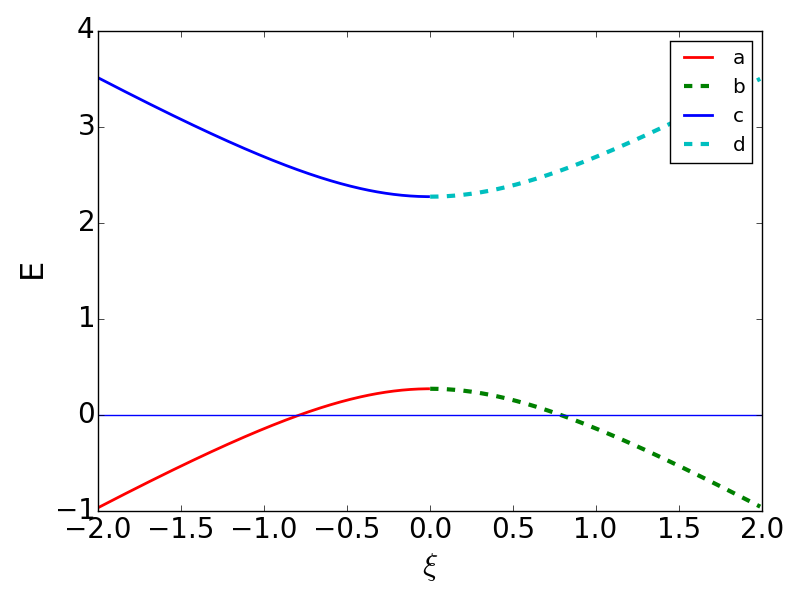}
  \caption{(color online) The four (Eq.~\ref{eq:fourspecies})
  branches [solid red ($a, \xi<0$), dashed
  green ($b, \xi>0$), solid blue ($c, \xi<0$), and dashed cyan ($d, \xi>0$)
  ] for $\mu=100$, $\Delta=1$MeV, $\delta\mu=1.4$MeV, $b=1.3$MeV, and
  $\cos\theta=-0.1$ (Eq.~\ref{eq:E1E2polar}). The gap between the lower and
  upper branches is $2\Delta$, and for $\Delta\ll T$ only excitations near
  $E=0$ participate in transport.~\label{fig:EBranches}
}
\end{figure}
\subsection{Spectrum of excitations}
\label{sec:spectrum}
The mean field lagrangian for $ur$, $ug$, $dr$, and $dg$ quarks
~\cite{Rajagopal:2000wf} quarks is given by
\begin{equation}
\begin{split}
{\calL} 
&= 
-2\frac{\Delta^*\Delta}{\lambda}+\\
&\frac{1}{2}\Psi^\dagger_{4L}\left(\begin{array}{cc}
   i\bar{\sigma}^{\mu}\partial_\mu + \mu_u 
   &-\Delta e^{2i\bfb\cdot\bfr}\\
  -\Delta^{*}e^{-2i\bfb\cdot\bfr}&i\sigma^{\mu}\partial_\mu - \mu_d 
  \end{array}
  \right)\Psi_{4L}+(L\rightarrow R)\;,~\label{eq:MFHelicity}
\end{split}
\end{equation}
where
\begin{equation}
\begin{split}
\Psi_{4L}(x) &= \left(
\begin{array}{c} 
  u_{rLa}(x) \\ 
  u_{gLa}(x) \\ 
  -\epsilon^{\dota\dotb}d^*_{gL\dotb}(x)\\ 
  +\epsilon^{\dota\dotb}d^*_{rL\dotb}(x)
\end{array}
\right)
\end{split}
\end{equation}
or more compactly as
\begin{equation}
\begin{split}
\Psi_{4L}(x) &= \left(
\begin{array}{c} 
  u_{L}(x) \\
  -[\epsilon^c]d^C_{L}(x)\\ 
\end{array}
\right)
\end{split}
\end{equation}
where, $[\epsilon^c]_{cc'} = \epsilon_{cc'}$ is the antisymmetric matrix in a two
dimensional sub-space of color.

For the $ur$, $dg$ quarks, it can be written as
\begin{equation}
\begin{split}
{\calL} 
&= 
-\frac{\Delta^*\Delta}{\lambda}+\\
&\frac{1}{2}\Psi^\dagger_L\left(\begin{array}{cc}
   i\bar{\sigma}^{\mu}\partial_\mu + \mu_u 
   &-\Delta e^{2i\bfb\cdot\bfr}\\
  -\Delta^{*}e^{-2i\bfb\cdot\bfr}&i\sigma^{\mu}\partial_\mu - \mu_d 
  \end{array}
  \right)\Psi_L+(L\rightarrow R)~\label{eq:MFHelicity_urdg}
\end{split}
\end{equation}
where~\cite{Rajagopal:2000wf},
\begin{equation}
\Delta=\lambda\Theta_3
\end{equation}
and the two dimensional Nambu-Gorkov spinors $\Psi$ are defined as
\begin{equation}
\begin{split}
\Psi_L(x) & = \left(
\begin{array}{c}u_{rLa}(x) \\ -\epsilon^{\dota\dotb}d^*_{gL\dotb}(x)\end{array}
\right)
= \left(\begin{array}{c}u_{rL}(x)\\ -d_{gL}^{C}(x)\end{array}\right)
\\
\Psi_L^\dagger(x) &= (u_{rL\dota}^*(x), -\epsilon^{ac}d_{gLc}(x))
= (u_{rL}^\dagger(x), -d_{gL}^{C\dagger}(x))\;,
\end{split}
\end{equation}
where $x=(t,\bfr)$.

Similarly, the Nambu-Gorkov spinor for $ug-dr$ 
\begin{equation}
\begin{split}
\Xi_L(x) & = \left(
\begin{array}{c}u_{gLa}(x) \\ \epsilon^{\dota\dotb}d^*_{rL\dotb}(x)\end{array}
\right)
= \left(\begin{array}{c}u_{gL}(x)\\ d_{rL}^{C}(x)\end{array}\right)
\\
\Xi_L^\dagger(x) &= (u_{gL\dota}^*(x), \epsilon^{ac}d_{rLc}(x))
= (u_{gL}^\dagger(x), d_{rL}^{C\dagger}(x))\;,
\end{split}
\end{equation}
has the same form as Eq.~\ref{eq:MFHelicity_urdg}. 

For $h=-1/2$, $\bfp\cdot\sigma=-p=-|{\bf{p}}|$, (This the correct helicity for $L$ handed
quarks. These are the ``large'' components in the Fourier decomposition of
the Dirac spinor~\cite{Mannarelli:2006}.) and the dispersion relation for the paired
quarks is obtained by diagonalising finding the energy eigenvalues of 
\begin{equation}
   \left(\begin{array}{cc}
   (E-|\bfp+\bfb|) + \mu_u  &-\Delta \\
  -\Delta^{*} & (E+|\bfp-\bfb|) - \mu_d
  \end{array}
  \right)\;.
\end{equation}

The eigenvalues and the eigenvectors are given by,
\begin{equation}
\begin{split}
E_1 =& (|\bfp+\bfb|-|\bfp-\bfb|+2\delta\mu)/2 \\
&- \sqrt{\xi^2+\Delta^2}\\
\Psi_1 = \left(\begin{array}{c}\Phi_{11}\\\Phi_{12}\end{array}
\right)
&=\left(\begin{array}{c}
 \frac{1}{\sqrt{2}}\sqrt{1-\frac{\xi}{\epsilon}}\\
 -e^{-i\phi}\frac{1}{\sqrt{2}}\sqrt{1+\frac{\xi}{\epsilon}}
 \end{array}
 \right)~\label{eq:E1Psi1}
\end{split}
\end{equation}
and,
\begin{equation}
\begin{split}
E_2 =& (|\bfp+\bfb|-|\bfp-\bfb|+2\delta\mu)/2 \\
&+ \sqrt{\xi^2+\Delta^2}\\
\Psi_2 = \left(\begin{array}{c}\Phi_{21}\\\Phi_{22}\end{array}
\right)
&=\left(\begin{array}{c}
 \frac{1}{\sqrt{2}}\sqrt{1+\frac{\xi}{\epsilon}}\\
 e^{-i\phi}\frac{1}{\sqrt{2}}\sqrt{1-\frac{\xi}{\epsilon}}
 \end{array}
 \right)~\label{eq:E2Psi2}
\end{split}
\end{equation}
$\xi=(|\bfp+\bfb|+|\bfp-\bfb|-2{\mu})/2$, $\epsilon=\sqrt{\xi^2+\Delta^2}$ and
$\phi$ is the phase of $\Delta$. $\mu=(\mu_u+\mu_d)/2$ is the mean of the
chemical potentials and $\mu_d-\mu_u=2\delta\mu$. 

The Bogoliubov coefficients can be arranged in a orthonormal matrix form,
\begin{equation}
\begin{split}
[\Phi] = \left(\begin{array}{cc}\Phi_{11} & \Phi_{21}\\
    \Phi_{12} & \Phi_{22}\end{array}
\right)
~\label{eq:ortho}
\end{split}
\end{equation}

We simplify the momentum integrals Eq.~\ref{eq:master} in
the limit $\mu\gg\delta\mu,\; b,\; \Delta$, and 
$\mu, \gg T$. Near the Fermi surface 
\begin{equation}
\begin{split}
\frac{d^3p}{(2\pi)^3} &= \frac{p^2dpd\Omega}{(2\pi)^3}\approx\mu^2d\xi
\frac{d\Omega}{(2\pi)^3}~\label{eq:FermiSurface}\;.
\end{split}
\end{equation} 

In this approximation  
\begin{equation}
\begin{split}
E_1(\bfp)&=\delta\mu + \bfb\cdot\bfv_F- \sqrt{\xi^2+\Delta^2}\\
E_2(\bfp)&=\delta\mu + \bfb\cdot\bfv_F+ \sqrt{\xi^2+\Delta^2}~\label{eq:E1E2}\;,
\end{split}
\end{equation}
or in polar coordinates with $\hat{b}=\hat{z}$,
\begin{equation}
\begin{split}
E_1(\xi, \theta)&=\delta\mu + b\cos\theta - \sqrt{\xi^2+\Delta^2}\\
E_2(\xi, \theta)&=\delta\mu + b\cos\theta + \sqrt{\xi^2+\Delta^2}
~\label{eq:E1E2polar}\;,
\end{split}
\end{equation}
where $\xi=p-\mu$ and $\bfv_F=(d\xi/dp)\hat{p}=\hat{p}$ is the Fermi velocity. 

The mode decomposition of $\Psi$ is
\begin{equation}
\begin{split}
\Psi_L(x) 
&=
\int \frac{d^4p}{(2\pi)^4}e^{-ip_\mu x^\mu}[\\
&\phantom{+}(2\pi)\delta(p^0-E_1) \left(
\begin{array}{c}\Phi_{11}e^{i\bfb\cdot\bfr}\xi_{-}(p) \\ 
\Phi_{12}e^{-i\bfb\cdot\bfr}\xi_{-}(p)\end{array}
\right)\gamma_L\\
&+(2\pi)\delta(p^0-E_2) \left(
\begin{array}{c}\Phi_{21}e^{i\bfb\cdot\bfr}\xi_{-}(p) \\ 
\Phi_{22}e^{-i\bfb\cdot\bfr}\xi_{-}(p)\end{array}
\right)\chi_L]~\label{eq:Bogoliubov}\;,
\end{split}
\end{equation}
where $\xi(-)$ is the two component spinor satisfying
\begin{equation}
\bfp\cdot\sigma \xi_{-}(p)= -p \xi_{-}(p)\;.
\end{equation}
Now, with the two energy eigenstates Eq.~\ref{eq:E1E2} in hand, it is tempting
to treat Eq.~\ref{eq:master} as a two species problem in the eigenstates
Eq.~\ref{eq:E1Psi1} and Eq.~\ref{eq:E2Psi2} which corresponds to an appropriate
linear combination of $u$ particles and $d$ holes (Eq.~\ref{eq:Bogoliubov}).

Since, 
$u$ and $d$ quarks have different couplings (which is the case for $\tilde{Q}$
``photons'') with the gauge fields, we treat it as a four species problem,
labelling the species as 
\begin{equation}
\begin{split}
a&\rightarrow E_1, \; \xi<0 \\
b&\rightarrow E_1, \; \xi>0 \\
c&\rightarrow E_2, \; \xi<0 \\
d&\rightarrow E_2, \; \xi>0
~\label{eq:fourspecies}\;.
\end{split}
\end{equation}
This labelling also clarifies the contributions to the shear viscosity from the
various branches of the Bogoliubov dispersions (Fig.(\ref{fig:EBranches})). 

The matrix equation, Eq.~\ref{eq:master}, is now a $4\times 4$ matrix equation
which gives the four relaxation times $\tau_i$ and the viscosities can be found
by using Eq.~\ref{eq:eta_ioftau_i}.

\subsection{Interactions}
\label{sec:interactions}
The interactions between the quarks are mediated by the gauge bosons: the
gluons and the photon. The gluon-quark vertex is 
\begin{equation}
S_{g} = (g)\int d^4 x \bar{\psi}\gamma^\mu t^m\psi A_\mu^m
\end{equation}
where $g$ is the strong coupling constant, and the photon-quark vertex is 
\begin{equation}
S_{e} = (-e)\int d^4 x \bar{\psi}\gamma^\mu Q\psi A_\mu
\end{equation}
where $t^m$ are the Gell-Mann matrices and $Q=\diag\{2/3,-1/3\}$ in flavor
space. 

To contrast with the properties of the mediators in the FF phase, we revise the
main features in the unpaired phase. Transport in the unpaired phase is dominated by 
the quarks and the electron. The mediators of their interactions have the
following properties.

The longitudinal components of the gluons as well as the photon are Debye
screened. The relevant propagators for the gauge boson are
\begin{equation}
\begin{split}
\frac{i}{-\bfq^2-\Pi_l(q)}[\hat{q}_i\hat{q}_j]\delta^{ab}~\label{eq:propL}
\end{split}
\end{equation}
where $(\omega,{\bf{q}})=p_3^\mu-p_1^\mu=p_2^\mu-p_4^\mu$ is the four momentum carried by the gauge field,
$\bfq^2=\bfq_i\bfq_i$, $q=|\bfq|$, and $\Pi_l$~\footnote{The projection operator, also called $\Pi$
in the previous section, always appears with indices $\Pi^{(n)}$ and can be easily distinguished from
the polarization.} is longitudinal polarization tensor. We have
neglected $\omega^2$ in the propagator. In the limit of small $\bfq$,
\begin{equation}
\begin{split}
\Pi_l(q) \approx \Pi_l(0) = m_D^2 = 2N_f (\frac{g}{2})^2
\frac{\mu^2}{\pi^2}~\label{eq:mD2SUN_f}\;,
\end{split}
\end{equation}
up to corrections of the order $(\delta\mu/\mu)^2$. $N_f=2$ in the two-flavor 
FF problem. 

In the absence of pairing the transverse gluons are dynamically screened
\begin{equation}
\begin{split}
\frac{i}{\omega^2-\bfq^2-\Pi_t(q)}[\delta_{ij}-\hat{q}_i\hat{q}_j]\delta^{ab}~\label{eq:propT}
\end{split}
\end{equation}
where $\Pi_t$ is transverse polarization function, which in the limit of small
$\omega, q$,
\begin{equation}
\begin{split}
\Pi_t(\omega, q) \approx (\frac{-i\pi\omega}{4q}) 2N_f (\frac{g}{2})^2
\frac{\mu^2}{\pi^2}~\label{eq:PitSUN_f}\;.
\end{split}
\end{equation}
Since the energy exchange, $\omega$, is governed by the temperature, while the
momentum exchange can be much larger (typically of the order of $g\mu$), the
transverse gluons have a longer range compared to the
longitudinal and therefore their exchange is the dominant 
scattering mechanism for the quarks. Consequently, the momentum exchange $q$
that dominates the collision integral for the transverse gauge bosons is $\sim
(T \Pi_l(0))^{1/3}$~\cite{Baym:1990}, while for the longitudinal gauge bosons it is $\sim
(\Pi_l(0))^{1/2}$. 

Similarly, for the photon the Debye screening mass can be obtained from
Eq.~\ref{eq:mD2SUN_f} by replacing $2N_f(\frac{g}{2})^2$ by $N_c
e^2(Q_u^2+Q_d^2)$ and the transverse polarization $\Pi_t$ can be obtained from
Eq.~\ref{eq:PitSUN_f} by replacing $2N_f(\frac{g}{2})^2$ by $N_c
e^2(Q_u^2+Q_d^2)$ with $N_c=3$. 

We now consider the gauge bosons in the two-flavor FF phase. As far as the Debye
screening masses are concerned, these are determined by the total density of
gapless states. For gluons that couple with the species that participate in
pairing, this density of states is affected by two competing effects.

First, there is a geometric factor associated with a reduction in size of the
surface of gapless excitations~\cite{Alford:2000ze} 
(Appendix~\ref{sec:blocking}), 
\begin{equation}
\begin{split}
\frac{\mu^2}{\pi^2}\rightarrow \frac{\mu^2}{\pi^2}(1-2\frac{\Delta}{b}),
\end{split}
\end{equation}
since the blocking region is absent for 
$\cos\theta\in[\frac{\delta\mu-\Delta}{b},\frac{\delta\mu+\Delta}{b}]$.

Second, the Fermi velocity on the surface of gapless excitations is reduced due
to pairing. In spherical coordinates, the Fermi velocity is given by 
\begin{equation}
\begin{split}
\frac{dE}{d\bfp}(\xi,\theta) &=
\hat{p}\frac{\xi}{\sqrt{\xi^2+\Delta^2}}|_{\xi=\sqrt{(\delta\mu+b\cos\theta)^2-\Delta^2}}+\frac{b}{p}\hat{\theta}\\
&\approx
\hat{p}\frac{\xi}{\sqrt{\xi^2+\Delta^2}}|_{\xi=\sqrt{(\delta\mu+b\cos\theta)^2-\Delta^2}}<\hat{p}\;.
\end{split}
\end{equation}
The reduction of the velocity enhances the density of states at the gapless
point~\cite{Alford:2005pooja} for certain values of $\theta$. 

Consequently, we expect that for the gluons
\begin{equation}
\begin{split}
\Pi_l(0) = 2N_f (\frac{g}{2})^2
\frac{\mu^2}{\pi^2} f(\frac{\Delta}{\delta\mu},\frac{b}{\delta\mu})~\label{eq:mD2SUN_fFF}
\end{split}
\end{equation}
where $f$ is expected to be of order $1$~\cite{Casalbuoni:2002my}. Similar
arguments hold for the photon.

We leave a detailed analysis of screening of the longitudinal modes for future
work. The two main points we want to emphasize are as follows. First the
longitudinal $t^1, t^2, t^3$ gluons are screened and hence unlike in the $2\SC$
phase~\cite{Rischke:2000cn} the $ur-dg-ug-dr$ quarks are not confined.  Second,
the longitudinal modes of all the mediators are screened and therefore can be
ignored compared to the Landau damped transverse mediators while calculating
the transport properties.  The transverse screening masses for the gauge bosons
in the two-flavor FF phase were analyzed in detail in
Ref.~\cite{Giannakis:2004pf,Giannakis:2005vw} and we summarize the main
conclusions here.

In contrast with unpaired phase, in the two-flavor FF phase, the Meissner
masses for the $t^4\dots t^7$ gluons are
non-zero~\cite{Giannakis:2004pf,Giannakis:2005vw,Ciminale:2006sm}. The FF
condensate cures the instability seen in the gapless $2\SC$ phase, and the
squares of all the four Meissner masses are positive, and are functions of
$\Delta, \delta\mu, \mu$, and $b$.

The Meissner masses tend to $0$ as $\Delta\rightarrow 0$ and hence naturally 
\begin{equation}
m_M^2\sim (\frac{g}{2})^2
\frac{\mu^2}{\pi^2}\frac{\Delta^2}{\delta\mu^2}\;.
\end{equation}
$\Delta^2$ tends to $0$ at $\delta\mu=0.754\Delta_{2\SC0}$, i.e., the transition
point. Away from the transition point, $\Delta$ is much larger
than $T$, and screening is strong if the Meissner mass is
non-zero.

The fate of the $t^8$ and the photon is more interesting.  As in the $2\SC$
phase, the condensate is neutral under the linear combination associated with
the $\tilde{Q}$ charge which is unscreened. The $\tilde{Q}$ photon is weakly
coupled to the quarks and is less important than the unscreened 
gluons~\cite{Rajagopal:2000wf}. The orthogonal linear
combination,
\begin{equation}
\begin{split}
A_\mu^X = \cos\varphi A_\mu^8 +\sin\varphi A_\mu^Q\;,
\end{split}
\end{equation}
with 
\begin{equation}
\begin{split}
\cos\varphi= \frac{\sqrt{3}g}{\sqrt{e^2+3g^2}}\;,
\end{split}
\end{equation}
is strongly coupled.

Because $\bfb$ chooses a particular direction, the transverse polarization
tensor is not invariant in the range of the transverse projection operator 
\begin{equation}
[\delta_{ij}-\hat{q}_i\hat{q}_j]~\label{eq:transverse}\;.
\end{equation}
Ref.~\cite{Giannakis:2005vw} showed that a further projection by $b^ib^j$ (note
that Ref.~\cite{Giannakis:2005vw} uses $q$ to denote what we call $\bfb$ and $k$ to
denote what we call $q$) gives a polarization tensor which vanishes in the
$0$ momentum limit (the ``longitudinal transverse gluon''). The
projection on to $\delta^{ij}-b^ib^j$ (the ``transverse transverse gluon'')
has a finite Meissner mass.

While the ``longitudinal transverse'' part of $A_\mu^X$ is long ranged as well
as strongly coupled, its contribution is smaller compared to the $t^1$, $t^2$,
$t^3$ gluons as we argue below (below Eq.~\ref{eq:M2SU2}). The transverse $t^1, t^2, t^3$ gluons are massless as in the $2\SC$
phase~\cite{Casalbuoni:2002my,Giannakis:2005vw}. What has not been appreciated
in the literature is that they are Landau damped. 

To express the transverse polarization tensor in a compact form we
choose an orthogonal basis for the range of Eq.~\ref{eq:transverse} as follows.
\begin{equation}
\begin{split}
\hat{y}'&= \frac{\hat{b}\times\hat{q}}{|\hat{b}\times\hat{q}|}\\
\hat{x}'&= \hat{y}'\times\hat{b}~\label{eq:xyprime}\;. 
\end{split}
\end{equation}

On general grounds, 
\begin{equation}
\begin{split}
\Pi^{i'j'}_t(\omega,\bfq) = (\frac{-i\pi\omega}{4q}) \Bigl[2N_f (\frac{g}{2})^2
\frac{\mu^2}{\pi^2}\Bigr]
h^{i'j'}_t(\frac{\Delta}{\delta\mu},\frac{b}{\delta\mu},\cos\theta_{bq})
~\label{eq:Pi_t123}\;.
\end{split}
\end{equation}
For unpaired quark matter $h^{i'j'} = \delta^{i'j'}$.

Numerical results for the Landau damping coefficient for $b/\delta\mu=\zeta$ 
are well described by the expressions
\begin{equation}
\begin{split}
h^{x'x'}_t(\frac{\Delta}{\delta\mu},\frac{b}{\delta\mu},\cos\theta_{bq}) 
&\approx 1-(\frac{\Delta}{b})^{1/4}\frac{1}{1.65}(1-\cos\theta_{bq}^4)^{1/2}\\
h^{y'y'}_t(\frac{\Delta}{\delta\mu},\frac{b}{\delta\mu},\cos\theta_{bq}) 
&\approx 1-(\frac{\Delta}{b})\frac{1}{1.75}(1-\cos\theta_{bq}^2)^{1/2}\\
h^{x'y'}_t(\frac{\Delta}{\delta\mu},\frac{b}{\delta\mu},\cos\theta_{bq})
&\approx 0\;.
~\label{eq:h_t123}
\end{split}
\end{equation}
We note that $h<1$, which is expected because the gapless surface (For a quick
refresher on the gapless modes in the FF phase see Appendix~\ref{sec:blocking}.
For details
see~\cite{Alford:2000ze,Bowers:2002xr,Bowers:2003ye,Mannarelli:2006}) in the FF
phase has a smaller surface area  compared to the unpaired phase The details of
the calculation will be given elsewhere~\cite{InPreparation}. 

To summarize, scattering between the $ur-dg-ug-dr$ quarks is dominated by
exchange of the transverse $t^1, t^2, t^3$ gluons. Their propagator is of the form  
\begin{equation}
\begin{split}
iD_{\mu\nu}^{ab} &= 
\frac{i}{\omega^2-q^2-\Pi_t^{i'j'}(\omega,\bfq)}
[\calP_{\mu \nu}^{i'j'}]\delta^{ab}~\label{eq:prop123}\;.
\end{split}
\end{equation}
The projection operator,
\begin{equation}
\calP_{\mu \nu}^{i'j'} = \delta_\mu^{i'}\delta_\nu^{j'}
\end{equation}
projects into the subspace spanned by the unit vectors $e_i', e_j'$ 
(Eq.~\ref{eq:xyprime}), $\Pi_t^{i'j'}(\omega,\bfq)$ are given by 
Eq.~\ref{eq:Pi_t123}.

The interaction can be written in terms of the Nambu-Gorkov
spinors~\footnote{Since we are considering transverse gluons there are no
vertex corrections.} as follows.
\begin{equation}
\begin{split}
S_{g} &= (g)\int d^4 x \bar{\psi}\gamma^\mu t^m\psi A_\mu^m\\
&=(g)\int d^4 x \bar{\psi}_L\bar{\sigma}^\mu t^m\psi_L A_\mu^m
+(g)\int d^4 x \bar{\psi}_R{\sigma}^\mu t^m\psi_R A_\mu^m\;.
\end{split}
\end{equation}

Going to momentum bases and using Eq.~\ref{eq:Bogoliubov} we obtain,
\begin{equation}
\begin{split}
&\calL_{AL}=\int \frac{d^4p_1}{(2\pi)^4} \frac{d^4p_3}{(2\pi)^4} \\
&
{g}\left(
  (\Phi^*_{11}\;\Phi^*_{12})\gamma_L^\dagger(2\pi)\delta(p_3^0-E_1)+
  (\Phi^*_{21}\;\Phi^*_{22})\chi_L^\dagger(2\pi)\delta(p_3^0-E_2)\right)  \\
  &
   \left(\begin{array}{cc}
   1\xi_{-}^\dagger(p_3)\bsigma^\mu \xi_{-}(p_1)t^m & 0\\
   0 & -1\xi_{-}^\dagger(p_3)C \bsigma^{T\mu} C^\dagger \xi_{-}(p_1)\epsilon_{c}^\dagger(t^m)^T \epsilon_c\\
  \end{array}
  \right)\\
&\left((\begin{array}{c}\Phi_{11}\\ \Phi_{12}\end{array})\gamma_L(2\pi)\delta(p_1^0-E_1)
      +(\begin{array}{c}\Phi_{21}\\ \Phi_{22}\end{array})\chi_L(2\pi)\delta(p_1^0-E_2)
      \right)\\
&      
A^a_{\mu}(p_3-p_1)\;.
\end{split}
\end{equation}

Now we can use the conjugation relation for $t^1,\;t^2,\;t^3$ generators,
\begin{equation}
\begin{split}
-1\epsilon_{c}^\dagger(t^m)^T \epsilon_c&=t^m\;,
\end{split}
\end{equation}
and the $t^m$ get decoupled from the Nambu-Gorkov structure. [This step won't
work for the other $\SU(3)$ generators and works because the $\SU(2)$ subgroup 
generated by $t^1\dots t^3$ is unbroken in two-flavor FF. See Eq.~\ref{eq:M2_8} for
an analysis of a broken generator.] We also use the conjugation relation for 
$\bsigma^\mu$,
\begin{equation}
\begin{split}
C \bsigma^{T\mu} C^\dagger = \sigma^{\mu}\;, 
\end{split}
\end{equation}
to simplify the spin structure. This gives,
\begin{equation}
\begin{split}
&\calL_{AL}=\int \frac{d^4p_1}{(2\pi)^4} \frac{d^4p_3}{(2\pi)^4} {g}\\
&
\left(
  (\Phi^*_{11}\;\Phi^*_{12})\gamma_L^\dagger(2\pi)\delta(p_3^0-E_1)+
  (\Phi^*_{21}\;\Phi^*_{22})\chi_L^\dagger(2\pi)\delta(p_3^0-E_2)\right)  \\
  &
   \left(\begin{array}{cc}
   1\xi_{-}^\dagger(p_3)\bsigma^\mu \xi_{-}(p_1) & 0\\
   0 & 1\xi_{-}^\dagger(p_3)\sigma^{\mu} \xi_{-}(p_1)\\
  \end{array}
  \right)\\
&\left((\begin{array}{c}\Phi_{11}\\ \Phi_{12}\end{array})\gamma_L(2\pi)\delta(p_1^0-E_1)
      +(\begin{array}{c}\Phi_{21}\\ \Phi_{22}\end{array})\chi_L(2\pi)\delta(p_1^0-E_2)
      \right)\\
&      
t^m A^m_{\mu}(p_3-p_1)~\label{eq:LALsimplified}\;.
\end{split}
\end{equation}

Similarly, for $R$ we obtain
\begin{equation}
\begin{split}
&\calL_{AR}=\int \frac{d^4p_1}{(2\pi)^4} \frac{d^4p_3}{(2\pi)^4} {g}\\
&
\left(
  (\Phi^*_{11}\;\Phi^*_{12})\gamma_R^\dagger(2\pi)\delta(p_3^0-E_1)+
  (\Phi^*_{21}\;\Phi^*_{22})\chi_R^\dagger(2\pi)\delta(p_3^0-E_2)\right)  \\
  &
   \left(\begin{array}{cc}
   1\xi_{+}^\dagger(p)\sigma^\mu \xi_{+}(p) & 0\\
   0 & 1\xi_{+}^\dagger(p)\bsigma^{\mu} \xi_{+}(p)\\
  \end{array}
  \right)\\
&\left((\begin{array}{c}\Phi_{11}\\ \Phi_{12}\end{array})\gamma_R(2\pi)\delta(p_1^0-E_1)
      +(\begin{array}{c}\Phi_{21}\\ \Phi_{22}\end{array})\chi_R(2\pi)\delta(p_1^0-E_2)
      \right)\\
&      
t^m A^m_{\mu}(p_3-p_1)~\label{eq:LARsimplified}\;.
\end{split}
\end{equation}

A nice way to separate the spinor and the Nambu-Gorkov structure is to re-combine
the $L$ (\ref{eq:LALsimplified}) and $R$ (\ref{eq:LALsimplified}) components 
\begin{equation}
\begin{split}
&\calL_{ALR}=\int \frac{d^4p_1}{(2\pi)^4} \frac{d^4p_3}{(2\pi)^4} \\
&
{g}\left(\bar{u}_s(p_3)\gamma^\mu u_s(p_1)
  \right)\\
& \left(
  (\Phi^*_{11}\;\Phi^*_{12})\gamma_s^\dagger(2\pi)\delta(p_3^0-E_1)+
  (\Phi^*_{21}\;\Phi^*_{22})\chi_s^\dagger(2\pi)\delta(p_3^0-E_2)\right)  \\
&\left((\begin{array}{c}\Phi_{11}\\ \Phi_{12}\end{array})\gamma_s(2\pi)\delta(p_1^0-E_1)
      +(\begin{array}{c}\Phi_{21}\\ \Phi_{22}\end{array})\chi_s(2\pi)\delta(p_1^0-E_2)
      \right)\\
&      
t^mA_{\mu}^m(p_3-p_1)~\label{eq:LALRsimplified}\;.
\end{split}
\end{equation}

The final ingredient we need is the simplification of the color structure in
the interaction. For this we use the relation ($a=1,2,3$)
\begin{equation}
\begin{split}
t^m_{ij} t^m_{kl} = [\frac{-1}{4}\delta_{ij}\delta_{kl} + \frac{1}{2}\delta_{il}\delta_{kj}]
\end{split}
\end{equation}
Summing over the final colors ($j,l$) and averaging over the initial colors
($i,k$) gives (the sum over colors runs over only two
colors $r$ and $g$)
\begin{equation}
\begin{split}
\frac{1}{4} t^m_{ij} t^m_{kl} t^{n*}_{ij} t^{n*}_{kl} &= 
\frac{1}{4} t^m_{ij} t^m_{kl} t^{n}_{ji} t^{n}_{lk}\\
&=\frac{3}{16}\;.
\end{split}
\end{equation}

Therefore, the square of the scattering matrix element averaged over initial
color and spin and summed over the final color and spin are given
by~\footnote{The only subtle step is noting 
\begin{equation}
\begin{split}
&\tr[\slsh{p}_3\gamma^0\gamma^\mu\gamma^0\slsh{p}_1\gamma^0\gamma^\nu\gamma^0]\\
=&\tr[\slsh{p}_3\gamma^\mu\slsh{p}_1\gamma^\nu]
\end{split}
\end{equation}
if $\mu, \nu$ are both spatial or both $0$.
}
\begin{equation}
\begin{split}
|i\bar{\cal{M}}|^2 &= 3(\frac{g}{2})^4\\
&
|(\Phi^*_{i_31}\;\Phi^*_{i_32})
\left(\begin{array}{c}\Phi_{i_11}\\ \Phi_{i_12}\end{array}\right)|^2
|(\Phi^*_{i_41}\;\Phi^*_{i_42})
\left(\begin{array}{c}\Phi_{i_21}\\ \Phi_{i_22}\end{array}\right)|^2\\
&
\frac{1}{4}\frac{1}{2p_12p_22p_32p_4}\tr[\slsh{p}_3\gamma^\mu\slsh{p}_1\gamma^\nu]
   \tr[\slsh{p}_4\gamma^\sigma\slsh{p}_2\gamma^\lambda]
   {D}_{\mu\sigma}{D}_{\nu\lambda}\\
   &= 3(\frac{g}{2})^4\delta_{i_1i_3}\delta_{i_2i_4}\\
&
\frac{1}{4}\frac{1}{2p_12p_22p_32p_4}\tr[\slsh{p}_3\gamma^\mu\slsh{p}_1\gamma^\nu]
   \tr[\slsh{p}_4\gamma^\sigma\slsh{p}_2\gamma^\lambda]
   {D}_{\mu\sigma}{D}_{\nu\lambda}
   ~\label{eq:M2SU2}\;,
\end{split}
\end{equation}
where $i_1,\; i_2,\; i_3,\;$ and $i_4$ run over $1,\; 2$ where $1$ corresponds to
$\gamma$ and $2$ to $\xi$. Note that the orthogonality of $[\Phi]$ 
(Eq.~\ref{eq:ortho}) ensures that $i_1=i_3$ and $i_2=i_4$, and the nature of the
Bologliubov particles doesn't change at the vertex. This can be traced to the
residual $\SU(2)$ symmetry. The factor ${1}/{(2p_12p_22p_32p_4)}$ appears in
$|\calM|^2$ due to the convention of the phase space integrals in
Eq.~\ref{eq:master}: the spinors $u_s$ are normalized to be dimensionless.

Eq.~\ref{eq:M2SU2} can also be used to complete the argument that we made
earlier about why the exchange of transverse $A^X$ is less important than the
exchange of $t^1$, $t^2$, $t^3$ even though they have $0$ Meissner mass. In matrix 
elements the exchange of $A^X$ comes with a coherence factor (Eq.~\ref{eq:M2_8}) where
two terms of similar size cancel. This is because $\Phi_{i_3 1},\; \Phi_{i_3 1},\; \Phi_{i_1
2}^*$, and $\Phi_{i_1 2}$ in Eq.~\ref{eq:M2_8} are all roughly $1/\sqrt{2}$ for
$\xi\approx 0$ and in Eq.~\ref{eq:M2_8} their products appear with a $-$ sign.  
On the other hand the coherence factors in Eq.~\ref{eq:M2SU2} add for $t^1,
t^2, t^3$ gluons. Therefore we expect the numerical
contribution from $A_\mu^X$ to be smaller than the contribution from $t^1, t^2,
t^3$ gluons. (There is an additional reduction by a factor of $\sim 1/2$ because
the ``transverse transverse gluon'' is massive.) Therefore we neglect the scatterings
mediated by $A_\mu^X$. This numerical suppression is not parametric and in a
future, more complete calculation, these scatterings should be included.  We
note that $A_\mu^X$ induces coupling between the $b$ quarks and the paired
quarks and complicates the Boltzmann equation Eq.~\ref{eq:master}
significantly.  

There are additional mediators of quark-quark interactions in the two-flavor FF
phase. Phonons~\cite{Casalbuoni:2001gt} associated with the periodicity of the
condensate~\cite{Mannarelli:2007bs,Radzihovsky:2009}, are derivatively coupled
to the fermion fields. 

The interaction between quark species $i$ and $j$, and phonon $\varphi^a$ can
generically be written as
\begin{equation}
\begin{split}
\calL_{\varphi\psi} = \frac{c_\mu}{f_\varphi^{ij}}
\partial_\mu\varphi^a\bar{\psi}_j \gamma^{\mu}
\psi_i~\label{eq:phonon_interaction}\;.
\end{split}
\end{equation}
where $c_\mu$ is naturally of the order of $v_F$.

Therefore, the scattering matrix in the absence of pairing can be written as
\begin{equation}
\begin{split}
i\calM \sim (\frac{c_\mu q_\mu c_\nu q_\nu}{(f_\varphi^{ij})^2})
\frac{i}{\omega^2-v_\varphi^2q^2}[\bar{u}_3\gamma^\mu u_1]
[\bar{u}_4\gamma^\mu u_2]
\end{split}
\end{equation}
where $q_\mu=(\omega,\bfq)_\mu$ is the four momentum carried by the phonon. For
$\omega\ll q$ 
\begin{equation}
\begin{split}
i\calM \sim (\frac{c_i^2}{(f_\varphi^{ij})^2})
\frac{i}{-v_\varphi^2}[\bar{u}_3\gamma^0 u_1]
[\bar{u}_4\gamma^0 u_2]~\label{eq:Mqqphonon}\;.
\end{split}
\end{equation}

This should be compared with the matrix element for the exchange of a Debye 
screened gauge field.  
\begin{equation}
\begin{split}
i\calM \sim (ig)^2
\frac{i}{q^2+m_D^2}[\bar{u}_3\gamma^0 u_1]
[\bar{u}_4\gamma^0 u_2]\\
\sim -i(g)^2
\frac{1}{m_D^2}[\bar{u}_3\gamma^0 u_1]
[\bar{u}_4\gamma^0 u_2]~\label{eq:MqqDebye}
\;.
\end{split}
\end{equation}

Noting that both $m_D$ and $f_\varphi$ can be related to thermodynamic
susceptibilities~\cite{Son:1999cm}, and that $v_\varphi\sim 1$ in relativistic systems
\begin{equation}
\begin{split}
m_D^2\sim g^2 f_\varphi^2
\end{split}
\end{equation}
we see that Eq.~\ref{eq:Mqqphonon} is of the same order as
Eq.~\ref{eq:MqqDebye}~\footnote{In non-relativistic systems~\cite{Ziman:1960},
the magnetic gauge bosons do not contribute due to the small speeds and the exchange of phonons and the
longitudinal gauge bosons comepete. For $v_\varphi\ll1$, the phonon exchange is
the dominant scattering mechanism. We thank Sanjay Reddy for his comment on
this point.}. Therefore, the contributions to quark-quark scattering from phonon exchanges
can be ignored in our calculation.

\subsection{Contribution of phonons}
\label{sec:phonons}
Phonons, the Goldstone modes associated with broken symmetries, are also low 
energy modes. Here we make a quick estimate of their contribution to transport
and to the collision integral. They are not relevant in the FF phase but play
an important role in gapped phases.

\subsubsection{Quark-Phonon scattering}
\label{sec:qph}
For quark-phonon scattering, the collision term is
\begin{equation}
\begin{split}
[\Gamma_{i}(p_i)]&=-\sum_3\int\;
\dthree{l}\dthree{p_3}(2\pi)^4\\
&[
\hatf_i\hatb_2(1-\hatf_3)\delta^{(4)}(p_i+l-p_3)|\calM(il\rightarrow3)|^2\\
&
+\hatf_i(1+\hatb_2)(1-\hatf_3)\delta^{(4)}(p_i-l-p_3)|\calM(i\rightarrow3l)|^2\\
&
-\hatf_3\hatb_2(1-\hatf_i)\delta^{(4)}(p_i-l-p_3)|\calM(i\rightarrow3l)|^2\\
&
-\hatf_3(1+\hatb_2)(1-\hatf_i)\delta^{(4)}(p_i+l-p_3)|\calM(il\rightarrow3)|^2]\;,
\end{split}
\end{equation}
where $\hatf$ and $\hatb$ are non-equilibrium distribution functions, and
$l^\mu=(\omega,\bfl)$ is the four momentum of the phonon satisfying
$(\omega)^2-v_{\varphi}^2(\bfl^2)=0$. 

To the lowest order in gradient of the fluid velocity $u_a$, 
$\hatf_i-f_i=\delta f_i=-\frac{df^i}{d\epsilon}[\Phi^i]$, where 
\begin{equation}
\begin{split}
\Phi^i=\sum_{(n)}\Phi^{(n)}_i=\sum_{(n)}3\tau^{(n)}_i
\psi^{(n)i ab}\frac{1}{2}(\partial_au_b+\partial_b u_a)
\label{eq:Phi}
\end{split}
\end{equation}
Substituting Eq.~\ref{eq:Phi} in the Boltzmann equation one can obtain the analogue of
Eq.~\ref{eq:master}
\begin{equation}
\begin{split}
&[R^{(n)}_{i}]=-\frac{1}{\gamma^{(n)}}\frac{1}{T}\nu_2 \int\;
\dthree{p_i}\dthree{l}\dthree{p_3}(2\pi)^4 3\phi_i\cdot\\
&[
f_ib_2(1-f_3)\delta^{(4)}(p_i+l-p_3)(\tau_i\psi_i^{(n)})|\calM(il\rightarrow3)|^2\\
&
+f_i(1+b_2)(1-f_3)\delta^{(4)}(p_i-l-p_3)(\tau_i\psi_i^{(n)})|\calM(i\rightarrow3l)|^2\\
&
-f_3b_2(1-f_i)\delta^{(4)}(p_i-l-p_3)(\tau_3\psi_3^{(n)})|\calM(i\rightarrow3l)|^2\\
&
-f_3(1+b_2)(1-f_i)\delta^{(4)}(p_i+l-p_3)(\tau_3\psi_3^{(n)})|\calM(il\rightarrow3)|^2]\;.
\end{split}
\end{equation}
where $b$ is the Bose distribution.

The scattering matrix element associated with the vertex Eq.~\ref{eq:phonon_interaction}
is given by
\begin{equation}
\begin{split}
|{\cal{M}}|^2 &= |\frac{i}{f^{ij}_\varphi}
\frac{1}{\sqrt{2\omega}\sqrt{2p_1}\sqrt{2p_3}}|^2
4[2p_i\cdot l p_3\cdot l-p_i\cdot p_3 l^2]|\\
&\sim \frac{l^2}{(f^{ij}_\varphi)^2\omega}
\sim \frac{\omega}{(f^{ij}_{\varphi})^2}
\end{split}
\end{equation}
where we have taken $c_\mu$ to be $1$ for convenience.

Simplifying the momentum integrals for the fermions ($d^3p_i$ and $d^3p_3$) as
in Eq.~\ref{eq:FermiSurface}, noting that $\xi_1$, $\xi_3$ and $\omega$ are all
of the order of $T$, and that $\phi$ and $\psi$ are of the order of $\mu$,  we
can see without evaluating the integrals that,
\begin{equation}
\begin{split}
[R_{(q-\ph) ij}]&\sim \frac{\mu^3T^4}{(f_\varphi^{ij})^2} \sim  \mu T^4,
~\label{eq:Rqph}
\end{split}
\end{equation}
where we have used a rough estimate for $f_\varphi$: $f_\varphi\sim\mu$.

When unpaired quarks participate in transport and $T$ is much
less than the chemical potential $\mu$, the contribution from Eq.~\ref{eq:Rqph}
is subleading compared to the collision term associated with quark-quark
scattering in Eq.~\ref{eq:RijunpairedDv2}.  This is simple to understand because
the phase space for fermions near the Fermi surface is enhanced. We shall see
in Sec.~\ref{subsubsec:dmuzero} that this is not true for paired systems.

\subsubsection{Momentum transport via phonons}
\label{sec:phonon_transport}
If phonons are present in the low energy theory then they can also transport
energy and momentum. While this is not the main topic of the paper, we make a
quick estimate to see how this contribution compares with the fermionic
contribution.

The kinetic theory estimate for the shear viscosity of the phonon gas is,
\begin{equation}
\begin{split}
\eta\sim \langle n\rangle \langle p\rangle v_\varphi \tau_\varphi\;.
\end{split}
\end{equation}

The density of phonons at temperature $T$ is given by $\langle n \rangle\sim
\frac{T^3}{v_\varphi^3} $ and $\langle p \rangle\sim \frac{T}{v_\varphi}$.
Consequently,
\begin{equation}
\begin{split}
\eta\sim \frac{1}{v_\varphi^3} T^4 \tau_\varphi\;.
\end{split}
\end{equation}

$\tau_\varphi$ is very sensitive to the nature of the excitations present in
the low energy theory. For example, if all the fermionic modes are gapped, then
the phonons only scatter with each other. Since the phonons are coupled
derivatively, the relaxation time in these cases is very long due to the small
density of phonons at low temperatures, and hence the viscosity is very large.
It is well known that in the absence of gapless fermions these ``phonons''
dominate the viscosity at low $T$
(Ref.~\cite{landau1949theory,maris1973hydrodynamics})

For example, if the dominant scattering rate is $2\rightarrow 2$ scattering
~\cite{Rupak:2007vp,Manuel:2011ed} 
\begin{equation}
\begin{split}
\eta\sim  \frac{1}{v_\varphi^3} \frac{f_\varphi^8}{T^5} \;.
\label{eq:etaphph}
\end{split}
\end{equation}

In both the unpaired phase and in the FF phase, phonons can scatter off gapless
fermionic excitations which have a large density of states near the Fermi
surface. This effect is simply the Landau damping of the phonons. The
scattering rate of the phonons is $\Gamma \sim 1/\tau \sim
\frac{\mu^2}{f_\varphi^2}T$~\cite{Aguilera:2008ed}.
A quick estimate gives
\begin{equation}
\begin{split}
\eta\sim \frac{1}{v_\varphi^3} \frac{f_\varphi^2}{\mu^2} T^3~\label{eq:etaph}\;.
\end{split}
\end{equation}
In the next section we will compare the phonon contribution~Eq.\ref{eq:etaph} with the 
fermion contribution.

\section{Results for a simple interaction for isotropic pairing}
\label{sec:simple}
As discussed in the previous section, pairing affects transport properties of
fermions in two important ways. First, it modifies the dispersion relations of
the fermions. Second, it changes the mediator interactions.

To get some understanding of how the modification of the dispersion relations
due to pairing affects transport (which is the dominant effect because of the
exponential thermal factors in Eq.~\ref{eq:master}), we solve the Boltzmann
equations with and without pairing for a simple system featuring two species of
quarks $1$ and $2$ interacting via a single abelian gauge field $A_\mu$ which
couples to the two species in the following manner
\begin{equation}
\begin{split}
\calL_{A\psi} = \frac{g}{2} \bar{\psi} A_\mu\gamma^{\mu}
\psi~\label{eq:simple_interaction}\;,
\end{split}
\end{equation}
where 
\begin{equation}
\begin{split}
\psi = (\psi_1,\psi_2)^T\;.
\end{split}
\end{equation}

Furthermore, we focus on scattering via longitudinal $A_\mu$, which is not
affected by pairing. We approximate the polarization tensor of the longitudinal
mode of $A_\mu$ by the Debye screened mass which has the standard form as given in 
Eq.(\ref{eq:mD2SUN_f}) with $N_f=1$.

The square of the matrix element averaged over initial spins and summed over
the final spins~\cite{Alford:2014doa} (after making some simplifying
approximations) is given by 
\begin{equation}
\begin{split}
|i\bar{\cal{M}}&(i_1i_2\rightarrow i_3i_4)|^2 
= (\frac{ig}{2})^4\times\\
&
[\frac{1}{\bfq^2+\Pi_l(0)}]^2[1-\frac{q^2}{4p_1p_3}][1-\frac{q^2}{4p_2p_4}]\;.
~\label{eq:M2_8_unp}
\end{split}
\end{equation}

We first review the results for the unpaired phase and then see how pairing
modifies them.

\subsection{Unpaired fermions}
\label{subsec:Unpaired}
The dispersions are given by Eq.~\ref{eq:E1E2polar} with $b=0$, $\Delta=0$.
Dropping the absolute sign in $\xi$, $E=\delta\mu\pm\xi$ and we don't need to
distinguish between the $\xi>0$ and $\xi<0$ modes. For convenience here we can
put $\delta\mu=0$ and the two species can be treated as identical. (The
corrections to the results are suppressed by $\delta\mu/\mu$.) 

In this case the left hand side of Eq.~\ref{eq:master} is simply given by the
integral,
\begin{equation}
\begin{split}
L_1^\un =  
\frac{1}{\gamma^{(n)}} \frac{2\pi\mu^2}{(2\pi)^3}
&\frac{1}{T}\int_{-\infty}^{\infty} d\xi
\frac{1}{(e^{\xi/T}+1)}\frac{1}{(e^{-\xi/T}+1)}\\ 
&\int d\cos\theta \mu^2 \frac{3}{2}(\cos^2\theta-\frac{1}{3})^2\;.
\end{split}
\end{equation}
Using $\gamma^{(0)}=1$, we obtain,
\begin{equation}
\begin{split}
[L_i^\un] =  \left(\begin{array}{c} 
-\frac{4}{15}\frac{\mu^4}{(2\pi)^2}\\
-\frac{4}{15}\frac{\mu^4}{(2\pi)^2}\end{array}
\right)~\label{eq:Liunpaired}\;.
\end{split}
\end{equation}
(We will use the superscript ``$\un$'' to denote the values of $L$, $R$, $\tau$ and
$\eta$ for one unpaired species.)

The right hand side of Eq.~\ref{eq:master} can be obtained following 
Refs.~\cite{Heiselberg:1993,Alford:2014doa}. The interaction Eq.~\ref{eq:simple_interaction} does not
change flavor, and hence the species index $3$ is the same as $i$, and $2$ the
same as $4$. There are two relevant integrals which give, 
\begin{equation}
\begin{split}
s_1^\un&=
 -\frac{1}{\gamma^{(n)}}\frac{1}{T}\nu \int\;
 \dthree{p_i}\dthree{p_2}\dthree{p_3}\dthree{p_4}\\
 &
 |\calM(i2\rightarrow34)|^2\\
 &
 \deltafour
 [f_if_2(1-f_3)(1-f_4)]\\
 &3\phi_i.[\psi_i^{(n)}-\psi_3^{(n)}]\\
&\approx-\nu\frac{g^4}{16\cdot 5}
\frac{\pi^3}{(2\pi)^5}
\frac{\mu^4T^2}{\sqrt{\Pi_l(0)}}\\
s_2^\un
 &=-\frac{1}{\gamma^{(n)}}\frac{1}{T}\nu \int\;
 \dthree{p_i}\dthree{p_2}\dthree{p_3}\dthree{p_4}\\
 &
 |\calM(i2\rightarrow34)|^2\\
 &
 \deltafour
 [f_if_2(1-f_3)(1-f_4)]\\
 &3\phi_i.[\psi_2^{(n)}-\psi_4^{(n)}]\\
&\approx 0\;.~\label{eq:sunpaired}
\end{split}
\end{equation}
The analytic approximations for the collision integrals are obtained by assuming
$q\ll\mu$ dominates the integral. (Only an interference between transverse and longitudinal gauge field 
exchange contributes to $s_2$.)
  
The matrix $R$ is related to $s^\un$ by,
\begin{equation}
\begin{split}
[R^\un_{ij}]&=
\left(\begin{array}{cc}
(2s^\un_1+s^\un_2)&s^\un_2\\
s^\un_2&(2s^\un_1+s^\un_2)
\end{array}\right)\\
&=-\frac{g^3T^2\mu^3\nu}{640\pi \sqrt{2}}
\left(\begin{array}{cc}
1&0\\
0&1
\end{array}\right)
~\label{eq:RijunpairedDv2}\;,
\end{split}
\end{equation}
where the final form is obtained by taking
Eq.~\ref{eq:mD2SUN_f} with $N_f=1$ in Eq.~\ref{eq:sunpaired}.

Eqs.~\ref{eq:RijunpairedDv2},~\ref{eq:Liunpaired} can be used to compute the
viscosity for unpaired quarks with which we can compare the results in the 
paired system. In the approximation $q\ll\mu$ one obtains 
\begin{equation}
\begin{split}
\tau_1^{\un}&=\tau_2^{\un} = \frac{L_1^\un}{2s_1^\un} =
\frac{256\sqrt{\Pi_l(0)}}{3\nu g^4T^2}\\
&=\frac{128\sqrt{2}\mu}{3 g^3\pi T^2\nu}\\
\eta_1^\un &= \eta_2^\un = -\frac{3}{2}\nu{L_1^\un}{\tau^\un} =
\frac{128\sqrt{\Pi_l(0)}\mu^4}{15g^4\pi^2T^2}\\
&=\frac{64\sqrt{2}\mu^5}{15g^3\pi^3T^2}
~\label{eq:etaunpairedD}\;,
\end{split}
\end{equation}
where the final forms are obtained by taking
Eq.~\ref{eq:mD2SUN_f} with $N_f=1$.

The total viscosity of the system is 
\begin{equation}
\eta^\un = \eta_1^\un+\eta_2^\un = 2\eta_1^\un\;. 
\end{equation}

Typically the system described above will feature additional low energy modes.
For example, to ensure the neutrality of the system a background of oppositely charged particles
is necessary, and fluctuations in their density is gapless. (A real world
example is the electron ``gas'' in a lattice of ions.) Quarks can scatter off
these ``phonons''. In Sec.~\ref{sec:qph} we made a quick estimate of how these
processes affect quark transport and found that $R_{ij}^{{\rm{q-ph}}}\sim
\mu T^4$, which is parametrically smaller than
Eq.~\ref{eq:RijunpairedDv2}. Therefore they can be ignored for unpaired quark
matter. However, these scattering processes will turn out to be important in
the next section. 

Finally, it is easy to see that the viscosity of unpaired quarks
(Eq.~\ref{eq:etaunpairedD}) is much larger than that of phonons in the presence
of unpaired fermions (Eq.~\ref{eq:etaph}).

\subsection{Paired fermions}
\label{subsec:Paired}
We now consider the effect of isotropic pairing on transport to get some
intuition into the anisotropic problem. For $b=0$, we
can simplify the integrals $R_{ij}$ (Eq.~\ref{eq:master}) using rotational
symmetry (Appendix~\ref{sec:numerics}).  In
Sec.~\ref{subsubsec:dmuzero} we take $\dmu=0$ and see how pairing affects the fermionic
contribution to viscosity. In Sec.~\ref{subsubsec:dmunonzero} we take $\delta\mu\neq0$
and see how the gapless modes that arise when $\Delta<\delta\mu$ contribute to
transport. In the FF phase, the fermions at the boundary of the blocking
regions are gapless and we expect to see that they share some features of the
system considered in Sec.~\ref{subsubsec:dmunonzero}.

The scattering matrix element for Bogoliubov quasi-particles (following the
steps used for obtaining Eq.~\ref{eq:M2SU2}) for an interaction of the form
Eq.~\ref{eq:simple_interaction} is given by
\begin{equation}
\begin{split}
|i\bar{\cal{M}}&(i_1i_2\rightarrow i_3i_4)|^2 
 = (\frac{ig}{2})^4\\
&|
[\Phi_{i_3 1}^*\Phi_{i_3 1}-\Phi_{i_1 2}^*\Phi_{i_1 2}]
[\Phi_{i_4 1}^*\Phi_{i_4 1}-\Phi_{i_2 2}^*\Phi_{i_2 2}]|^2\\
&
[\frac{1}{\bfq^2+\Pi_l(0)}]^2[1-\frac{q^2}{4p_1p_3}][1-\frac{q^2}{4p_2p_4}]
~\label{eq:M2_8}
\end{split}
\end{equation}
where $i$'s run over $1,\;2$ corresponding to the two eigenstates
Eq.~\ref{eq:E1E2polar}. $\Phi$'s are the coherence factors Eq.~\ref{eq:E1Psi1},
Eq.~\ref{eq:E2Psi2}. There are vertex
corrections~\cite{schrieffer1983theory} for the longitudinal
mode but since we are are only looking for qualitative insight for the simple
interaction in this section, we do not consider these here.

\begin{figure*}[tbp]
\includegraphics[width=0.49\textwidth,clip=]{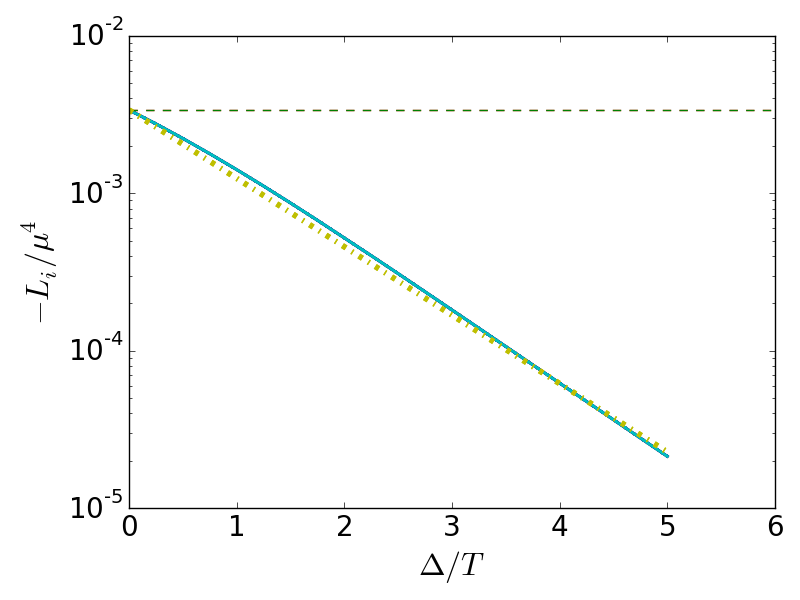}
\includegraphics[width=0.49\textwidth,clip=]{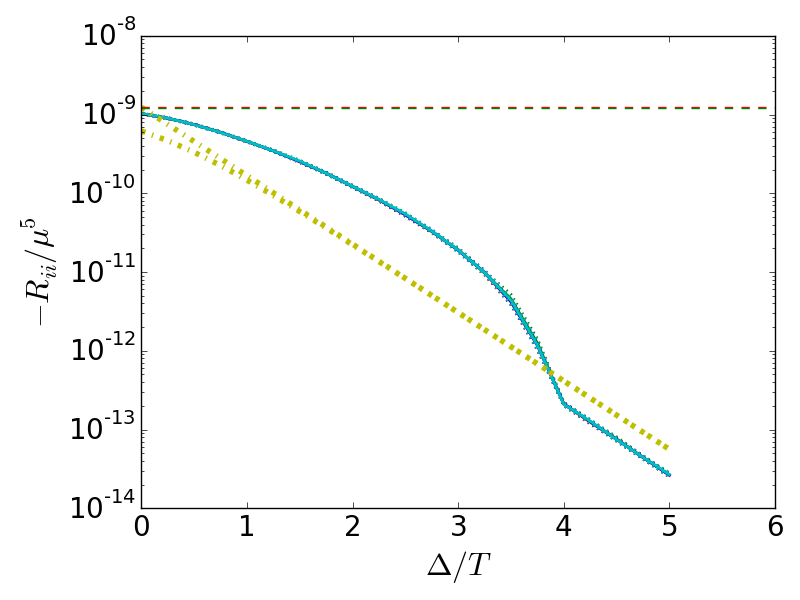}
\includegraphics[width=0.49\textwidth,clip=]{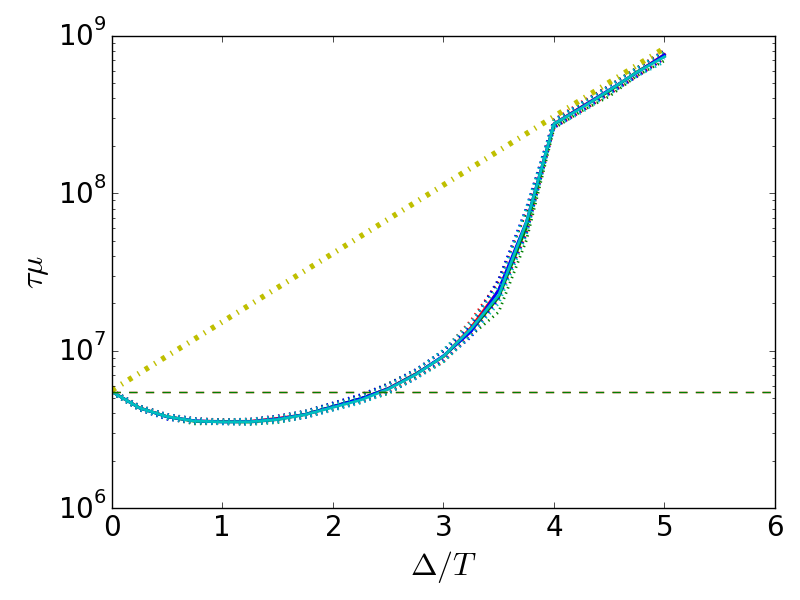}
\includegraphics[width=0.49\textwidth,clip=]{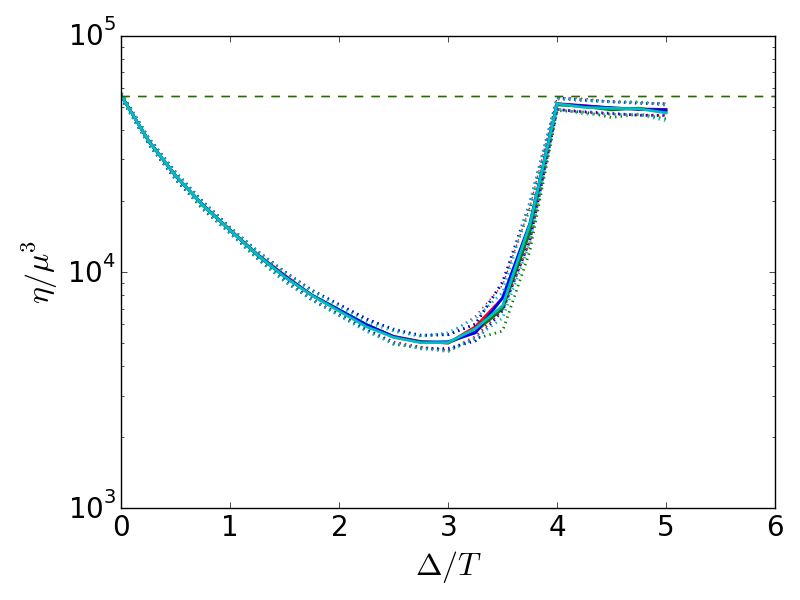}
  \caption{(color online) Plots of $L_i$, the diagonal entries of $R$, $\eta_i$
  and $\tau_i$ (anticlockwise from top left) for $|\bar{\calM}|^2$ given in
  Eq.~\ref{eq:M2_8}. The overall scale is set by
  $\mu$. Keeping $T/\mu=3.34\times10^{-4}$ fixed and $\delta\mu=0$,  we plot
  these as a function of $\Delta/T$ for the ``four species'' $i=$, $a$, $b$,
  $c$, and $d$ (Eq.~\ref{eq:fourspecies}). The four solid curves [red ($a$),
  green ($b$), blue ($c$), and cyan ($d$) online which are indistinguishable in
  the plots] denote the results for the four species.  The dotted curves (not
  visibly distinguishable in the plots) signify the errors in the numerical
  integration for $R$ (Eq.~\ref{eq:RSphericalSimplified}).  The dashed
  horizontal curves (green online) are proportional to values for unpaired
  quarks (see text). The dot dashed curves (yellow online) show an exponential
  fall off, $\propto\exp(-\Delta/T)$, for $L_i$ (Eq.~\ref{eq:LDelta}), an exponential
  fall off, $\propto\exp(-2\Delta/T)$, for $R_{ii}$ (Eq.~\ref{eq:R2Delta}), and an exponential
  increase, $\propto\exp(\Delta/T)$, for $\tau_{i}$. The horizontal dashed line for $L$ [$R$, $\tau$,
  $\eta$] corresponds to $L^\un/2$ (Eq.~\ref{eq:Liunpaired}) [$R_{11}^\un$
  (Eq.~\ref{eq:RijunpairedDv2}), $\tau_1^\un$ and
  $\eta_1^\un/2$ (Eq.~\ref{eq:etaunpairedD})].
  ~\label{fig:Deta_vs_Deltadmu0_4}
}
\end{figure*}

\subsubsection{BCS pairing}
\label{subsubsec:dmuzero}
We first consider the standard BCS pairing with $\delta\mu=0$. The results are
shown in Fig.~\ref{fig:Deta_vs_Deltadmu0_4}. The top left panel shows $L_i$
$i=a,\;b,\;c,\;d$ (Eq.~\ref{eq:fourspecies}). In this symmetric situation,
$L_i$ are equal for all the species and are represented by four overlapping
curves (red, green, blue and cyan online). Similarly, $R_{ii}$ ($i$ not summed)
are all equal. (This is shown on the top right panel. We don't show the cross
terms.) Results for $\tau$ (Eq.~\ref{eq:matrix}) and $\eta_i$
(Eq.~\ref{eq:eta_ioftau_i}) are shown in the bottom left and right panel
respectively.

{\textit{$\Delta\rightarrow 0$---}} When the pairing parameter
$\Delta\rightarrow 0$ ($0$ of the $x$-axis in
Fig.~\ref{fig:Deta_vs_Deltadmu0_4}), we get back a system of unpaired fermions
and one should obtain the result in Sec.~\ref{subsec:Unpaired} in the language
of Bogoliubov quasi-particles (Eq.~\ref{eq:fourspecies}).

$L_i$ is given by half the values given in Eq.~\ref{eq:Liunpaired} (the factor
of $1/2$ arises because we restrict the integrals in $\xi$ to $\xi>0$ or $<0$
corresponding to
Eq.~\ref{eq:fourspecies}) 
\begin{equation}
\begin{split}
[L](\Delta=0) = \frac{1}{2}L_1^{\un} 
\left(1, 1, 1, 1\right)\label{eq:matrixL_i0}
\end{split}
\end{equation}
The dashed horizontal line (green online) on the top left panel of
Fig.~\ref{fig:Deta_vs_Deltadmu0_4} corresponds to $L_i=\frac{1}{2}L_1^{\un}$
(Eq.~\ref{eq:matrixL_i0}, Eq.~\ref{eq:Liunpaired}). A numerical evaluation of
the integral for $L_i$ in Eq.~\ref{eq:master} agrees with the analytic result
Eq.~\ref{eq:Liunpaired} to a very high accuracy. For the collision integral, 
we numerically find that to a high accuracy the matrix $R_{ij}$ has the form 
\begin{widetext}
\begin{equation}
\begin{split}
R_{ij}(\Delta=0) = 2s_1^\un
\left(
\begin{array}{cccc}
\frac{1}{2}+\f(\frac{T}{\mu},\frac{1}{g}) & 0 & 0 & \frac{1}{2}-\f(\frac{T}{\mu},\frac{1}{g}) \\
0 & \frac{1}{2}+\f(\frac{T}{\mu},\frac{1}{g}) & \frac{1}{2}-\f(\frac{T}{\mu},\frac{1}{g}) & 0  \\
0 & \frac{1}{2}-\f(\frac{T}{\mu},\frac{1}{g})  & \frac{1}{2}+\f(\frac{T}{\mu},\frac{1}{g}) & 0  \\
\frac{1}{2}-\f(\frac{T}{\mu},\frac{1}{g}) & 0 & 0 & \frac{1}{2}+\f(\frac{T}{\mu},\frac{1}{g})
\end{array}
\right)\label{eq:matrixR_ij0}
\end{split}
\end{equation}
\end{widetext}
(with $s^\un_2=0$ and $s^\un_1$ given in Eq.~\ref{eq:sunpaired}~\footnote{For
the parameters of Fig.~\ref{fig:Deta_vs_Deltadmu0_4} numerical result for
$R^\un_{11}/\mu^5=-1.23\times10^{-9}$. The analytic expressions
(Eq.~\ref{eq:sunpaired}) give $R^\un_{11}/\mu^5=-1.31\times10^{-9}$.}). The
dashed horizontal lines (green online) on the top right panel of
Fig.~\ref{fig:Deta_vs_Deltadmu0_4} corresponds to $R_{11}^\un=2s_1^{\un}$
(Eq.~\ref{eq:matrixR_ij0}, Eq.~\ref{eq:RijunpairedDv2}).

The structure of the matrix Eq.~\ref{eq:matrixR_ij0} is easy to understand. The
diagonal entries correspond to scattering of species $i$ with $i$. For
$\Delta=0$ the branch $a$ is connected to $d$ and $b$ to $c$, and these
scattering contributions are finite and they add up to $2s_1^{\un}$.  In a wide
range of $T\ll\mu$, $\f(\frac{T}{\mu},\frac{1}{g})$ is relatively
insensitive to $T/\mu$ and increases with increasing $\frac{1}{g}$ (weak coupling).
This is because $\f(\frac{T}{\mu},\frac{1}{g})$ is related to the
scatterings between $a$ and $d$ (or $b$ and $c$) species which is more
prominent if the scatterings that dominate the collision integral are small
angle ($q\sim g\mu \ll\mu$). 

Finally, the contribution to the collisional integral from scattering of
particles in the branch $a$ with $b$ or $c$ is $0$ from rotational symmetry
(just like $s_2^{\un}=0$ in Eq.~\ref{eq:sunpaired}). For $g=1$ in 
Eq.~\ref{eq:mD2SUN_f}, $\f(\frac{T}{\mu},\frac{1}{g})\approx 0.32$. 

From Eq.~\ref{eq:matrixL_i0} and Eq.~\ref{eq:matrixR_ij0} one can easily obtain
relaxation time $\tau_i(\Delta=0) = \tau_1^{\un}$ and hence the shear viscosity
is $\eta_i = \frac{1}{2}\eta_1^{\un}$ for all four species. The total viscosity
is given by
\begin{equation}
\eta(\Delta=0) = \sum_i \eta_i(\Delta=0) = 4\eta_i(\Delta=0) =
2\eta_1^{\un}~\label{eq:eta_i0_tot}\;.
\end{equation}
The dashed horizontal line (green online) on the bottom left (bottom right) 
panel of Fig.~\ref{fig:Deta_vs_Deltadmu0_4} corresponds to $\tau_i$ 
($\eta_i=\eta_1^{\un}/2$) (Eq.~\ref{eq:etaunpairedD}).

{\textit{$\Delta\gg T$---}} As $\Delta$ is increased, the participation of
fermions in transport is thermally suppressed. Since $L_i$ involves single
particle excitations, it is easy to see that 
\begin{equation}
L_i(\Delta)\approx L_i(\Delta=0)e^{-\Delta/T}~\label{eq:LDelta}\;.
\end{equation}
This is shown in Fig.~\ref{fig:Deta_vs_Deltadmu0_4} by the dot-dashed curve (yellow
online). Similarly, since $R_i$ involve two particle excitations, we expect
that 
\begin{equation}
R_{ij}(\Delta)\sim R_{ij}(\Delta=0)e^{-2\Delta/T}~\label{eq:R2Delta}\;.
\end{equation}
We see in Fig.~\ref{fig:Deta_vs_Deltadmu0_4} that this turns out to be true for
$\Delta/T$ larger than $4$ and the suppression for $\Delta/T\lesssim4$ while present,
is a little weaker. Consequently, one can quickly deduce that
$\tau_i(\Delta)\sim\tau_i(\Delta=0)e^{\Delta/T}$: the few thermally excited quarks
rarely scatter with each other. The large relaxation time compensates for the
small number of momentum carrying fermions and for $\Delta/T\gtrsim 4$ the
viscosity converges back to the value for unpaired quark matter.

This result is puzzling since we expect the paired fermions to be frozen
at temperatures smaller than the pairing gap and hence not contribute to the
viscosity.  We expect only the low energy phonons to participate in transport
at low energies~\cite{Rupak:2007vp}. 

We argued in the previous section (Sec.~\ref{subsec:Unpaired}) that in the absence of
pairing for $T\ll\mu$, the contribution to the quark collision integral $R$
from quark-phonon scattering (Eq.~\ref{eq:RijunpairedDv2}) is sub-dominant to the
contribution from quark-quark scattering (Eq.~\ref{eq:Rqph}). Pairing, however, 
affects these two contributions differently. Since quark-phonon scattering involves only one gapped mode, we expect the
$R_{ij(q-{\rm{ph}})}(\Delta)\sim R_{ij(q-{\rm{ph}})}(0)e^{-\Delta/T}$ rather
than as in Eq.~\ref{eq:R2Delta} and dominates scattering. Then $\tau_i$ doesn't
grow exponentially and $\eta_i$ is exponentially suppressed.

More systematically, for $\Delta\gg T$
\begin{equation}
\begin{split}
L_i\approx \frac{-2}{15} \frac{\mu^4}{(2\pi)^2} 
e^{-\Delta/T},\;\;\;\;\;\; {i=a,b,c,d}~\label{eq:LDeltav2}
\end{split}
\end{equation}
and (Eqs.~\ref{eq:R2Delta},~\ref{eq:Rqph}),
\begin{equation}
\begin{split}
R_{ii}(\Delta)\approx&  [-\frac{1}{2}\frac{g^3T^2\mu^3\nu}{640 \pi\sqrt{2}}
e^{-2\Delta/T}- c\mu T^4 e^{-\Delta/T}]\\
\approx&- c\mu T^4 e^{-\Delta/T}\;\;\;\;\;\; {i=a,b,c,d}~\label{eq:RDelta}
\end{split}
\end{equation}
where we have taken $\Pi_l(0)=(g\mu/(2\pi))^2$ and $c$ is a number 
${\cal{O}}(1)$. Hence, 
\begin{equation}
\begin{split}
\tau_i\approx \frac{2c}{15(2\pi)^2}\frac{\mu^3}{T^4}\;\;\;\;\;\; {i=a,b,c,d}\;.
~\label{eq:tauph}
\end{split}
\end{equation}

Therefore, the fermionic contribution to the shear viscosity is given by
\begin{equation}
\begin{split}
\eta_i=-3\tau_iL_i\approx
\frac{4}{75(2\pi)^4}\frac{\mu^7}{T^4}e^{-\Delta/T}
\;\;\;\;\;\; {i=a,b,c,d}\;,~\label{eq:etagappedWphonons} 
\end{split}
\end{equation}
which is subdominant to the phonon contribution (Eq.~\ref{eq:etaphph}, since we
are assuming no other gapless fermions are present).

This entire argument relies on the existence of a gapless mode in the low
energy theory, but in most of the paired systems we know such a mode is present. If
the symmetry broken by the fermion condensate is global or has a global
component~\footnote{For the quark pairing the condensate breaks baryon number
conservation. For cold atoms fermion number conservation is a global symmetry.
In both these cases the dispersion of the resultant mode is $v_F/\sqrt{3}$ and
hence absorption of phonons by fermions is kinematically allowed.} then the
pairing itself gives rise to a Goldstone mode which can scatter off fermions.
If the symmetry broken by the condensate is local rather than global, then
pairing does not by itself give rise to a phonon mode. For example in ordinary
BCS superconductors the local $U(1)\rightarrow Z_2$ gives a mass to the
transverse photons (the Meissner effect). However even in this case there is a
Goldstone mode associated with the breaking of translational symmetry by
the underlying lattice. ~\footnote{The sound speed of the lattice phonons is
much smaller than the Fermi speed of the fermions and fermion phonon scattering
is kinematically allowed. However hypothetically one can consider a situation
where this is not the case. Then the statement that gapped contributions do not
contribute to transport will not hold. Since this is not germane to our paper we will not explore this
further here.}

Therefore the common statement that the paired fermions don't contribute to
transport at low temperatures is generically true, but subtle.  Things are
cleaner if there are fermionic modes that are gapless, in which case they
dominate transport when $\mu\gg T$. This is the situation we shall explore
next.

In drawing Figs.~\ref{fig:Deta_vs_Deltadmu0_4} we have taken $g=1$. Obtaining
results for arbitrary $g$ is simple. The top left panel ($L_i$) doesn't depend
on the collisions and is not modified. The square of the matrix element,
$|{\cal{M}}|^2$, scales as ${g^4}$ and $\sqrt{\Pi_l}$ scales as $g$.
Consequently $\tau_i$ and $\eta_i$ scale as ${1}/{g^3}$.

\begin{figure*}[tbp]
\includegraphics[width=0.49\textwidth,clip=]{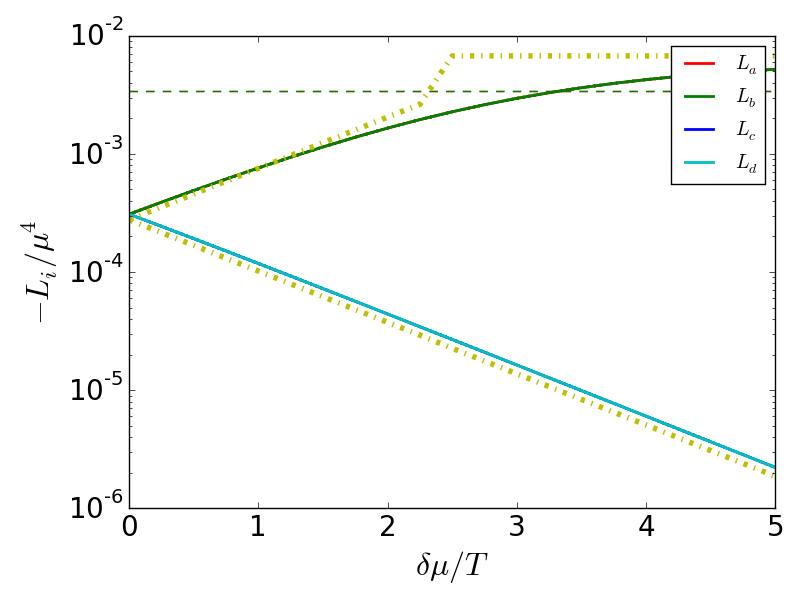}
\includegraphics[width=0.49\textwidth,clip=]{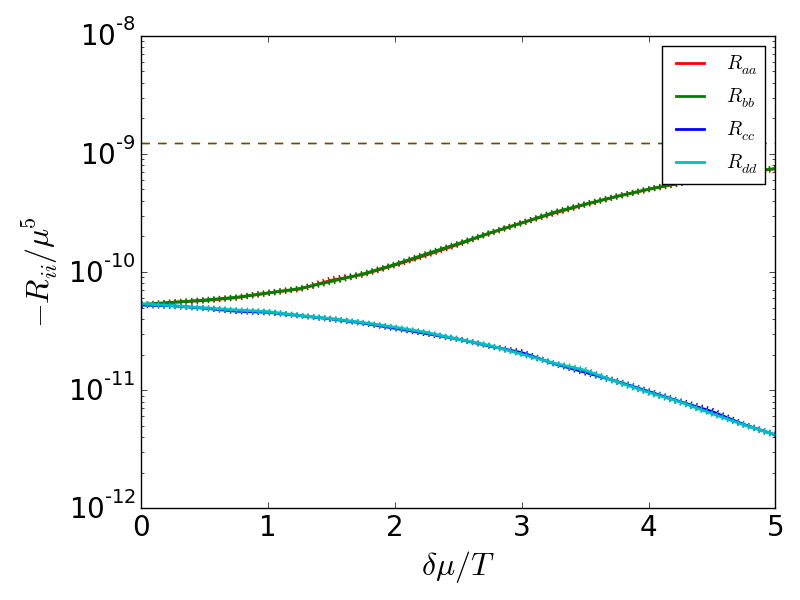}
\includegraphics[width=0.49\textwidth,clip=]{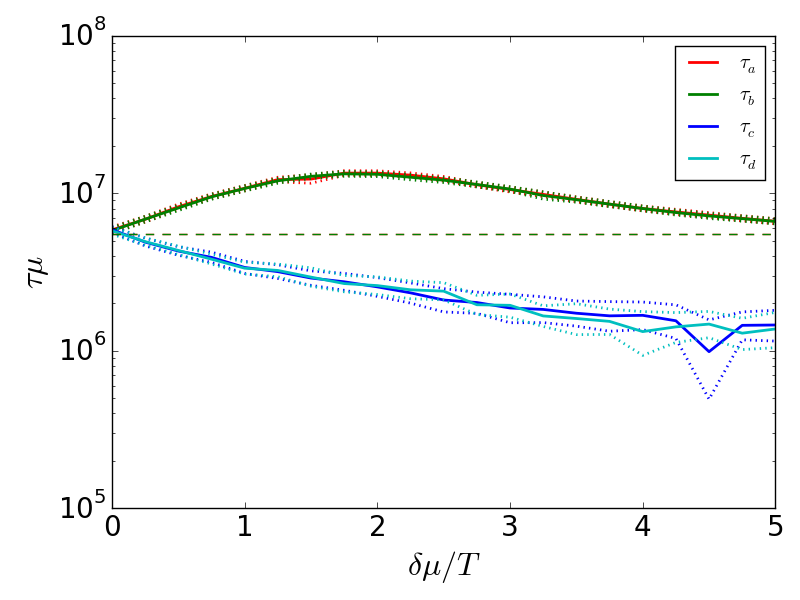}
\includegraphics[width=0.49\textwidth,clip=]{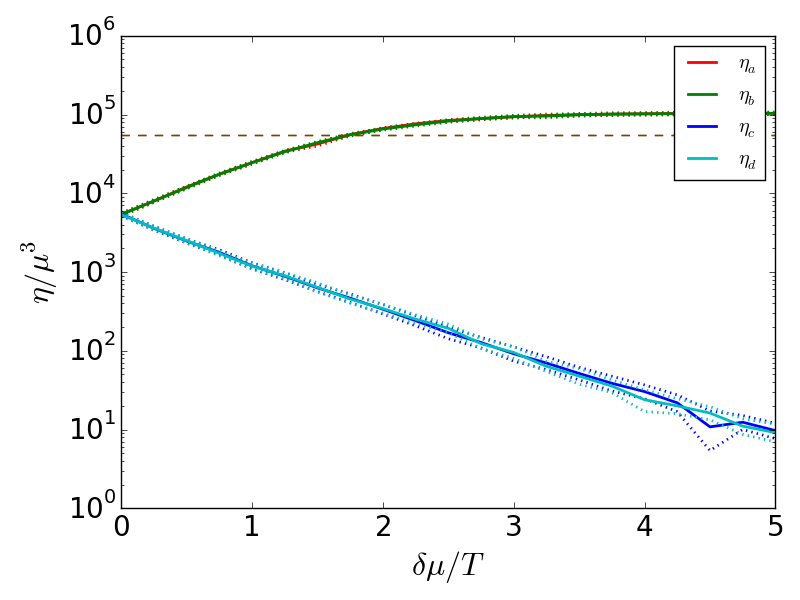}
  \caption{(color online) Plots of $L_i$, the diagonal entries of $R$, $\eta_i$
  and $\tau_i$ (anticlockwise from top left) with $|\bar{\calM}|^2$ given in
  Eq.~\ref{eq:M2_8} for the four species $a$, $b$, $c$, and $d$
  (Eq.~\ref{eq:fourspecies}). The dashed horizontal lines correspond to the
  values for unpaired matter (see the caption of
  Fig.~\ref{fig:Deta_vs_Deltadmu0_4} for details). The pairing is isotropic
  ($\bfb=0$).  $T/\mu=3.34\times10^{-4}$ and $\Delta/T=2.5$ are held fixed, and
  $\delta\mu$ is varied from $0$ to $2\Delta$. For $\Delta>\delta\mu$
  ($\delta\mu/T<2.5$ in all the plots) all fermionic excitations are gapped and
  all components of $R$ are exponentially suppressed. For $\Delta<\delta\mu$,
  branches $a$ and $b$ feature gapless fermionic excitations. The asymptotic
  value ($\delta\mu\gg\Delta$) for $\eta_a=\eta_b=2\eta^\un_1$. The dot dashed
  curves (yellow online) for $L$ are the simple forms given in
  Eq.~\ref{eq:Ldmularge} for $\delta\mu>\Delta$, and Eq.~\ref{eq:Ldmusmall} for
  $\delta\mu<\Delta$. The error bands are shown by the dashed curves of the
  color of the corresponding solid curves, and are associated with errors in
  the five dimensional Monte Carlo integration used for evaluating $R_{ij}$
  (Eq.~\ref{eq:RSphericalSimplified}).  $\tau_c$, $\tau_d$ are noisy but don't
  affect the final result for $\eta$.  ~\label{fig:Deta_vs_dmuDelta0250_4}
}
\end{figure*}

\subsubsection{Isotropic gapless pairing}
~\label{subsubsec:dmunonzero}
To analyze the effect of gapless fermions in this simple system let us consider an
isotropic gapless paired phase ($\bfb\rightarrow 0$, $\Delta>\delta\mu$). As
discussed in Sec.~\ref{sec:review},
this phase is unstable, but the analyses will give us insight into the
anisotropic calculation. In Fig.~\ref{fig:Deta_vs_dmuDelta0250_4} we keep 
$\Delta>T$ fixed ($\Delta/T=2.5$), and consider the effect of increasing 
$\delta\mu$ keeping $b$ equal to $0$.

Based on the discussion in Sec.~\ref{subsubsec:dmuzero}, we expect that for
$\Delta>\delta\mu$ both $L_i$ and $R_{ij}$ to be exponentially suppressed from
the unpaired value.  Whereas for $\Delta<\delta\mu$ (Eq.~\ref{eq:E1E2}) $E_1=0$
for $\xi=\pm\sqrt{\delta\mu^2-\Delta^2}$ and therefore the branches $a$ and $b$
in Eq.~\ref{eq:fourspecies} are gapless while the branches $c$ and $d$ are
gapped. Therefore, for $\delta\mu>\Delta$ we expect $L_i$, and $R_{ij}$
corresponding $a$ and $b$ to be unsuppressed compared to the unpaired
value.

More specifically, for $\Delta-\delta\mu\gg T$
\begin{equation}
\begin{split}
L_{a,b}(\Delta,\delta\mu) &\approx \frac{1}{2}L_1^{\un} 
e^{-\frac{(\Delta-\delta\mu)}{T}}
=L_a(\Delta=0, \delta\mu=0)e^{-\frac{(\Delta-\delta\mu)}{T}}\\
L_{c,d}(\Delta,\delta\mu) &\approx \frac{1}{2}L_1^{\un}
e^{-\frac{(\Delta+\delta\mu)}{T}}
=L_a(\Delta=0,
\delta\mu=0)e^{-\frac{(\Delta+\delta\mu)}{T}}~\label{eq:Ldmusmall}\;.
\end{split}
\end{equation}
In the top left panel of Fig.~\ref{fig:Deta_vs_dmuDelta0250_4}, the curves
for $L_i$ for the $a$ and $b$ branches (red and green online) split from the
$c$ and $d$ branches (blue and cyan online) on switching on a small
$\delta\mu$. The splitting increases as we increase $\delta\mu$ and for
$\delta\mu-\Delta\gg T$, near the gapless surfaces
$\xi=\pm\sqrt{\delta\mu^2-\Delta^2}$, both $a$ and $b$ branches resemble
unpaired fermions. Therefore,
\begin{equation}
\begin{split}
L_{a,b}(\Delta,\delta\mu)_{\delta\mu\gg\Delta} &\rightarrow L_1^{\un} = 2L_a(\Delta=0, \delta\mu=0)\\
L_{c,d}(\Delta,\delta\mu) &\approx \frac{1}{2}L_1^{\un}
e^{-\frac{(\Delta+\delta\mu)}{T}}~\label{eq:Ldmularge}\;.
\end{split}
\end{equation}
The limiting behaviors Eq.~\ref{eq:Ldmusmall} and Eq.~\ref{eq:Ldmularge} are
shown by dot dashed curves (yellow online) on the top left panel of
Fig.~\ref{fig:Deta_vs_dmuDelta0250_4}.

Similarly, for $\Delta-\delta\mu\gg T$ we expect $R_{ii}$ for each $i$ to be
suppressed compared to $R^{\un}_{ii}$. For example, for $\delta\mu=0$, we see
that for $\Delta/T=2.5$, $R_{ii}(\Delta=2.5T,\delta\mu=0)\approx
R^{\un}_{ii}/15$.  The suppression factor of $15$ is consistent with
$R_{ii}(\Delta=2.5T,\delta\mu=0)/R^{\un}_{ii}$ in
Fig.~\ref{fig:Deta_vs_Deltadmu0_4}.) 

As $\delta\mu$ is increased, the gapless branches $a$ (green online), $b$ (red
online) split from $c$ (blue online) and $d$ (cyan online), and eventully for $\delta\mu-\Delta\gg T$
\begin{equation}
\begin{split}
R_{aa,bb}(\Delta,\delta\mu)|_{\delta\mu\gg \Delta}\rightarrow R_{1}^{\un}
\end{split}
\end{equation}
the top right panel of Fig.~\ref{fig:Deta_vs_dmuDelta0250_4} shows this
behavior clearly. $R_{cc,dd}(\Delta,\delta\mu)\sim
R_{1}^{\un}\exp(-2(\Delta+\delta\mu)/T)$. The off-diagonal terms of $R$ are
also exponentially suppressed. 

This pattern is repeated for $\tau$ and $\eta$: $\tau_a$ ($\eta_a$), $\tau_b$
($\eta_b$) tend towards $\tau_1^\un$ ($\eta_1^\un$) for $\delta\mu-\Delta\gg T$
while $\tau_c$ ($\eta_c$), $\tau_d$ ($\eta_d$) are weakly (exponentially) suppressed.
All this is just a complicated way to obtain the well understood result (for eg. see Ref.~\cite{Alford:2005}) that
the transport in gapless superfluids is dominated by fermionic modes near the
gapless surfaces (Eq.~\ref{eq:GaplessXi})  and the result for the
total viscosity in the limit $\delta\mu-\Delta\gg T$ is the same as for an
unpaired system,
\begin{equation}
\eta(\delta\mu\gg\Delta) = \sum_i \eta_i \approx \eta_a+\eta_b \approx 2\eta_1^{\un}
~\label{eq:etagapless}\;.
\end{equation}

In light of the simple and intuitive result Eq.~\ref{eq:etagapless}, the  
analysis of this section seems needlessly complicated: one could restrict
to modes near the gapless spheres (mode $a$ near
$\xi=-\sqrt{\delta\mu^2-\Delta^2}$ and mode $b$ near
$\xi=+\sqrt{\delta\mu^2-\Delta^2}$) and neglect modes $c$ and $d$. Near the gapless
$\xi$, the dispersion of the modes can be approximated as linear, which means
that standard Fermi liquid techniques would lead to Eq.~\ref{eq:etagapless} for
gapless fermions if $\delta\mu\gg\Delta$. 

While the discussion of the isotropic gapless phase clarifies some aspects of
the calculation of the collision integrals for the FF phase, the details of the
analysis is more subtle because the pairing pattern is anisotropic. In the
following section we present the results for anisotropic 
pairing. 

\section{Results for anisotropic pairing}
\label{sec:anisotropic}
 From the dispersions Eq.~\ref{eq:E1E2polar}, one can think of the problem in
terms of an angle dependent Fermi surface splitting,
\begin{equation}
\delta\mu_{\rm{eff}}(\cos\theta) = \delta\mu +b \cos\theta\;.
\end{equation}

\begin{figure}[tbp]
\includegraphics[width=0.49\textwidth,clip=]{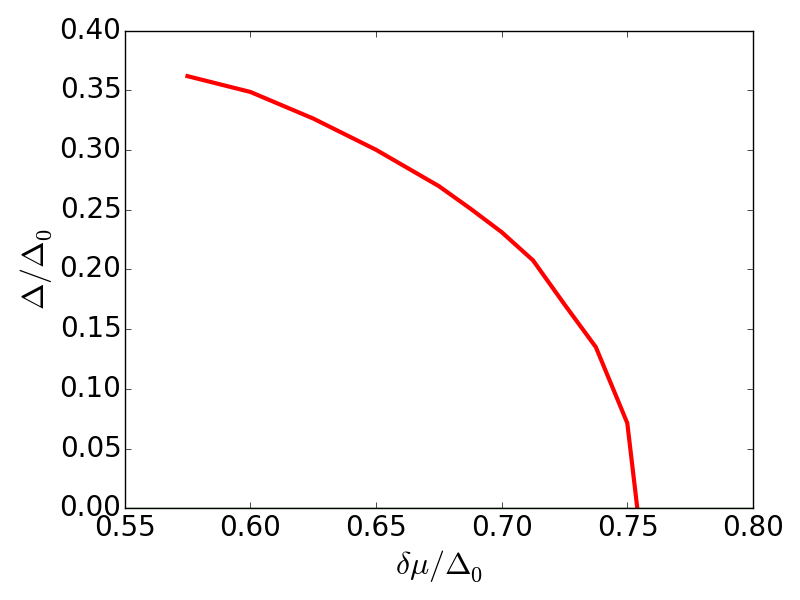}
  \caption{(color online) Plot of $\Delta/\Delta_{0}$ versus $\dmu/\Delta_{0}$
  for $b=\zeta\delta\mu$ from~\cite{Mannarelli:2006}.  $\Delta_{0}$ is the gap
  for the $2\SC$ phase in the absence of the Fermi surface split. At
  $\delta\mu=0.754\Delta_{0}$, $\Delta=0$. 
  ~\label{fig:Delta_vs_dmu}
}
\end{figure}

For $\delta\mu +b \cos\theta>\Delta$, species $a$ and $b$ are gapless, for
$\Delta>\delta\mu +b \cos\theta>-\Delta$ all four modes are gapped, and for
$\delta\mu +b \cos\theta<-\Delta$ modes $c$ and $d$ are gapless
(Eq.~\ref{eq:E1E2polar})~\cite{Alford:2000ze,Bowers:2003ye}. Therefore, the
shape of the gapless surface depends on the values of $\delta\mu$ and the gap
parameter $\Delta$ which is a function of $\delta\mu$. Furthermore,
even the nature of the gapless modes changes with the angle depending upon
whether $|\delta\mu_{\rm{eff}}(\cos\theta)|\gg \Delta$ (in which case the
dispersion near the gapless modes is linear and the mode velocity $v\approx 1$)
or $|\delta\mu_{\rm{eff}}(\cos\theta)|\approx \Delta$ (in which case the
dispersion near the gapless modes is quadratic and the mode velocity $v\approx
0$). Consequently, the results for the FF phase can not be obtained by a simple
extension of the isotropic analysis and a detailed calculation of the collision
integral is necessary. We perform this analysis for a simple model for quark
interactions: exchange of Debye screened, longitudinal gluons described by
Eqs.~\ref{eq:MqqDebye}~\ref{eq:simple_interaction}, in the next section
(Sec.~\ref{sec:simple_anisotropic}). In Sec.~\ref{sec:t123} we show the
analysis for the two-flavor FF phase with the realistic interaction: exchange
of dynamically screened, transverse $t^1$, $t^2$, and $t^3$ gauge bosons.

\begin{figure*}[tbp]
\includegraphics[width=0.49\textwidth,clip=]{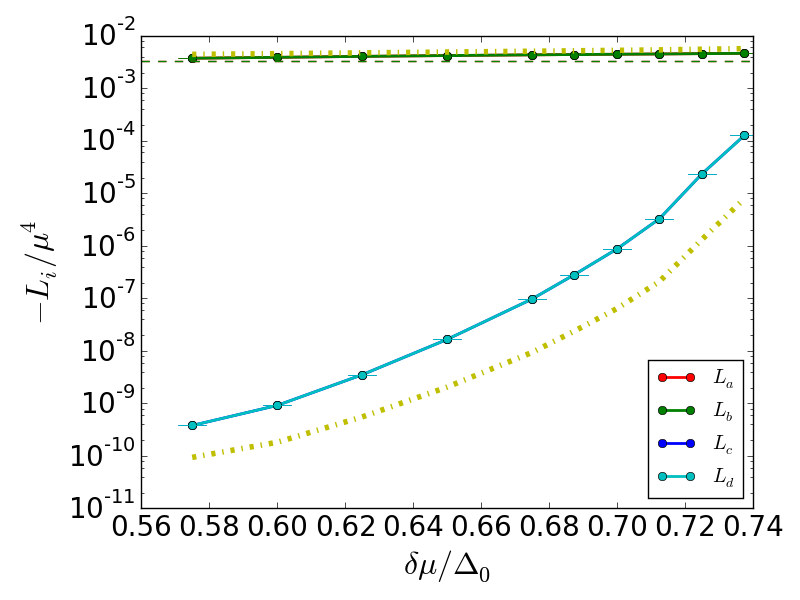}
\includegraphics[width=0.49\textwidth,clip=]{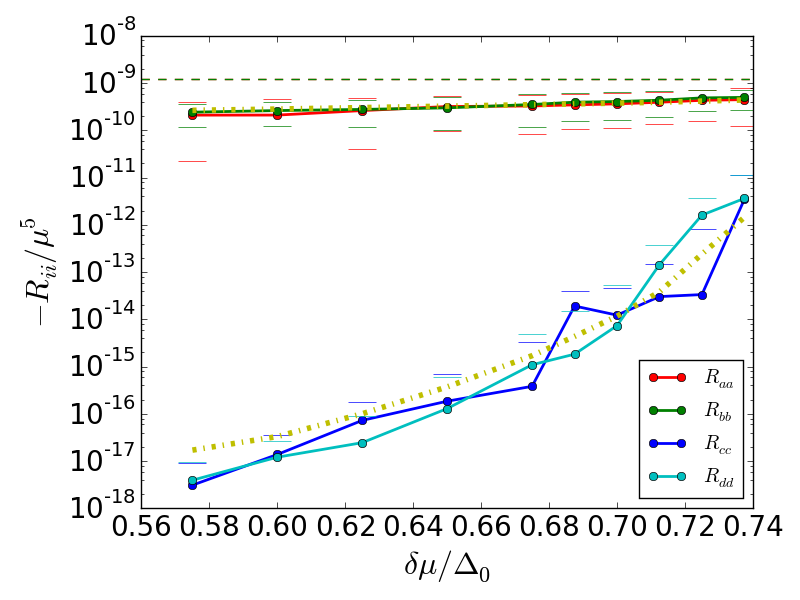}
\includegraphics[width=0.49\textwidth,clip=]{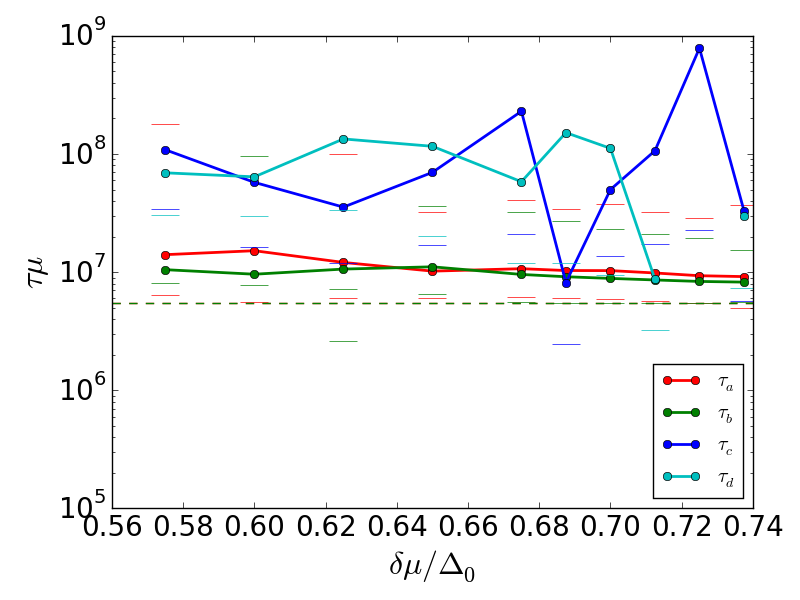}
\includegraphics[width=0.49\textwidth,clip=]{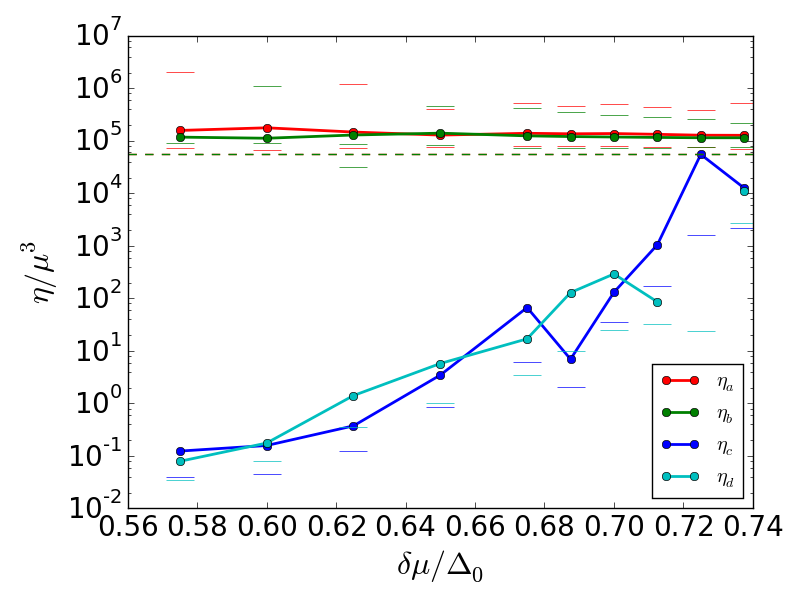}
  \caption{(color online) Plots of $L_i$, the diagonal entries of $R$, $\eta_i$
  and $\tau_i$ (anticlockwise from top left)  with $|\bar{\calM}|^2$ given in
  Eq.~\ref{eq:M2_8} for the four species $a$, $b$, $c$, and $d$
  (Eq.~\ref{eq:fourspecies}). The pairing is anisotropic with $b=1.19\delta\mu$
  and $\Delta$ is taken from Fig.~\ref{fig:Delta_vs_dmu}.
  $T/\mu=3.34\times10^{-4}$ is held fixed and $T/\Delta_0=0.02$. The central
  values are given by filled circles and the error bars are are shown by the
  dashes of the corresponding color. The error bars for $R_{ij}$ are associated
  with errors in the seven dimensional Monte Carlo integration used for
  evaluating $R_{ij}$ (Eq.~\ref{eq:RnonSpherical}) and are propagated to $\tau$
  and $\eta$. The large errors in $\tau_{c}$ and $\tau_{d}$ (blue and cyan
  online) do not affect the
  final $\eta$. The dashed horizontal lines correspond to the values for
  unpaired matter (see the caption of Fig.~\ref{fig:Deta_vs_Deltadmu0_4} for
  details). The upper dot dashed curves in the panels for $L_a$ and $L_b$
  ($R_{aa}$, $R_{bb}$) [yellow online] are associated with geometric reduction
  due to a smaller gapless surface as described in Eq.~\ref{eq:Lab_FFD}
  (Eq.~\ref{eq:Rab_FFD}). The lower dot dashed curves in the panels for $L_c$
  and $L_d$ ($R_{cc}$, $R_{dd}$) [yellow online] are associated with
  exponential reduction due to pairing. The results are discussed from Eqs.~\ref{eq:Lab_FFD}
  to Eq.~\ref{eq:eta_FFD} in the text.  ~\label{fig:Deta_vs_dmu_aniso} }
  \end{figure*}

\subsection{Debye screened gluon exchange}
\label{sec:simple_anisotropic}
To evaluate the integrals $L_i$ and $R_{ij}$ appearing in Eq.~\ref{eq:master} with the
dispersions Eq.~\ref{eq:E1E2polar} for any given $T$ and $\delta\mu$, we need
$\Delta$ and $b$ as a function of $\delta\mu$.  For a given $b$ and
$\delta\mu$, $\Delta$ can be found by solving the gap equation for the FF phase~\cite{Mannarelli:2006}.

We take the solution of the gap equation, $\Delta$ as a function of
$\delta\mu$, from Fig.~$3$ in Ref.~\cite{Mannarelli:2006}.  The calculations in
Ref.~\cite{Mannarelli:2006} were performed for three-flavor pairing with an FF
pairing pattern for $ud$ and $sd$ pairing. When the angle $\phi$ between the
two plane waves is $0$, the two pairing rings on the $u$ interfere minimally
and hence we use the corresponding solution to the gap equation (green curve in
Fig.~$3$ in Ref.~\cite{Mannarelli:2006}), reproduced in
Fig.~\ref{fig:Delta_vs_dmu}.

Near the second order phase transition at $\delta\mu<0.754\Delta_{0}$, 
$\Delta/\Delta_{0}\ll1$ and $b=\zeta\delta\mu$
minimizes the free energy. Here, 
$\Delta_0$ is the gap in the $2\SC$ phase in the absence of Fermi surface
splitting and the precise value of $\zeta$ is
$1.1996786...$~\cite{larkin1964nonuniform,fulde1964superconductivity,
Bowers:2002xr,Bowers:2003ye,Mannarelli:2006}. $\Delta$ is zero at the transition and increases with
decreasing $\delta\mu$ Fig.~\ref{fig:Delta_vs_dmu}. In this paper we will use
$b/\delta\mu=1.19$ for the entire range $\delta\mu/\Delta_{0}\in(0.575,
0.75)$.  

Even though the FF phase is not favoured compared to the homogeneous
pairing phase (in which the paired fermions are gapped since
$\Delta=\Delta_0>\delta\mu$) for $\delta\mu<0.707\Delta$, we will explore this
range because of the possibility that the higher plane wave states may be
favoured in a wider region and may be governed by similar physics. (Also see
Ref.~\cite{Leibovich:2001xr}.) 

In Fig.~\ref{fig:Deta_vs_dmu_aniso} we plot the results for $L$, $R$, $\tau$
and $\eta$ as a function of $\delta\mu$ with $b$ and $\Delta$ chosen as
described above. For a fixed $\mu$, which sets the overall scale, there are two dimensionless ratios 
that are needed to specify the transport properties of the FF phase as a
function of $\delta\mu$, $T/\mu$, and $\Delta_0/\mu$. We show the results for 
$T/\mu=3.34\times10^{-4}$ and $\Delta_0/\mu=1.67\times10^{-2}$ in
Fig.~\ref{fig:Deta_vs_dmu_aniso} though the results should be unchanged as long as
the hierarchy of scales $T\ll\Delta_0$ and $\Delta_0\ll\mu$ are satisfied. 

To get a concrete feel for numbers, one can take $\mu=300$MeV, $\Delta_0=5$MeV
(which is on the lower edge of $\Delta_0$ for model studies) and $T=0.1$MeV. We
choose $\Delta_0=5$MeV so that the exponential suppression $\exp(-\Delta/T)$
is small enough to be clearly visible in the results, but still large enough to
be accessible within numerical errors.

First considering $L_i$ (top left panel of Fig.~\ref{fig:Deta_vs_dmu_aniso}) as a function of $\delta\mu$, we note that the branches
$a$ and $b$ are gapless for 
\begin{equation}
\begin{split}
\cos\theta\in[\frac{-\delta\mu+\Delta}{b}, 1]~\label{eq:cthgapless_ab}
\end{split}
\end{equation}
throughout the range $\delta\mu/\Delta_{0}\in(0.575, 0.75)$
(Appendix~\ref{sec:blocking}). The gapless surface is the boundary of a crescent
with arc-length $1+\frac{\delta\mu-\Delta}{b}$ instead of $2$. Therefore, we expect
$L_a=L_b=L^{\un}\times{\rm{g}}(\frac{\delta\mu}{b},\frac{\Delta}{b})$ where
${\rm{g}}$ is a dimensionless function smaller than $1$ corresponding to the
limited range for which the modes are gapless. The following geometric 
estimate turns out to be reasonably accurate
\begin{equation}
\begin{split}
L_a=L_b&\approx
L^{\un}\times\frac{1}{2}(1+\frac{\delta\mu}{b}-\frac{\Delta}{b})
~\label{eq:Lab_FFD}\;.
\end{split}
\end{equation}
This is shown by the upper dot dashed line (yellow online).

On the other hand, the branches $c$ and $d$ are gapped for
$\delta\mu/\Delta_0<0.735$ and hence $L_c$ and $L_d$ are exponentially
suppressed. A rough estimate is  $L_c=L_d\sim  e^{-\Delta/T}$.
The lower dot dashed curve (yellow online) is $L^{\un}e^{-\Delta/T}$ and shows
the right form up to a scale. The proportionality factor depends
upon $T$. 

Similarly, $R_{ij}$ is expected to be suppressed by the square of the phase
space factor, 
\begin{equation}
\begin{split}
R_{aa}=R_{bb}&\approx
\frac{1}{2}R^{\un}_{11}\times(\frac{1}{2}(1+\frac{\delta\mu}{b}-\frac{\Delta}{b}))^2
~\label{eq:Rab_FFD}\;.
\end{split}
\end{equation}
This is shown by the upper dot dashed line (yellow online) on the top right
panel of Fig.~\ref{fig:Deta_vs_dmu_aniso}. $R_{cc}$ and $R_{dd}$ are exponentially suppressed and their evaluation is
noisy (see Appendix~\ref{sec:numerics} for details). However, a reasonable
estimate is $R_{cc}=R_{cc}\approx R^{\un}_{11}\times e^{-\Delta/T}$.
We stop the results for $\delta\mu/\Delta_0=0.74$ since for larger $\delta\mu$,
$\Delta\rightarrow 0$ and the numerical evaluation of the integral is noisy.

From Eqs.~\ref{eq:Lab_FFD} and ~\ref{eq:Rab_FFD} we expect,
\begin{equation}
\begin{split}
\tau_{a}&=\tau_b \approx 2\tau^{\un}\\
\eta_{a}&=\eta_b \approx \eta^{\un}~\label{eq:etaab_FFD}\;.
\end{split}
\end{equation}
$\tau_{c}$ and $\tau_d$ are noisy but do not contribute significantly to the
final $\eta$ and can be ignored.

Consequently,
\begin{equation}
\eta(b\neq 0) = \sum_i \eta_i \approx \eta_a+\eta_b \approx 2\eta_1^{\un}
~\label{eq:eta_FFD}\;.
\end{equation}

This is a remarkable result and is once again a consequence of the intricate
interplay between $\tau$ and $\eta$ that we saw in
Sec.~\ref{subsubsec:dmuzero}. The reduced phase space due to pairing
increases $\tau_a$ and $\tau_b$ by the same factor as it decreases $L_{a}$ and
$L_{b}$ because $R$ goes as the square of this factor. Consequently, in the
product, the two effects cancel out.

The key results that we obtained in this section are that
\begin{enumerate}
\item{$L_a$ and $L_b$ are only geometrically suppressed by the smaller gapless
surface (Eq.~\ref{eq:Lab_FFD})}
\item{$L_c$ and $L_d$ are exponentially suppressed }
\item{For Debye screened mediators, $R_{aa}$ and $R_{bb}$ are only geometrically suppressed
(Eq.~\ref{eq:Rab_FFD})}
\item{$R_{cc}$ and $R_{dd}$ are exponentially suppressed }
\end{enumerate}

\begin{figure}[tbp]
\includegraphics[width=0.49\textwidth,clip=]{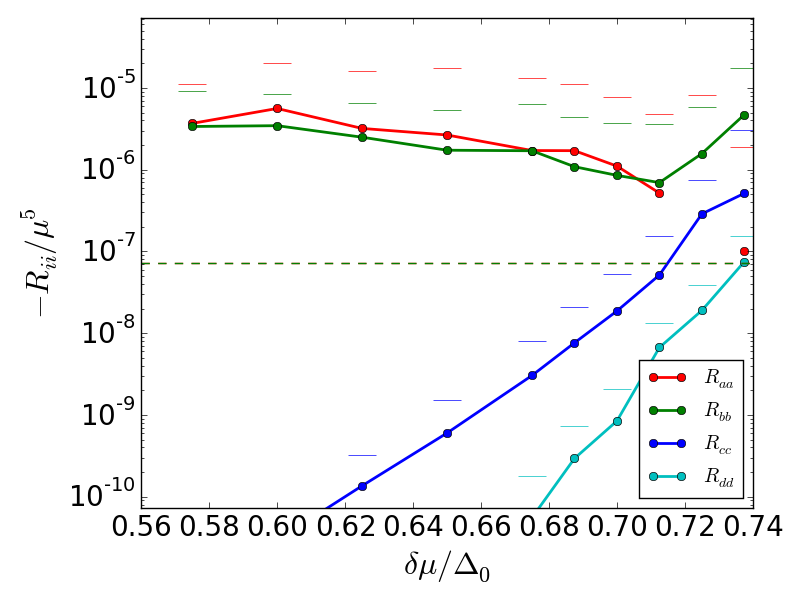}
  \caption{(color online) Plots of the diagonal entries of $R$ for the four
  species $a$, $b$, $c$, and $d$ (Eq.~\ref{eq:fourspecies}) with
  $|\bar{\calM}|^2$ given by Eq.~\ref{eq:M2SU2v2}. The pairing is anisotropic
  with $b=1.19\delta\mu$ and $\Delta$ is taken from
  Fig.~\ref{fig:Delta_vs_dmu}. $T/\mu=3.34\times10^{-4}$ is held fixed and
  $T/\Delta_0=0.02$ (the same as in Fig.~\ref{fig:Deta_vs_dmu_aniso}). The
  central values are given by filled circles and the error bars (from the Monte
  Carlo integration for $R$) are are shown by the dashes of the corresponding
  color. The dashed horizontal line corresponds to $R_{11}$ for unpaired
  matter with the interaction specified by
  Eqs.~\ref{eq:propT},~\ref{eq:PitSUN_f}.  \label{fig:Leta_vs_dmu_aniso} }
  \end{figure}

\begin{figure*}[tbp]
\includegraphics[width=0.49\textwidth,clip=]{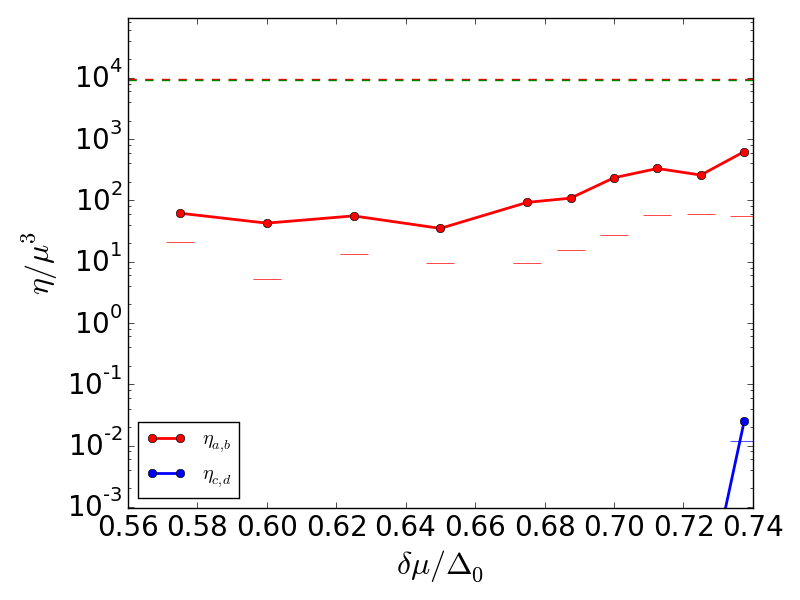}
\includegraphics[width=0.49\textwidth,clip=]{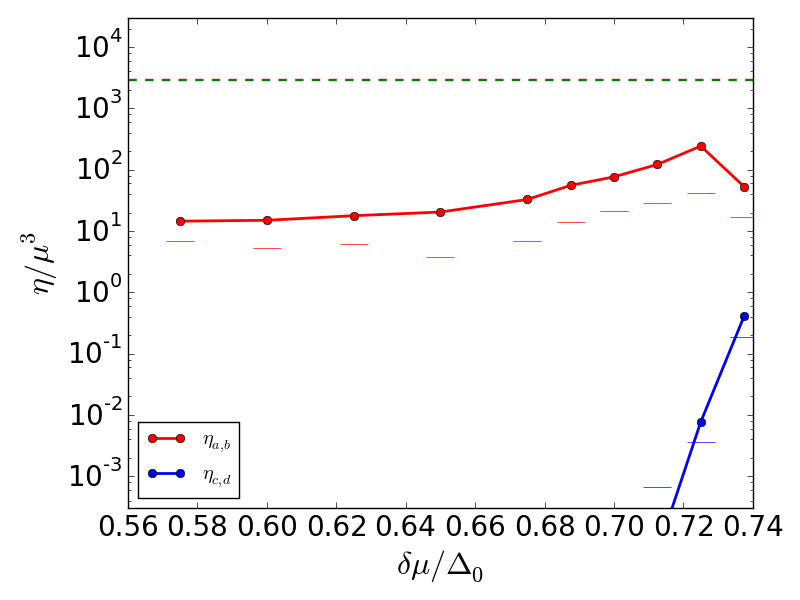}
\includegraphics[width=0.49\textwidth,clip=]{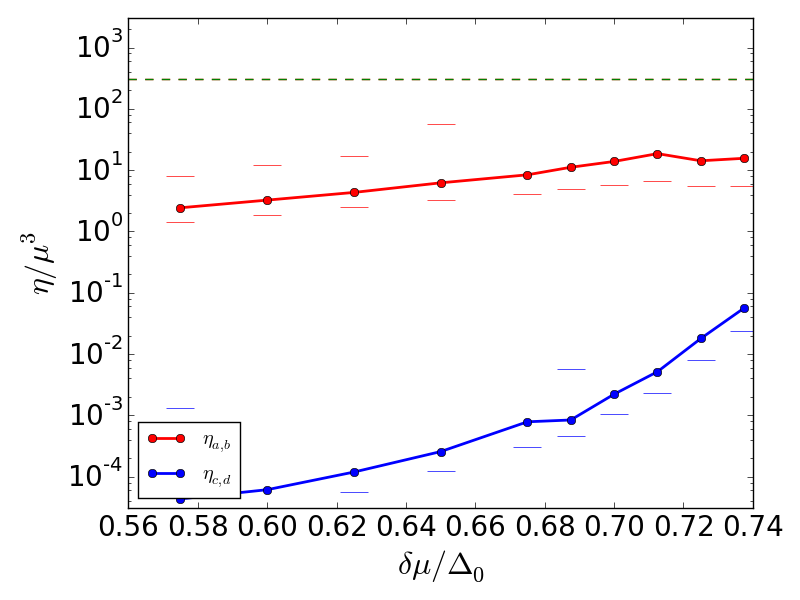}
\includegraphics[width=0.49\textwidth,clip=]{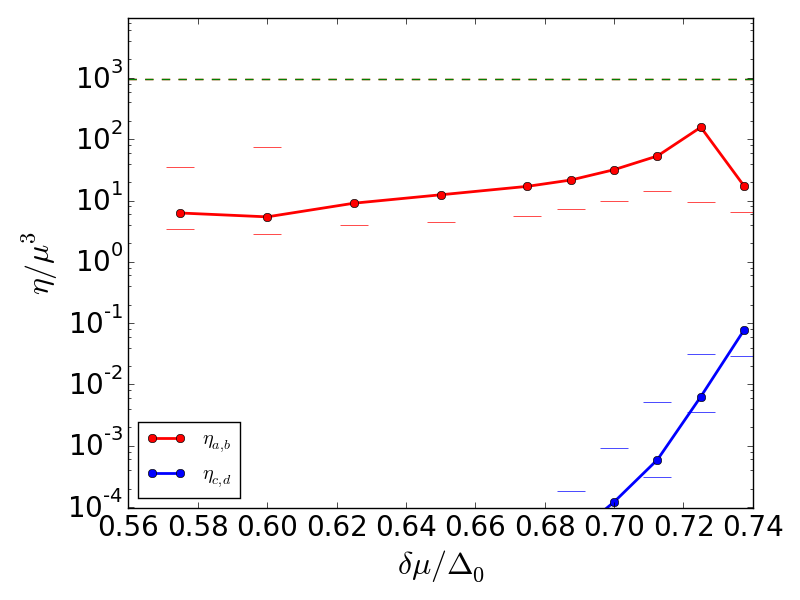}
  \caption{(color online) Plots of $\eta_i$
  (anticlockwise from top left) for the four species $a$, $b$,
  $c$, and $d$ (Eq.~\ref{eq:fourspecies}) for anisotropic pairing with
  $\bfb=1.19\delta\mu$. $T/\mu=3.34\times10^{-4}$ is held fixed and
  $T/\Delta_0=0.025$, $T/\Delta_0=0.05$, $T/\Delta_0=0.1$, $T/\Delta_0=0.2$.
  ~\label{fig:Leta_vs_dmu_aniso_vs_T} } \end{figure*}

\subsection{Two-flavor FF phase using $t^1$, $t^2$, $t^3$ exchange}
~\label{sec:t123}
Now we use the interaction mediated by the Landau damped $t^1$, $t^2$,
$t^3$ to calculate $\eta$ in the two-flavor FF phase. In the
Eq.~\ref{eq:master} the left hand sides, $L_i$, depend only on the spectrum of
quasi-particles and not the interaction. Therefore, they are not modified. For
$T/\Delta_0=0.02$ they are as shown in the top left panel of
Fig.~\ref{fig:Deta_vs_dmu_aniso} in Sec.~\ref{sec:simple_anisotropic}. 

The difference from Sec.~\ref{sec:simple_anisotropic} appears in the collision
integral $[R_{ij}]$, where the square of the matrix element
$|\calM(12\rightarrow34)|^2$ in Eq.~\ref{eq:master} (or
Eq.~\ref{eq:RnonSpherical}) is given by Eq.~\ref{eq:M2SU2} instead of
Eq.~\ref{eq:M2_8}. As discussed in Sec.~\ref{sec:interactions}, the matrix element in the collision
integral is (Eq.~\ref{eq:M2SU2})
\begin{equation}
\begin{split}
|i\bar{\cal{M}}|^2 &= 3(\frac{g}{2})^4\delta_{i_2i_4}\delta_{i_1i_3}\\
&
\frac{1}{4}\frac{1}{2p_12p_22p_32p_4}\tr[\slsh{p}_3\gamma^\mu\slsh{p}_1\gamma^\nu]
   \tr[\slsh{p}_4\gamma^\sigma\slsh{p}_2\gamma^\lambda]
   {D}_{\mu\sigma}{D}_{\nu\lambda}~\label{eq:M2SU2v2}\;.
\end{split}
\end{equation}

Evaluating the Dirac traces we obtain, 
\begin{equation}
\begin{split}
|\bar{\calM}|^2 &= 3 \bigl(\frac{g}{2}\bigr)^4  \Bigl( 
    L_t^{xx}\frac{1}{|\bfq^2-w^2+\Pi^{xx}_t|^2}
  + L_t^{yy}\frac{1}{|\bfq^2-w^2+\Pi^{yy}_t|^2}\\
  &
  + 2\Re e[L_t^{xy}\frac{1}{\bfq^2-w^2+\Pi^{xx}_t}\frac{1}{\bfq^2-w^2+(\Pi^{yy}_t)^*}] 
    \Bigr)\;,~\label{eq:M2SU2v3}
\end{split}
\end{equation}
where,
\begin{equation}
\begin{split}
L_t^{xx} &= ( \cos(\phi_1)\cos(\phi_2) )^2\\
L_t^{yy} &= ( \sin(\phi_1)\sin(\phi_2) )^2\\
L_t^{xy} &= \frac{1}{4}( \sin(2\phi_1)\sin(2\phi_2) )\\
L_t^{yx} &= L_t^{xy}\;,~\label{eq:M2SU2v4}
\end{split}
\end{equation}
and $\Pi^{xx}_t$, $\Pi^{yy}_t$ are specified by 
Eqs.~\ref{eq:Pi_t123}~\ref{eq:h_t123}.  In an isotropic system 
$\Pi^{xx}_t=\Pi^{yy}_t=\Pi$ for which Eq.~\ref{eq:M2SU2v3} matches the
expressions in Refs.~\cite{Heiselberg:1993,Alford:2014doa}.

Before exploring the main results of anisotropic pairing with anisotropic
Landau damping, we quickly review well known results for the the simpler
unpaired system. For isotropic Landau damping (Eq.~\ref{eq:PitSUN_f}) a rough
estimate~\cite{Heiselberg:1993,Alford:2014doa} is 
\begin{equation}
\begin{split}
\frac{R_{11}^{t\;\un}}{R_{11}^{\un}} 
&\approx  3\bigl(\frac{4\sqrt{2}g\mu}{\pi^2 T}\bigr)^{1/3}~\label{eq:Rl_vs_Rt}\;,
\end{split}
\end{equation}
where $R_{11}^{\un}$ is given by Eq.~\ref{eq:RijunpairedDv2}.  For
$T/\mu=3.34\times10^{-4}$ and $g=1$ the estimated enhancment factor is
numerically about $36$. Evaluating the collision integral numerically,
one obtains $R_{11}^{t\;\un}/\mu^5\approx-7.2\times10^{-8}$ shown by the 
dashed horizontal line (green online) in Fig.~\ref{fig:Leta_vs_dmu_aniso} .  Comparing with the numerical
result for the longitudinal gluon exchange, $R_{11}^{\un}/\mu^5\approx
-1.23\times10^{-9}$ from the dashed horizontal line in (green online) on 
the top right column of Fig.~\ref{fig:Deta_vs_dmu_aniso}, we see that numerically the
enhancement factor is $\sim58$, which shows that the estimate
(Eq.~\ref{eq:Rl_vs_Rt}) is in the right ballpark. This also impies we can
ignore the longitudinal gluons. This also applies to the FF phase.

The non-trivial results shown in Fig.~\ref{fig:Leta_vs_dmu_aniso} are the
values of $R_{aa}$ and $R_{bb}$ for the FF phase. The pairing is anisotropic
with $b=1.19\delta\mu$ and $\Delta$ is taken from Fig.~\ref{fig:Delta_vs_dmu}.
$T/\mu=3.34\times10^{-4}$ is held fixed and $T/\Delta_0=0.02$ (the same as the
parameters used in Fig.~\ref{fig:Deta_vs_dmu_aniso}). The geometric suppression
due to the smaller gapless surface (Eq.~\ref{eq:Rab_FFD}) would lead to a
reduction in $R_{aa}$ and $R_{bb}$. The actual
numerical evaluation for $R_{aa}$ and $R_{bb}$ shows that they are
{\it{enhanced}} over the unpaired isotropic result. This can be understood as
follows.

For small $\bfq$,
\begin{equation}
\begin{split}
E_{\bfp_3}-E_{\bfp_1} 
&\approx \frac{d E_{\bfp_1}}{d\xi_{p_1}}\delta\xi_{p_1} 
+ \frac{d E_{\bfp_1}}{d\cos\theta_{p_1}}\delta\cos\theta_{p_1} 
\approx  v_{p_1}\delta\xi_{p_1}~\label{eq:dE}\;,
\end{split}
\end{equation}
where
\begin{equation}
\begin{split}
v_{p_1} &= \frac{d E_{\bfp_1}}{d\xi_{p_1}}=\frac{\xi_{\bfp}}{\sqrt{\xi_{\bfp}^2+\Delta^2}}\;,\;\;
\delta\xi_{p_1} = |\bfp_1+\bfq|-|\bfp_1|\;.
\end{split}
\end{equation}
Therefore, the energy conserving $\delta$ functions can be approximately written as
$\delta({v_{p_1}}\delta\xi_{p_1}-\omega)\delta(-{v_{p_2}}\delta\xi_{p_2}-\omega)$.
For $|\delta\mu+b\cos\theta|\approx\Delta$, the dispersion is gapless for
$\xi\approx 0$ (Eq.~\ref{eq:excitations_dblocking}) implying $v_p\rightarrow 0$ 
and the jacobian for the $\delta$ functions diverges. Higher order terms in the
Taylor expansion of $\xi_p$ prevent $R_{ij}$ from diverging, but this shows up as an increase in $R_{ij}$. A similar 
phenomenon for the isotropic gapless CFL phase was seen earlier in
Ref.~\cite{Alford:2005pooja}.

There are two reasons why this effect is not seen in
Fig.~\ref{fig:Deta_vs_dmu_aniso} where the interaction is mediated by
Eq.~\ref{eq:simple_interaction}. First, the relative $-$ sign between the
coherence factor in Eq.~\ref{eq:M2_8} compared with the $+$ sign in
Eq.~\ref{eq:M2SU2} implies that the matrix element Eq.~\ref{eq:M2_8}
tends to $0$ if $\xi\rightarrow 0$ while Eq.~\ref{eq:M2SU2} does not (we also
discussed a similar effect in Sec.~\ref{sec:interactions}). Second, this effect
is more important if the collision integral is dominated by small $\bfq$
compared to $\mu$ and
the linear expansion (Eq.~\ref{eq:dE}) is accurate: The effect is therefore more
pronounced where the exchanged gauge boson is Landau damped.~\footnote{There is
an additional source of enhancement when the gauge boson polarization is given
by Eq.~\ref{eq:PitSUN_f} rather than Eq.~\ref{eq:Pi_t123}. Since $h<1$,
$|\calM|^2$ is larger in the anisotropic paired phase than the isotropic
unpaired phase. However, since $h$ is not $\ll 1$ for all $\cos\theta$, this is
not the dominant effect.} 

With the collision integral $R$ in hand, we can calculate $\tau$ and $\eta$.
The results for $\eta$ for four different values of the temperatures,
$T/\Delta_0=0.025$, $T/\Delta_0=0.05$, $T/\Delta_0=0.1$, $T/\Delta_0=0.2$, are
shown in Fig.~\ref{fig:Leta_vs_dmu_aniso_vs_T} and clearly show a reduction in
$\eta$ by a factor of roughly $100$ associated with the enhancement in the
collision integral. 

One technical comment about the numerical evaluation is that because of the 
more peaked nature of the integrand due to the two reasons mentioned above, the Monte Carlo integration 
for $R_{ij}$ (Eq.~\ref{eq:RnonSpherical}) converges very slowly.
Therefore to improve the statistics, we have averaged $R_{aa}$ and $R_{bb}$
(which should be equal), and $R_{ab}$ and $R_{ba}$ (which should be equal)
while making Fig.~\ref{fig:Leta_vs_dmu_aniso_vs_T} and added the errors in
quadrature. Similarly, we have combined the data for the $c$ and $d$ branches
in Fig.~\ref{fig:Leta_vs_dmu_aniso_vs_T}  

\begin{figure*}[tbp]
\includegraphics[width=0.49\textwidth,clip=]{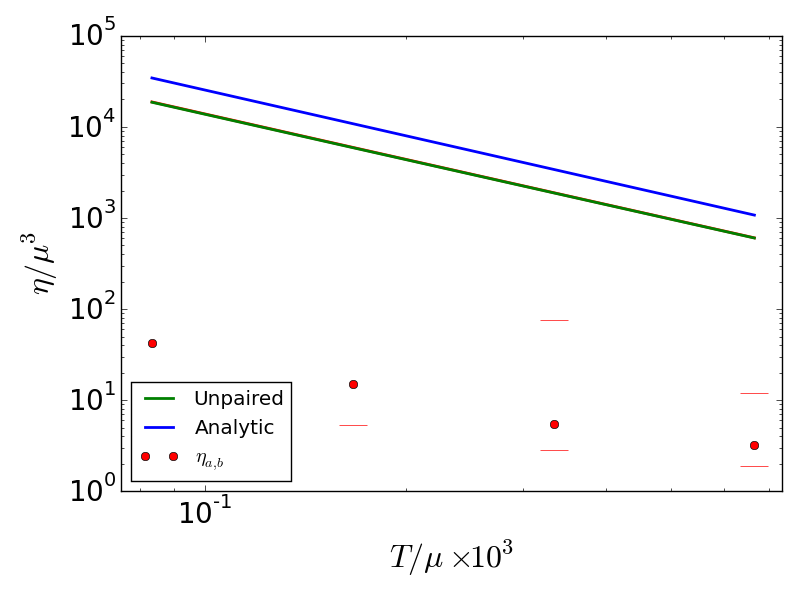}
  \caption{(color online) $\eta_i$ for species $a$ and $b$ for anisotropic
  pairing with $\Delta/\Delta_0=0.35$, $\delta\mu/\Delta_0=0.6$, and
  $\bfb=1.19\delta\mu$ as a function of $T$. The viscosities are obtained from
  Fig.~\ref{fig:Leta_vs_dmu_aniso_vs_T}.
  ~\label{fig:Leta_vs_T_aniso} } \end{figure*}

Fig.~\ref{fig:Leta_vs_T_aniso} shows viscosity as a function of $T$ for
anisotropic phases for $\delta\mu/\mu=10^{-2}$. The blue curve is the analytic
estimate
\begin{equation}
\begin{split}
\eta^{t\;\un}_{1} = \eta_{1}^{\un} \frac{1}{3}\bigl(\frac{4\sqrt{2}g\mu}{\pi^2
T}\bigr)^{-1/3}
~\label{eq:etal_vs_etat}
\end{split}
\end{equation}
based on Eq.~\ref{eq:Rl_vs_Rt}, where $\eta_{1}^{\un}$ is given in
Eq.~\ref{eq:etaunpairedD}.

The green curve is the numerical evaluation of the viscosity in the unpaired
phase using Eq.~\ref{eq:RSphericalSimplified}. The result for the FF phase is
denoted by the solid points with errors denoted by error bars. The error bars
are large enough that we do not attempt a fit but a rough description of the
data in this $T$ range is given by
\begin{equation}
\begin{split}
\eta \sim 10^{-2}\eta^{t\;\un}_{1}~\label{eq:Leta_vs_T_aniso}\;.
\end{split}
\end{equation}
Since $\eta$ is relatively flat with respect to $\delta\mu$ for all the $T$'s in
a wide range of $T\ll\Delta$ (Fig.~\ref{fig:Leta_vs_dmu_aniso}), we propose 
Eq.~\ref{eq:Leta_vs_T_aniso} as a fair parameterization of the shear
viscosity in the FF phase for $T\ll\Delta$ throughout the two-flavor FF window.
Eq.~\ref{eq:Leta_vs_T_aniso} is a concise summary of our main result.

\section{Conclusions}
~\label{sec:conclusions}
We present the first calculation of the shear viscosity of the two-flavor FF
phase of quark matter. 

We identify the low energy quasi-particles that play an important role in
transporting momentum and energy at low $T$. Due to the large density of states
near the Fermi surface, the $u$ and $d$ quarks, and the electrons dominate
transport properties if they are gapless. The ``blue'' $u$ and $d$ quarks, and
the electrons do not participate in pairing and their viscosity is the same as
in the $2\SC$ phase, calculated in Ref.~\cite{Alford:2014doa}.

The $ur-dg-ug-dr$ quarks pair and form Bogoliubov quasi-particles. The main
difference between the two-flavor FF and the $2\SC$ phase is that the spectra of
Bogoliubov quasi-particles feature gapless modes near surfaces that form the boundaries of
crescent shaped blocking regions. The technical advance made in the paper is
the calculation of their viscosity. 

The other low energy modes, the phonons associated with the compressions and
rarefactions of the iso-phase surfaces of the order
parameter~\cite{Mannarelli:2007bs}, are Landau
damped and do not contribute significantly to energy-momentum transport at low
temperatures.

By comparing the strength and the ranges of the particles that mediate quark
interactions (see Sec.~\ref{sec:interactions} for details) we conclude that the
dominant mechanism of scattering of the $ur-dg-ug-dr$ Bogoliubov
quasi-particles in the two-flavor FF phase is the exchange of transverse $t^1$,
$t^2$ and $t^3$ gluons which are Landau damped. Note in particular that the
longitudinal $t^1$, $t^2$ and $t^3$ gluons are Debye screened and can be
ignored. The Landau damping is anisotropic. The gluon polarization tensor is
given in Eqs.~\ref{eq:Pi_t123}~\ref{eq:h_t123}.  (More details about the
calculation of the gluon polarization will be given in a forthcoming
publication~\cite{InPreparation}.) We also show that the scattering of the
Bogoliubov quasi-particles via exchange of the Goldstone modes, and
due to their absorption and emission is subdominant for $T\ll \mu$ and can
be ignored.

We give a novel formalism to describe the scattering of Bogoliubov
quasi-particles. We separate the two branches of the quasi-particle dispersions
(Eq.~\ref{eq:E1E2polar}) into $\xi>0$ and $\xi<0$ (Eq.~\ref{eq:fourspecies},
Fig.~\ref{fig:EBranches}) modes.  This
doubles the dimension of the collision integral matrix $[R_{ij}]$, with four modes
$a$, $b$, $c$ and $d$ (Eq.~\ref{eq:fourspecies}). The utility of this formalism is
that it interpolates between two pairing regimes. When $T$ is comparable to
$\Delta$ (near the superconducting phase transition) the collision integral
includes processes involving $a+c\rightarrow a+c$ (``inter-band'' processes).
When $T\ll\Delta$ the collision integral only features $a+a\rightarrow a+a$,
$b+b\rightarrow b+b$, and $a+b\rightarrow a+b$ (``intra-band'' processes).
Pair breaking processes are frozen. For isotropic gapless pairing in this
regime ($b=0$, $\delta\mu>\Delta$) a simpler
formalism involving only the $E_1$ branch (Eq.~\ref{eq:E1E2polar}) would be
sufficient. The subtlety in the FF phases is that both $E_1$ and $E_2$
branches can become gapless depending on the values of $b$, $\delta\mu$,
$\Delta$ and the angle $\theta$ of the momentum with the $\hat{b}$ direction.
Our formalism allows for all these possibilities.

Our main result is given in
Fig.~\ref{fig:Leta_vs_dmu_aniso_vs_T} and Fig.~\ref{fig:Leta_vs_T_aniso}. The key result is
that the viscosity of the $ur-dg-ug-dr$ quarks for a wide range of $\delta\mu$
in the LOFF window is reduced by a factor of roughly $10^{-2}$ compared to the
viscosity of unpaired quarks interacting via Landau damped transverse gluons.
This is summarized in a compact parameterization of the viscosity
Eq.~\ref{eq:Leta_vs_T_aniso}.

This is a surprising result. In the $2\SC$ phase the $ur-dg-ug-dr$ quarks are
fully gapped and are frozen. In the FF phase the geometric area of the gapless
surface is reduced by pairing. But at the same time the phase space for
collisions is also reduced by the square of the geometric factor. Hence this
simple argument suggests that the shear viscosity should be comparable to that
for unpaired quarks. Indeed this is precisely what happens if the interaction
between the quarks is assumed to be mediated by Debye screened longitudinal
gluons corresponding to the broken generators, as shown in
Fig.~\ref{fig:Deta_vs_dmu_aniso}. For long range interactions (dominated by
smaller momentum exchanges), however there is an additional effect due to the
increase of the density of states satisfying the energy conservation equation
due to small velocities over a part of the Fermi surface. The collision
integral is enhanced and the shear viscosity is reduced
(Eq.~\ref{eq:Leta_vs_T_aniso}). This effect is particularly pronounced for
$t^1$, $t^2$, $t^3$ gluons because the coherence factors in the matrix element 
don't cancel (Eq.~\ref{eq:M2SU2}).

 Comparing the shear viscosity of the paired quarks in the FF phase with the
contribution of the $ub$ and $db$ quarks Ref.~\cite{Alford:2014doa} we
note that it is suppressed due to two effects.  First, the paired quarks
dominantly scatter via Landau damped gluons. As discussed above
(Eq.~\ref{eq:Leta_vs_T_aniso}), the viscosity is further reduced by a factor
$100$ due to pairing effects. In contrast the transverse gluons exchanged by
$b$ quarks all have a Meissner mass and only the transverse $\tilde{Q}$ photon is
Landau damped. For 
$(\mu/T)^{1/3}\gtrsim (\alpha_s^{3/2}/\alpha^{5/3})$, $\tilde{Q}$ exchange 
dominates $b-b$ scattering and $\eta_{\rm{paired}}/\eta_b\sim
10^{-2}\alpha_s^{5/3}/\alpha^{5/3}$. For 
$(\mu/T)^{1/3}\lesssim (\alpha_s^{3/2}/\alpha^{5/3})$, gluon exchange 
dominates $b-b$ scattering and $\eta_{\rm{paired}}/\eta_b\sim
10^{-2}(T/\mu)^{1/3}$. This implies that the sum of the viscosities
due to the $b$ quarks and the electrons calculated for the $2\SC$ phase
in~\cite{Alford:2014doa} gives the shear viscosity of the two-flavor FF
phase to a very good approximation.

In this paper we have only given results for the projection operator
$\Pi^{(0)}$. It will be interesting to repeat the calculation for the other
projection operators $\Pi^{(1)}$ and $\Pi^{(2)}$ ($\Pi^{(3)}$ and $\Pi^{(4)}$
are Hall projections and the associated viscosities are expected to be zero in a system without magnetic
fields). The difference between the three is related to the anisotropy in the
shear viscosity tensor and might have interesting implications, although
condensation in multiple direction will tend to isotropise the shear viscosity.

Looking ahead, one can think of several advances that can improve upon our
calculation; for example considering more complicated pairing patterns and
including the strange quark. In the following discussion we attempt to present
a plausible picture of how the shear viscosity of these more realistic phases
might behave based on the intuition gained from our calculation, and make some 
speculations about the physical implications for neutron star phenomenology.

For example, one can consider more realistic two-flavor LOFF
structures~\cite{Bowers:2002xr} involving multiple plane waves. Even these more
complex condensates~\cite{Casalbuoni:2001gt} feature gapless fermionic
excitations, and while the details are more complicated, the two main features
(a) gapless quasi-particle excitations over a Fermi surface with a complicated
shape (b) transverse $t^1$, $t^3$, and $t^3$ gluons are Landau damped, are
expected to be present also in these more complicated phases. Consequently the
shear viscosity of the $ur-dg-ug-dr$ quarks can be ignored as in the FF phase.  

Depending on the strange quark mass and the coupling strength between quarks,
quark matter in neutron stars may also feature strange quarks. In the $2\SC+s$
phase, the electron number is suppressed and numerous unpaired strange quarks
contribute to transport. One expects their contributions to be comparable to
that of the $ub$ and $db$ quarks in the $2\SC$ phase. The same is also expected
for the two FF$+s$ phase. ~\footnote{Some details will be modified. The $t^1$,
$t^2$ and $t^3$ will get additional Landau damping contributions from the $s$
quarks. The qualitative answers, however, are not expected to change.} In all these cases our calculation suggests that whether unpaired $s$ quarks
are present or not it is impossible to distinguish two flavor LOFF pairing from
$2\SC$ pairing by comparing the shear viscosity of the two phases.  The paired
quarks are suppressed, though not exponentially.
  
Qualitative differences, however, are expected to arise if the strange quarks
are also paired. That is, the three-flavor FF~\cite{Mannarelli:2006} or
three-flavor LOFF phases~\cite{Rajagopal:2006ig}. In these phases the electrons
are few in number and can be ignored as a first approximation. The fermionic
excitations are gapless on non-trivial surfaces~\cite{Mannarelli:2006}, as in
the two flavor case. But importantly all Meissner masses are
finite~\cite{Ciminale:2006sm} [even if they are smaller by a factor
$\sim(\Delta/\delta\mu)^2$ compared to their values in the CFL phase].
Scattering of the Bogoliubov quasi-particles is carried out by long ranged
$\tilde{Q}$ photon (weakly coupled) and short ranged gluons (screened).
Furthermore, for statically screened gauge boson exchanges
(Fig.~\ref{fig:Deta_vs_dmu_aniso}) we find that that the geometric reduction in
the size of the Fermi surface does not lead to the suppression of the shear
viscosity relative to unpaired matter since the geometric factor cancels out in
$\eta$. If this intuition holds for the three-flavor FF, it would imply that
three-flavor FF where all the gluons are screened
(and perhaps even three flavor LOFF), has a significantly larger shear viscosity compared to
unpaired quark matter. 

For example for $(\mu/T)^{1/3}\gtrsim (\alpha_s^{3/2}/\alpha^{5/3})$, we expect
that the viscosity of three-flavor FF phases will be larger than the results
for unpaired quark matter by a factor $\sim(\mu\frac{\Delta}{\delta\mu}/T)^{1/3}$.
On the $T$ versus $\nu$ (rotational frequency) plot (for example, see the left 
panel of Fig.~$1$ in \cite{Alford:2013pma}), it implies that the stability edge
determined by shear viscosity \{left most curve (blue online) in Fig.~$1$ in
\cite{Alford:2013pma}\} for three-flavor FF will be on the right of the curve
shown for unpaired (but interacting) quark matter. This may affect the observed
distribution of the neutron stars in the $T-\Omega$ plane~\cite{Levin:1998wa}.

Currently, the temperatures of several fast spinning neutron stars are not well
known (they are simply upper bounds), and no neutron stars are known which lie
close to the shear viscosity stability edge (cooler than $10^7$ K). A discovery
of such a star could in principle distinguish between three-flavor paired and unpaired
quark matter as the source of damping of $r-$modes, if one can simultaneously
pin down the damping by other mechanisms (for example phase boundaries or 
Eckman layers).

Making this speculation more quantitative will require finding a better
estimate of the LOFF window in three-flavor quark matter, making models of
hybrid neutron stars with quark matter and LOFF cores with equations of states
compatible with recent constraints on masses and radii of neutron stars, and a
calculation of the shear viscosity in three-flavor LOFF phases as a
function of the $T$ and $\mu$.   

Presently, stronger constraints on the viscosities of dense matter come from
hotter, fast rotating neutron stars. The bulk viscosity provides the damping
mechanism in this regime and only selected phases of dense matter are
consistent with the observations unless the $r-$ modes saturate at small
amplitudes~\cite{Alford:2013pma}. Since the bulk viscosity in quark matter does
not involve the scattering between two quarks it is not sensitive to the nature
of screening of the gluons but it is sensitive to the presence of
gapless quark modes. Therefore it may be interesting to calculate the bulk
viscosity in these phases to find out how the geometric reduction in the
gapless surface affects the bulk viscosity in LOFF phases.

\section{Acknowledgements}
We thank the workshop on the Phases of Dense Matter organized at the INT in
University of Washington, Seattle, where part of this work was completed. We
thank Mark Alford, Nils Andersson, Sophia Han, Sanjay Reddy, Andras Schmitt and especially Kai
Schwenzer for discussions. Sreemoyee Sarkar acknowledges the support of DST under INSPIRE Faculty award.
We also thank Prashanth Jaikumar for comments.

\appendix

\section{Pairing and blocking regions}
\label{sec:blocking}
At $T=0$, all energy eigenstates with $E<0$ are filled and the $E>0$
eigenstates are empty. This defines the pairing and the blocking
regions~\cite{Alford:2000ze,Bowers:2003ye}. (For a geometrical description 
see Fig. $2$ in Ref.~\cite{Alford:2000ze}.)

In the pairing region $E_1<0$ and $E_2>0$ (Eq.~\ref{eq:E1E2polar}) and the
quasi-particle excitation energies (the magnitude of the dispersion relations)
as a function momenta (in terms of $\xi$ and $\cos\theta$) are
\begin{equation}
\begin{split}
E_-(\xi, \theta) &= E_1 = -\delta\mu - b\cos\theta +
\sqrt{\xi^2+\Delta^2}\\
E_+(\xi, \theta) &= E_2 = \delta\mu + b\cos\theta + \sqrt{\xi^2+\Delta^2}\;.
~\label{eq:excitation_pairing_region}
\end{split}
\end{equation}
The pairing region is expressed by the relation
\begin{equation}
\begin{split}
\cos\theta &\in [-1, {\rm{max}}(\frac{-\delta\mu-\Delta}{b},-1)] \\
\xi &\in(-\infty,-\sqrt{(\delta\mu+b\cos\theta)^2-\Delta^2})\\
&
\cup(\sqrt{(\delta\mu+b\cos\theta)^2-\Delta^2},\infty)\\
&{\rm{or}}\\
\cos\theta &\in [\frac{-\delta\mu-\Delta}{b},
\frac{-\delta\mu+\Delta}{b}] \\
\xi &\in (-\infty, +\infty)\\
&{\rm{or}}\\
\cos\theta &\in [{\rm{min}}(\frac{-\delta\mu+\Delta}{b},1), 1]\ \\
\xi &\in(-\infty,-\sqrt{(\delta\mu+b\cos\theta)^2-\Delta^2})\\
&
\cup(\sqrt{(\delta\mu+b\cos\theta)^2-\Delta^2},\infty)
\;.
\end{split}
\end{equation}
The system is cyllindrically symmetric and polar angle $\phi\in[0,2\pi]$.

The complementary region in momentum space is the blocking region consists of
two disconnected regions crescent shaped regions near the Fermi sphere.  The boundaries 
of the blocking regions are the place where the dispersions
Eq.~\ref{eq:E1E2polar} are gapless.

In the $d$ (larger) blocking region, $E_1>0$, $E_2>0$. Then,
\begin{equation}
\begin{split}
E_-(\xi, \theta) &= -E_1 = \delta\mu + b\cos\theta 
  - \sqrt{\xi^2+\Delta^2}\\
E_+(\xi, \theta) &= E_2 =\delta\mu + b\cos\theta + \sqrt{\xi^2+\Delta^2}\;.
~\label{eq:excitations_dblocking}
\end{split}
\end{equation}
This requires,
\begin{equation}
\begin{split}
\cos\theta &\in [{\rm{min}}(\frac{-\delta\mu+\Delta}{b},1), 1] \\
\xi &\in(-\sqrt{(\delta\mu+b\cos\theta)^2-\Delta^2},
\sqrt{(\delta\mu+b\cos\theta)^2-\Delta^2})\;.
\end{split}
\end{equation}
At the edge of the $d$ blocking region, $E_-$ is gapless. From
Fig.~\ref{fig:Delta_vs_dmu}, $\Delta<0.6\delta\mu$ for
$\delta\mu\in[0.55,0.754]\Delta_{0}$ and therefore
$(-\delta\mu+\Delta)/b=(-\delta\mu+\Delta)/1.19\delta\mu<0$,
and hence the $d$ blocking region never closes.

The $u$ (smaller) blocking region is defined as the momenta where $E_1<0$, $E_2<0$. Then,
\begin{equation}
\begin{split}
E_-(\xi, \theta) &= -\delta\mu - b\cos\theta 
   - \sqrt{\xi^2+\Delta^2}\\
E_+(\xi, \theta) &= -\delta\mu - b\cos\theta + \sqrt{\xi^2+\Delta^2}
~\label{eq:excitations_ublocking}
\end{split}
\end{equation}
This requires,
\begin{equation}
\begin{split}
\cos\theta &\in [-1, {\rm{max}}(\frac{-\delta\mu-\Delta}{b},-1)] \\
\xi &\in
{-\sqrt{(\delta\mu+b\cos\theta)^2-\Delta^2},\sqrt{(\delta\mu+b\cos\theta)^2-\Delta^2}}\;.
\end{split}
\end{equation}
At the edge of the $u$ blocking region, $E_+$ is gapless. For 
$(-\delta\mu-\Delta)/b=(-\delta\mu-\Delta)/1.19\delta\mu<-1$,
the $u$ blocking region closes and the associated gapless surface disappears.
This happens for $\Delta/\delta\mu>0.19$ which corresponds to
$\delta\mu/\Delta_{0}<0.735$.

\section{Evaluation of the collision integral}
\label{sec:numerics}

The evaluation of the left hand side ($L_i$) of the Boltzmann equation
(Eq.~\ref{eq:master}) is straightforward. For $\Delta=0, \delta\mu=0, b=0$ the
integral can be performed analytically for
$T\ll\mu$~\cite{Heiselberg:1993,Alford:2014doa} and gives $L^\un$
(Eq.~\ref{eq:Liunpaired}). For $\delta\mu\gg\Delta$, Fermi liquid theory
predicts that the result is $2L^\un$. For generic $\Delta, \delta\mu, b$ one
can use Azimuthal symmetry to write the integral as a two dimensional integral
which can be evaluated easily numerically.

The general evaluation of the collision integral $R$ is more difficult. For $b=0$,
spherical symmetry can be used to simplify the
integral~\cite{Heiselberg:1993,Alford:2014doa}. Performing $d^3\bfp_4$ integration
using the momentum $\delta$ function, changing variables from $\bfp_3$ to
$\bfq=\bfp_3-\bfp_1$, and using Eq.~\ref{eq:FermiSurface}
\begin{equation}
\begin{split}
&R_{ij}^{(n)}=-\frac{1}{\gamma^{(n)}}\frac{1}{T} \nu_2\\ 
&\frac{(2\pi)^2}{(2\pi)^9}\mu_i^2\mu_j^2 \int\;
d\xi_{p_1} d\xi_{p_2} d\phi_{p_2} dq (4\pi) 
\int d\omega\\
&
|\calM(12\rightarrow34)|^2
[f_1f_2(1-f_3)(1-f_4)]\\
&
3\Bigl[\phi^{1}\cdot(\psi^{(0)1}\tau^{(0)1}-\psi^{(0)3}\tau^{(0)3}) \\
 &
 \phantom{+}+
 \phi^{1}\cdot(\psi^{(0)2}\tau^{(0)2}-\psi^{(0)4}\tau^{(0)4})
 \Bigr]|_{E_{p_3}-E_{p_1}=\omega=E_{p_2}-E_{p_4}}~\label{eq:RSphericalSimplified}\;.
\end{split}
\end{equation}
The azimuthal angle $\phi_{p_1}$ can be set to be $0$.

The five dimensional integration can be done easily using Monte Carlo techniques and
we find converged answers with $10^5-10^6$ points. The results for $R_{ij}$ shown in
Figs.~\ref{fig:Deta_vs_Deltadmu0_4},~\ref{fig:Deta_vs_dmuDelta0250_4} are
obtained using $10^6$ points. The error bars are the estimated error
in the Monte Carlo integration. More points are required for
Landau damped exchange bosons since the $|\calM(12\rightarrow34)|^2$ and hence
the integrand is more sharply peaked at $\bfq\rightarrow 0$.

For the anisotropic case, simplifications associated with spherical symmetry
are not applicable and one is left with a seven dimensional integral. The
direction $z$ is taken as the direction of the unit vector parallel to $\bfb$.
\begin{equation}
\begin{split}
&R_{ij}^{(n)}=-\frac{1}{\gamma^{(n)}}\frac{1}{T} \nu_2\\ 
&\frac{(2\pi)^2}{(2\pi)^9}\mu_i^2\mu_j^2 \int\;
d\xi_{p_1} d\cos\theta_{p_1} d\xi_{p_2} d\cos\theta_{p_2} d\phi_{p_2} dq_z
\int d\omega\\
&
|\calM(12\rightarrow34)|^2
\frac{1}{J_q}[f_1f_2(1-f_3)(1-f_4)]\\
&
3\Bigl[\phi^{1}\cdot(\psi^{(0)1}\tau^{(0)1}-\psi^{(0)3}\tau^{(0)3}) \\
 &
 \phantom{+}+
 \phi^{1}\cdot(\psi^{(0)2}\tau^{(0)2}-\psi^{(0)4}\tau^{(0)4})
 \Bigr]|_{E_{p_3}-E_{p_1}=\omega=E_{p_2}-E_{p_4}}
\;.~\label{eq:RnonSpherical}
\end{split}
\end{equation}
The azimuthal angle $\phi_{p_1}$ can be set to be $0$.

Because of the higher dimensions the convergence of the Monte Carlo evaluation
of Eq.~\ref{eq:RnonSpherical} is much slower compared to
Eq.~\ref{eq:RSphericalSimplified}. In making Fig.~\ref{fig:Deta_vs_dmu_aniso} where the
mediator is Debye screened, we used $7\times 10^7$ points and obtained
reasonably converged results. The evaluation of $R_{ij}$ for
Figs.~\ref{fig:Leta_vs_dmu_aniso},~\ref{fig:Leta_vs_dmu_aniso_vs_T} was more
computationally involved because the dispersions as well as the
interactions are anisotropic and the interactions are mediated by Landau damped
gluons. Thus the integrand is sharply peaked at small $q$. To evaluate $R_{ij}$
for Figs.~\ref{fig:Leta_vs_dmu_aniso}~\ref{fig:Leta_vs_dmu_aniso_vs_T} we used
$2.2\times10^8$ Monte Carlo points which took about a week on a modern cluster
with $100$ nodes. The most challenging part of the computation is
simultaneously solving for the momentum energy conservation constraint
$E_{p_3}-E_{p_1}=\omega=E_{p_2}-E_{p_4}$ and required writing a robust solver
in $c$.
The convergence of $R_{ij}$ is poor, as seen by the large error bars in
$R_{ij}$ and $\eta$. Substantial improvements would require significantly higher
computing resources and/or a better algorithm which we leave for future.

\bibliographystyle{apsrev4-1}
\bibliography{local}

\begin{thebibliography}{139}%
\makeatletter
\providecommand \@ifxundefined [1]{%
 \@ifx{#1\undefined}
}%
\providecommand \@ifnum [1]{%
 \ifnum #1\expandafter \@firstoftwo
 \else \expandafter \@secondoftwo
 \fi
}%
\providecommand \@ifx [1]{%
 \ifx #1\expandafter \@firstoftwo
 \else \expandafter \@secondoftwo
 \fi
}%
\providecommand \natexlab [1]{#1}%
\providecommand \enquote  [1]{``#1''}%
\providecommand \bibnamefont  [1]{#1}%
\providecommand \bibfnamefont [1]{#1}%
\providecommand \citenamefont [1]{#1}%
\providecommand \href@noop [0]{\@secondoftwo}%
\providecommand \href [0]{\begingroup \@sanitize@url \@href}%
\providecommand \@href[1]{\@@startlink{#1}\@@href}%
\providecommand \@@href[1]{\endgroup#1\@@endlink}%
\providecommand \@sanitize@url [0]{\catcode `\\12\catcode `\$12\catcode
  `\&12\catcode `\#12\catcode `\^12\catcode `\_12\catcode `\%12\relax}%
\providecommand \@@startlink[1]{}%
\providecommand \@@endlink[0]{}%
\providecommand \url  [0]{\begingroup\@sanitize@url \@url }%
\providecommand \@url [1]{\endgroup\@href {#1}{\urlprefix }}%
\providecommand \urlprefix  [0]{URL }%
\providecommand \Eprint [0]{\href }%
\providecommand \doibase [0]{http://dx.doi.org/}%
\providecommand \selectlanguage [0]{\@gobble}%
\providecommand \bibinfo  [0]{\@secondoftwo}%
\providecommand \bibfield  [0]{\@secondoftwo}%
\providecommand \translation [1]{[#1]}%
\providecommand \BibitemOpen [0]{}%
\providecommand \bibitemStop [0]{}%
\providecommand \bibitemNoStop [0]{.\EOS\space}%
\providecommand \EOS [0]{\spacefactor3000\relax}%
\providecommand \BibitemShut  [1]{\csname bibitem#1\endcsname}%
\let\auto@bib@innerbib\@empty
\bibitem [{\citenamefont {Demorest}\ \emph {et~al.}(2010)\citenamefont
  {Demorest}, \citenamefont {Pennucci}, \citenamefont {Ransom}, \citenamefont
  {Roberts},\ and\ \citenamefont {Hessels}}]{Demorest:2010bx}%
  \BibitemOpen
  \bibfield  {author} {\bibinfo {author} {\bibfnamefont {P.}~\bibnamefont
  {Demorest}}, \bibinfo {author} {\bibfnamefont {T.}~\bibnamefont {Pennucci}},
  \bibinfo {author} {\bibfnamefont {S.}~\bibnamefont {Ransom}}, \bibinfo
  {author} {\bibfnamefont {M.}~\bibnamefont {Roberts}}, \ and\ \bibinfo
  {author} {\bibfnamefont {J.}~\bibnamefont {Hessels}},\ }\href {\doibase
  10.1038/nature09466} {\bibfield  {journal} {\bibinfo  {journal} {Nature}\
  }\textbf {\bibinfo {volume} {467}},\ \bibinfo {pages} {1081} (\bibinfo {year}
  {2010})},\ \Eprint {http://arxiv.org/abs/1010.5788} {arXiv:1010.5788
  [astro-ph.HE]} \BibitemShut {NoStop}%
\bibitem [{\citenamefont {Antoniadis}\ \emph {et~al.}(2013)\citenamefont
  {Antoniadis} \emph {et~al.}}]{Antoniadis:2013pzd}%
  \BibitemOpen
  \bibfield  {author} {\bibinfo {author} {\bibfnamefont {J.}~\bibnamefont
  {Antoniadis}} \emph {et~al.},\ }\href {\doibase 10.1126/science.1233232}
  {\bibfield  {journal} {\bibinfo  {journal} {Science}\ }\textbf {\bibinfo
  {volume} {340}},\ \bibinfo {pages} {6131} (\bibinfo {year} {2013})},\ \Eprint
  {http://arxiv.org/abs/1304.6875} {arXiv:1304.6875 [astro-ph.HE]} \BibitemShut
  {NoStop}%
\bibitem [{\citenamefont {Ozel}\ and\ \citenamefont
  {Freire}(2016)}]{Ozel:2016oaf}%
  \BibitemOpen
  \bibfield  {author} {\bibinfo {author} {\bibfnamefont {F.}~\bibnamefont
  {Ozel}}\ and\ \bibinfo {author} {\bibfnamefont {P.}~\bibnamefont {Freire}},\
  }\href {\doibase 10.1146/annurev-astro-081915-023322} {\  (\bibinfo {year}
  {2016}),\ 10.1146/annurev-astro-081915-023322},\ \Eprint
  {http://arxiv.org/abs/1603.02698} {arXiv:1603.02698 [astro-ph.HE]}
  \BibitemShut {NoStop}%
\bibitem [{\citenamefont {Page}\ and\ \citenamefont
  {Reddy}(2006)}]{page2006dense}%
  \BibitemOpen
  \bibfield  {author} {\bibinfo {author} {\bibfnamefont {D.}~\bibnamefont
  {Page}}\ and\ \bibinfo {author} {\bibfnamefont {S.}~\bibnamefont {Reddy}},\
  }\href@noop {} {\bibfield  {journal} {\bibinfo  {journal} {Annual Review of
  Nuclear and Particle Science}\ }\textbf {\bibinfo {volume} {56}},\ \bibinfo
  {pages} {327} (\bibinfo {year} {2006})}\BibitemShut {NoStop}%
\bibitem [{\citenamefont {Ozel}\ \emph {et~al.}(2010)\citenamefont {Ozel},
  \citenamefont {Psaltis}, \citenamefont {Ransom}, \citenamefont {Demorest},\
  and\ \citenamefont {Alford}}]{Ozel:2010bz}%
  \BibitemOpen
  \bibfield  {author} {\bibinfo {author} {\bibfnamefont {F.}~\bibnamefont
  {Ozel}}, \bibinfo {author} {\bibfnamefont {D.}~\bibnamefont {Psaltis}},
  \bibinfo {author} {\bibfnamefont {S.}~\bibnamefont {Ransom}}, \bibinfo
  {author} {\bibfnamefont {P.}~\bibnamefont {Demorest}}, \ and\ \bibinfo
  {author} {\bibfnamefont {M.}~\bibnamefont {Alford}},\ }\href {\doibase
  10.1088/2041-8205/724/2/L199} {\bibfield  {journal} {\bibinfo  {journal}
  {Astrophys. J.}\ }\textbf {\bibinfo {volume} {724}},\ \bibinfo {pages} {L199}
  (\bibinfo {year} {2010})},\ \Eprint {http://arxiv.org/abs/1010.5790}
  {arXiv:1010.5790 [astro-ph.HE]} \BibitemShut {NoStop}%
\bibitem [{\citenamefont {{Y Potekhin}}(2010)}]{Potekhin:2010}%
  \BibitemOpen
  \bibfield  {author} {\bibinfo {author} {\bibfnamefont {A.}~\bibnamefont {{Y
  Potekhin}}},\ }\href {\doibase 10.3367/UFNe.0180.201012c.1279} {\bibfield
  {journal} {\bibinfo  {journal} {Physics Uspekhi}\ }\textbf {\bibinfo {volume}
  {53}},\ \bibinfo {pages} {1235} (\bibinfo {year} {2010})},\ \Eprint
  {http://arxiv.org/abs/1102.5735} {arXiv:1102.5735 [astro-ph.SR]} \BibitemShut
  {NoStop}%
\bibitem [{\citenamefont {Lattimer}(2012)}]{Lattimer:2012nd}%
  \BibitemOpen
  \bibfield  {author} {\bibinfo {author} {\bibfnamefont {J.~M.}\ \bibnamefont
  {Lattimer}},\ }\href {\doibase 10.1146/annurev-nucl-102711-095018} {\bibfield
   {journal} {\bibinfo  {journal} {Ann. Rev. Nucl. Part. Sci.}\ }\textbf
  {\bibinfo {volume} {62}},\ \bibinfo {pages} {485} (\bibinfo {year} {2012})},\
  \Eprint {http://arxiv.org/abs/1305.3510} {arXiv:1305.3510 [nucl-th]}
  \BibitemShut {NoStop}%
\bibitem [{\citenamefont {Prakash}(2015)}]{Prakash:2014tva}%
  \BibitemOpen
  \bibfield  {author} {\bibinfo {author} {\bibfnamefont {M.}~\bibnamefont
  {Prakash}},\ }\href {\doibase 10.1007/s12043-015-0979-7} {\bibfield
  {journal} {\bibinfo  {journal} {Pramana}\ }\textbf {\bibinfo {volume} {84}},\
  \bibinfo {pages} {927} (\bibinfo {year} {2015})},\ \Eprint
  {http://arxiv.org/abs/1404.1966} {arXiv:1404.1966 [astro-ph.SR]} \BibitemShut
  {NoStop}%
\bibitem [{\citenamefont {Ranea-Sandoval}\ \emph {et~al.}(2016)\citenamefont
  {Ranea-Sandoval}, \citenamefont {Han}, \citenamefont {Orsaria}, \citenamefont
  {Contrera}, \citenamefont {Weber},\ and\ \citenamefont
  {Alford}}]{Ranea-Sandoval:2015ldr}%
  \BibitemOpen
  \bibfield  {author} {\bibinfo {author} {\bibfnamefont {I.~F.}\ \bibnamefont
  {Ranea-Sandoval}}, \bibinfo {author} {\bibfnamefont {S.}~\bibnamefont {Han}},
  \bibinfo {author} {\bibfnamefont {M.~G.}\ \bibnamefont {Orsaria}}, \bibinfo
  {author} {\bibfnamefont {G.~A.}\ \bibnamefont {Contrera}}, \bibinfo {author}
  {\bibfnamefont {F.}~\bibnamefont {Weber}}, \ and\ \bibinfo {author}
  {\bibfnamefont {M.~G.}\ \bibnamefont {Alford}},\ }\href {\doibase
  10.1103/PhysRevC.93.045812} {\bibfield  {journal} {\bibinfo  {journal} {Phys.
  Rev.}\ }\textbf {\bibinfo {volume} {C93}},\ \bibinfo {pages} {045812}
  (\bibinfo {year} {2016})},\ \Eprint {http://arxiv.org/abs/1512.09183}
  {arXiv:1512.09183 [nucl-th]} \BibitemShut {NoStop}%
\bibitem [{\citenamefont {Steiner}\ \emph {et~al.}(2013)\citenamefont
  {Steiner}, \citenamefont {Lattimer},\ and\ \citenamefont
  {Brown}}]{Steiner:2012xt}%
  \BibitemOpen
  \bibfield  {author} {\bibinfo {author} {\bibfnamefont {A.~W.}\ \bibnamefont
  {Steiner}}, \bibinfo {author} {\bibfnamefont {J.~M.}\ \bibnamefont
  {Lattimer}}, \ and\ \bibinfo {author} {\bibfnamefont {E.~F.}\ \bibnamefont
  {Brown}},\ }\href {\doibase 10.1088/2041-8205/765/1/L5} {\bibfield  {journal}
  {\bibinfo  {journal} {Astrophys. J.}\ }\textbf {\bibinfo {volume} {765}},\
  \bibinfo {pages} {L5} (\bibinfo {year} {2013})},\ \Eprint
  {http://arxiv.org/abs/1205.6871} {arXiv:1205.6871 [nucl-th]} \BibitemShut
  {NoStop}%
\bibitem [{\citenamefont {Lattimer}\ and\ \citenamefont
  {Steiner}(2014)}]{Lattimer:2013hma}%
  \BibitemOpen
  \bibfield  {author} {\bibinfo {author} {\bibfnamefont {J.~M.}\ \bibnamefont
  {Lattimer}}\ and\ \bibinfo {author} {\bibfnamefont {A.~W.}\ \bibnamefont
  {Steiner}},\ }\href {\doibase 10.1088/0004-637X/784/2/123} {\bibfield
  {journal} {\bibinfo  {journal} {Astrophys. J.}\ }\textbf {\bibinfo {volume}
  {784}},\ \bibinfo {pages} {123} (\bibinfo {year} {2014})},\ \Eprint
  {http://arxiv.org/abs/1305.3242} {arXiv:1305.3242 [astro-ph.HE]} \BibitemShut
  {NoStop}%
\bibitem [{\citenamefont {Chamel}\ and\ \citenamefont
  {Haensel}(2008)}]{Chamel:2008ca}%
  \BibitemOpen
  \bibfield  {author} {\bibinfo {author} {\bibfnamefont {N.}~\bibnamefont
  {Chamel}}\ and\ \bibinfo {author} {\bibfnamefont {P.}~\bibnamefont
  {Haensel}},\ }\href {\doibase 10.12942/lrr-2008-10} {\bibfield  {journal}
  {\bibinfo  {journal} {Living Rev. Rel.}\ }\textbf {\bibinfo {volume} {11}},\
  \bibinfo {pages} {10} (\bibinfo {year} {2008})},\ \Eprint
  {http://arxiv.org/abs/0812.3955} {arXiv:0812.3955 [astro-ph]} \BibitemShut
  {NoStop}%
\bibitem [{\citenamefont {Page}\ and\ \citenamefont
  {Reddy}(2012)}]{Page:2012zt}%
  \BibitemOpen
  \bibfield  {author} {\bibinfo {author} {\bibfnamefont {D.}~\bibnamefont
  {Page}}\ and\ \bibinfo {author} {\bibfnamefont {S.}~\bibnamefont {Reddy}},\
  }\href@noop {} {\  (\bibinfo {year} {2012})},\ \Eprint
  {http://arxiv.org/abs/1201.5602} {arXiv:1201.5602 [nucl-th]} \BibitemShut
  {NoStop}%
\bibitem [{\citenamefont {Pethick}(1992)}]{Pethick:1992}%
  \BibitemOpen
  \bibfield  {author} {\bibinfo {author} {\bibfnamefont {C.~J.}\ \bibnamefont
  {Pethick}},\ }\href {\doibase 10.1103/RevModPhys.64.1133} {\bibfield
  {journal} {\bibinfo  {journal} {Rev. Mod. Phys.}\ }\textbf {\bibinfo {volume}
  {64}},\ \bibinfo {pages} {1133} (\bibinfo {year} {1992})}\BibitemShut
  {NoStop}%
\bibitem [{\citenamefont {Yakovlev}\ \emph {et~al.}(2001)\citenamefont
  {Yakovlev}, \citenamefont {Kaminker}, \citenamefont {Gnedin},\ and\
  \citenamefont {Haensel}}]{Yakovlev:2000jp}%
  \BibitemOpen
  \bibfield  {author} {\bibinfo {author} {\bibfnamefont {D.~G.}\ \bibnamefont
  {Yakovlev}}, \bibinfo {author} {\bibfnamefont {A.~D.}\ \bibnamefont
  {Kaminker}}, \bibinfo {author} {\bibfnamefont {O.~Y.}\ \bibnamefont
  {Gnedin}}, \ and\ \bibinfo {author} {\bibfnamefont {P.}~\bibnamefont
  {Haensel}},\ }\href {\doibase 10.1016/S0370-1573(00)00131-9} {\bibfield
  {journal} {\bibinfo  {journal} {Phys. Rept.}\ }\textbf {\bibinfo {volume}
  {354}},\ \bibinfo {pages} {1} (\bibinfo {year} {2001})},\ \Eprint
  {http://arxiv.org/abs/astro-ph/0012122} {arXiv:astro-ph/0012122 [astro-ph]}
  \BibitemShut {NoStop}%
\bibitem [{\citenamefont {Yakovlev}\ \emph {et~al.}(2003)\citenamefont
  {Yakovlev}, \citenamefont {Gnedin}, \citenamefont {Kaminker}, \citenamefont
  {Levenfish},\ and\ \citenamefont {Potekhin}}]{Yakovlev:2003}%
  \BibitemOpen
  \bibfield  {author} {\bibinfo {author} {\bibfnamefont {D.~G.}\ \bibnamefont
  {Yakovlev}}, \bibinfo {author} {\bibfnamefont {O.~Y.}\ \bibnamefont
  {Gnedin}}, \bibinfo {author} {\bibfnamefont {A.~D.}\ \bibnamefont
  {Kaminker}}, \bibinfo {author} {\bibfnamefont {K.~P.}\ \bibnamefont
  {Levenfish}}, \ and\ \bibinfo {author} {\bibfnamefont {A.~Y.}\ \bibnamefont
  {Potekhin}},\ }\href {http://cds.cern.ch/record/620491} {\bibfield  {journal}
  {\bibinfo  {journal} {Adv. Space Res.}\ }\textbf {\bibinfo {volume} {33}},\
  \bibinfo {pages} {523} (\bibinfo {year} {2003})}\BibitemShut {NoStop}%
\bibitem [{\citenamefont {Alford}\ and\ \citenamefont
  {Schwenzer}(2014)}]{Alford:2013pma}%
  \BibitemOpen
  \bibfield  {author} {\bibinfo {author} {\bibfnamefont {M.~G.}\ \bibnamefont
  {Alford}}\ and\ \bibinfo {author} {\bibfnamefont {K.}~\bibnamefont
  {Schwenzer}},\ }\href {\doibase 10.1103/PhysRevLett.113.251102} {\bibfield
  {journal} {\bibinfo  {journal} {Phys. Rev. Lett.}\ }\textbf {\bibinfo
  {volume} {113}},\ \bibinfo {pages} {251102} (\bibinfo {year} {2014})},\
  \Eprint {http://arxiv.org/abs/1310.3524} {arXiv:1310.3524 [astro-ph.HE]}
  \BibitemShut {NoStop}%
\bibitem [{\citenamefont {Andersson}(1998)}]{Andersson:1997xt}%
  \BibitemOpen
  \bibfield  {author} {\bibinfo {author} {\bibfnamefont {N.}~\bibnamefont
  {Andersson}},\ }\href {\doibase 10.1086/305919} {\bibfield  {journal}
  {\bibinfo  {journal} {Astrophys. J.}\ }\textbf {\bibinfo {volume} {502}},\
  \bibinfo {pages} {708} (\bibinfo {year} {1998})},\ \Eprint
  {http://arxiv.org/abs/gr-qc/9706075} {arXiv:gr-qc/9706075 [gr-qc]}
  \BibitemShut {NoStop}%
\bibitem [{\citenamefont {Andersson}\ and\ \citenamefont
  {Kokkotas}(1998)}]{Andersson:1997rn}%
  \BibitemOpen
  \bibfield  {author} {\bibinfo {author} {\bibfnamefont {N.}~\bibnamefont
  {Andersson}}\ and\ \bibinfo {author} {\bibfnamefont {K.~D.}\ \bibnamefont
  {Kokkotas}},\ }\href {\doibase 10.1046/j.1365-8711.1998.01840.x} {\bibfield
  {journal} {\bibinfo  {journal} {Mon. Not. Roy. Astron. Soc.}\ }\textbf
  {\bibinfo {volume} {299}},\ \bibinfo {pages} {1059} (\bibinfo {year}
  {1998})},\ \Eprint {http://arxiv.org/abs/gr-qc/9711088} {arXiv:gr-qc/9711088
  [gr-qc]} \BibitemShut {NoStop}%
\bibitem [{\citenamefont {Lindblom}\ \emph {et~al.}(2001)\citenamefont
  {Lindblom}, \citenamefont {Tohline},\ and\ \citenamefont
  {Vallisneri}}]{Lindblom:2000az}%
  \BibitemOpen
  \bibfield  {author} {\bibinfo {author} {\bibfnamefont {L.}~\bibnamefont
  {Lindblom}}, \bibinfo {author} {\bibfnamefont {J.~E.}\ \bibnamefont
  {Tohline}}, \ and\ \bibinfo {author} {\bibfnamefont {M.}~\bibnamefont
  {Vallisneri}},\ }\href {\doibase 10.1103/PhysRevLett.86.1152} {\bibfield
  {journal} {\bibinfo  {journal} {Phys. Rev. Lett.}\ }\textbf {\bibinfo
  {volume} {86}},\ \bibinfo {pages} {1152} (\bibinfo {year} {2001})},\ \Eprint
  {http://arxiv.org/abs/astro-ph/0010653} {arXiv:astro-ph/0010653 [astro-ph]}
  \BibitemShut {NoStop}%
\bibitem [{\citenamefont {Alford}\ \emph
  {et~al.}(2012{\natexlab{a}})\citenamefont {Alford}, \citenamefont
  {Mahmoodifar},\ and\ \citenamefont {Schwenzer}}]{Alford:2011pi}%
  \BibitemOpen
  \bibfield  {author} {\bibinfo {author} {\bibfnamefont {M.~G.}\ \bibnamefont
  {Alford}}, \bibinfo {author} {\bibfnamefont {S.}~\bibnamefont {Mahmoodifar}},
  \ and\ \bibinfo {author} {\bibfnamefont {K.}~\bibnamefont {Schwenzer}},\
  }\href {\doibase 10.1103/PhysRevD.85.044051} {\bibfield  {journal} {\bibinfo
  {journal} {Phys. Rev.}\ }\textbf {\bibinfo {volume} {D85}},\ \bibinfo {pages}
  {044051} (\bibinfo {year} {2012}{\natexlab{a}})},\ \Eprint
  {http://arxiv.org/abs/1103.3521} {arXiv:1103.3521 [astro-ph.HE]} \BibitemShut
  {NoStop}%
\bibitem [{\citenamefont {Alford}\ \emph
  {et~al.}(2012{\natexlab{b}})\citenamefont {Alford}, \citenamefont
  {Mahmoodifar},\ and\ \citenamefont {Schwenzer}}]{Alford:2010fd}%
  \BibitemOpen
  \bibfield  {author} {\bibinfo {author} {\bibfnamefont {M.}~\bibnamefont
  {Alford}}, \bibinfo {author} {\bibfnamefont {S.}~\bibnamefont {Mahmoodifar}},
  \ and\ \bibinfo {author} {\bibfnamefont {K.}~\bibnamefont {Schwenzer}},\
  }\href {\doibase 10.1103/PhysRevD.85.024007} {\bibfield  {journal} {\bibinfo
  {journal} {Phys. Rev.}\ }\textbf {\bibinfo {volume} {D85}},\ \bibinfo {pages}
  {024007} (\bibinfo {year} {2012}{\natexlab{b}})},\ \Eprint
  {http://arxiv.org/abs/1012.4883} {arXiv:1012.4883 [astro-ph.HE]} \BibitemShut
  {NoStop}%
\bibitem [{\citenamefont {{Flowers}}\ and\ \citenamefont
  {{Itoh}}(1976)}]{Flowers:1976}%
  \BibitemOpen
  \bibfield  {author} {\bibinfo {author} {\bibfnamefont {E.}~\bibnamefont
  {{Flowers}}}\ and\ \bibinfo {author} {\bibfnamefont {N.}~\bibnamefont
  {{Itoh}}},\ }\href {\doibase 10.1086/154375} {\bibfield  {journal} {\bibinfo
  {journal} {\apj}\ }\textbf {\bibinfo {volume} {206}},\ \bibinfo {pages} {218}
  (\bibinfo {year} {1976})}\BibitemShut {NoStop}%
\bibitem [{\citenamefont {{Flowers}}\ and\ \citenamefont
  {{Itoh}}(1979)}]{Flowers:1979}%
  \BibitemOpen
  \bibfield  {author} {\bibinfo {author} {\bibfnamefont {E.}~\bibnamefont
  {{Flowers}}}\ and\ \bibinfo {author} {\bibfnamefont {N.}~\bibnamefont
  {{Itoh}}},\ }\href {\doibase 10.1086/157145} {\bibfield  {journal} {\bibinfo
  {journal} {\apj}\ }\textbf {\bibinfo {volume} {230}},\ \bibinfo {pages} {847}
  (\bibinfo {year} {1979})}\BibitemShut {NoStop}%
\bibitem [{\citenamefont {Andersson}\ \emph {et~al.}(2000)\citenamefont
  {Andersson}, \citenamefont {Jones}, \citenamefont {Kokkotas},\ and\
  \citenamefont {Stergioulas}}]{Andersson:2000pt}%
  \BibitemOpen
  \bibfield  {author} {\bibinfo {author} {\bibfnamefont {N.}~\bibnamefont
  {Andersson}}, \bibinfo {author} {\bibfnamefont {D.~I.}\ \bibnamefont
  {Jones}}, \bibinfo {author} {\bibfnamefont {K.~D.}\ \bibnamefont {Kokkotas}},
  \ and\ \bibinfo {author} {\bibfnamefont {N.}~\bibnamefont {Stergioulas}},\
  }\bibfield  {booktitle} {\emph {\bibinfo {booktitle} {{Gravitational waves: A
  challenge to theoretical astrophysics. Proceedings, Trieste, Italy, June 6-9,
  2000}}},\ }\href {\doibase 10.1086/312643} {\bibfield  {journal} {\bibinfo
  {journal} {Astrophys. J.}\ }\textbf {\bibinfo {volume} {534}},\ \bibinfo
  {pages} {L75} (\bibinfo {year} {2000})},\ \bibinfo {note} {[,297(2000)]},\
  \Eprint {http://arxiv.org/abs/astro-ph/0002114} {arXiv:astro-ph/0002114
  [astro-ph]} \BibitemShut {NoStop}%
\bibitem [{\citenamefont {Bildsten}\ and\ \citenamefont
  {Ushomirsky}(2000)}]{Bildsten:1999zn}%
  \BibitemOpen
  \bibfield  {author} {\bibinfo {author} {\bibfnamefont {L.}~\bibnamefont
  {Bildsten}}\ and\ \bibinfo {author} {\bibfnamefont {G.}~\bibnamefont
  {Ushomirsky}},\ }\href {\doibase 10.1086/312454} {\bibfield  {journal}
  {\bibinfo  {journal} {Astrophys. J.}\ }\textbf {\bibinfo {volume} {529}},\
  \bibinfo {pages} {L33} (\bibinfo {year} {2000})},\ \Eprint
  {http://arxiv.org/abs/astro-ph/9911155} {arXiv:astro-ph/9911155 [astro-ph]}
  \BibitemShut {NoStop}%
\bibitem [{\citenamefont {Jaikumar}\ \emph {et~al.}(2008)\citenamefont
  {Jaikumar}, \citenamefont {Rupak},\ and\ \citenamefont
  {Steiner}}]{Jaikumar:2008kh}%
  \BibitemOpen
  \bibfield  {author} {\bibinfo {author} {\bibfnamefont {P.}~\bibnamefont
  {Jaikumar}}, \bibinfo {author} {\bibfnamefont {G.}~\bibnamefont {Rupak}}, \
  and\ \bibinfo {author} {\bibfnamefont {A.~W.}\ \bibnamefont {Steiner}},\
  }\href {\doibase 10.1103/PhysRevD.78.123007} {\bibfield  {journal} {\bibinfo
  {journal} {Phys. Rev.}\ }\textbf {\bibinfo {volume} {D78}},\ \bibinfo {pages}
  {123007} (\bibinfo {year} {2008})},\ \Eprint {http://arxiv.org/abs/0806.1005}
  {arXiv:0806.1005 [nucl-th]} \BibitemShut {NoStop}%
\bibitem [{\citenamefont {Levin}\ and\ \citenamefont
  {Ushomirsky}(2001)}]{Levin:2000vq}%
  \BibitemOpen
  \bibfield  {author} {\bibinfo {author} {\bibfnamefont {Y.}~\bibnamefont
  {Levin}}\ and\ \bibinfo {author} {\bibfnamefont {G.}~\bibnamefont
  {Ushomirsky}},\ }\href {\doibase 10.1046/j.1365-8711.2001.04323.x} {\bibfield
   {journal} {\bibinfo  {journal} {Mon. Not. Roy. Astron. Soc.}\ }\textbf
  {\bibinfo {volume} {324}},\ \bibinfo {pages} {917} (\bibinfo {year}
  {2001})},\ \Eprint {http://arxiv.org/abs/astro-ph/0006028}
  {arXiv:astro-ph/0006028 [astro-ph]} \BibitemShut {NoStop}%
\bibitem [{\citenamefont {Lindblom}\ \emph {et~al.}(2000)\citenamefont
  {Lindblom}, \citenamefont {Owen},\ and\ \citenamefont
  {Ushomirsky}}]{Lindblom:2000gu}%
  \BibitemOpen
  \bibfield  {author} {\bibinfo {author} {\bibfnamefont {L.}~\bibnamefont
  {Lindblom}}, \bibinfo {author} {\bibfnamefont {B.~J.}\ \bibnamefont {Owen}},
  \ and\ \bibinfo {author} {\bibfnamefont {G.}~\bibnamefont {Ushomirsky}},\
  }\href {\doibase 10.1103/PhysRevD.62.084030} {\bibfield  {journal} {\bibinfo
  {journal} {Phys. Rev.}\ }\textbf {\bibinfo {volume} {D62}},\ \bibinfo {pages}
  {084030} (\bibinfo {year} {2000})},\ \Eprint
  {http://arxiv.org/abs/astro-ph/0006242} {arXiv:astro-ph/0006242 [astro-ph]}
  \BibitemShut {NoStop}%
\bibitem [{\citenamefont {Rupak}\ and\ \citenamefont
  {Jaikumar}(2013)}]{Rupak:2012wk}%
  \BibitemOpen
  \bibfield  {author} {\bibinfo {author} {\bibfnamefont {G.}~\bibnamefont
  {Rupak}}\ and\ \bibinfo {author} {\bibfnamefont {P.}~\bibnamefont
  {Jaikumar}},\ }\href {\doibase 10.1103/PhysRevC.88.065801} {\bibfield
  {journal} {\bibinfo  {journal} {Phys. Rev.}\ }\textbf {\bibinfo {volume}
  {C88}},\ \bibinfo {pages} {065801} (\bibinfo {year} {2013})},\ \Eprint
  {http://arxiv.org/abs/1209.4343} {arXiv:1209.4343 [nucl-th]} \BibitemShut
  {NoStop}%
\bibitem [{\citenamefont {Lindblom}\ and\ \citenamefont
  {Mendell}(2000)}]{Lindblom:1999wi}%
  \BibitemOpen
  \bibfield  {author} {\bibinfo {author} {\bibfnamefont {L.}~\bibnamefont
  {Lindblom}}\ and\ \bibinfo {author} {\bibfnamefont {G.}~\bibnamefont
  {Mendell}},\ }\href {\doibase 10.1103/PhysRevD.61.104003} {\bibfield
  {journal} {\bibinfo  {journal} {Phys. Rev.}\ }\textbf {\bibinfo {volume}
  {D61}},\ \bibinfo {pages} {104003} (\bibinfo {year} {2000})},\ \Eprint
  {http://arxiv.org/abs/gr-qc/9909084} {arXiv:gr-qc/9909084 [gr-qc]}
  \BibitemShut {NoStop}%
\bibitem [{\citenamefont {Lindblom}\ and\ \citenamefont
  {Owen}(2002)}]{Lindblom:2002}%
  \BibitemOpen
  \bibfield  {author} {\bibinfo {author} {\bibfnamefont {L.}~\bibnamefont
  {Lindblom}}\ and\ \bibinfo {author} {\bibfnamefont {B.~J.}\ \bibnamefont
  {Owen}},\ }\href {\doibase 10.1103/PhysRevD.65.063006} {\bibfield  {journal}
  {\bibinfo  {journal} {Phys. Rev. D}\ }\textbf {\bibinfo {volume} {65}},\
  \bibinfo {pages} {063006} (\bibinfo {year} {2002})}\BibitemShut {NoStop}%
\bibitem [{\citenamefont {Haensel}\ \emph {et~al.}(2000)\citenamefont
  {Haensel}, \citenamefont {Levenfish},\ and\ \citenamefont
  {Yakovlev}}]{Haensel:2000vz}%
  \BibitemOpen
  \bibfield  {author} {\bibinfo {author} {\bibfnamefont {P.}~\bibnamefont
  {Haensel}}, \bibinfo {author} {\bibfnamefont {K.~P.}\ \bibnamefont
  {Levenfish}}, \ and\ \bibinfo {author} {\bibfnamefont {D.~G.}\ \bibnamefont
  {Yakovlev}},\ }\href@noop {} {\bibfield  {journal} {\bibinfo  {journal}
  {Astron. Astrophys.}\ }\textbf {\bibinfo {volume} {357}},\ \bibinfo {pages}
  {1157} (\bibinfo {year} {2000})},\ \Eprint
  {http://arxiv.org/abs/astro-ph/0004183} {arXiv:astro-ph/0004183 [astro-ph]}
  \BibitemShut {NoStop}%
\bibitem [{\citenamefont {{Haensel}}\ \emph {et~al.}(2001)\citenamefont
  {{Haensel}}, \citenamefont {{Levenfish}},\ and\ \citenamefont
  {{Yakovlev}}}]{Haensel:2001}%
  \BibitemOpen
  \bibfield  {author} {\bibinfo {author} {\bibfnamefont {P.}~\bibnamefont
  {{Haensel}}}, \bibinfo {author} {\bibfnamefont {K.~P.}\ \bibnamefont
  {{Levenfish}}}, \ and\ \bibinfo {author} {\bibfnamefont {D.~G.}\ \bibnamefont
  {{Yakovlev}}},\ }\href {\doibase 10.1051/0004-6361:20010383} {\bibfield
  {journal} {\bibinfo  {journal} {"Astron. Astrophys."}\ }\textbf {\bibinfo
  {volume} {372}},\ \bibinfo {pages} {130} (\bibinfo {year} {2001})},\ \Eprint
  {http://arxiv.org/abs/astro-ph/0103290} {astro-ph/0103290} \BibitemShut
  {NoStop}%
\bibitem [{\citenamefont {{Haensel}}\ \emph {et~al.}(2002)\citenamefont
  {{Haensel}}, \citenamefont {{Levenfish}},\ and\ \citenamefont
  {{Yakovlev}}}]{Haensel:2002}%
  \BibitemOpen
  \bibfield  {author} {\bibinfo {author} {\bibfnamefont {P.}~\bibnamefont
  {{Haensel}}}, \bibinfo {author} {\bibfnamefont {K.~P.}\ \bibnamefont
  {{Levenfish}}}, \ and\ \bibinfo {author} {\bibfnamefont {D.~G.}\ \bibnamefont
  {{Yakovlev}}},\ }\href {\doibase 10.1051/0004-6361:20011532} {\bibfield
  {journal} {\bibinfo  {journal} {"Astron. Astrophys."}\ }\textbf {\bibinfo
  {volume} {381}},\ \bibinfo {pages} {1080} (\bibinfo {year} {2002})},\ \Eprint
  {http://arxiv.org/abs/astro-ph/0110575} {astro-ph/0110575} \BibitemShut
  {NoStop}%
\bibitem [{\citenamefont {Shternin}\ and\ \citenamefont
  {Yakovlev}(2008)}]{Shternin:2008}%
  \BibitemOpen
  \bibfield  {author} {\bibinfo {author} {\bibfnamefont {P.~S.}\ \bibnamefont
  {Shternin}}\ and\ \bibinfo {author} {\bibfnamefont {D.~G.}\ \bibnamefont
  {Yakovlev}},\ }\href {\doibase 10.1103/PhysRevD.78.063006} {\bibfield
  {journal} {\bibinfo  {journal} {Phys. Rev. D}\ }\textbf {\bibinfo {volume}
  {78}},\ \bibinfo {pages} {063006} (\bibinfo {year} {2008})}\BibitemShut
  {NoStop}%
\bibitem [{\citenamefont {{Haskell}}\ and\ \citenamefont
  {{Andersson}}(2010)}]{Haskell:2010}%
  \BibitemOpen
  \bibfield  {author} {\bibinfo {author} {\bibfnamefont {B.}~\bibnamefont
  {{Haskell}}}\ and\ \bibinfo {author} {\bibfnamefont {N.}~\bibnamefont
  {{Andersson}}},\ }\href {\doibase 10.1111/j.1365-2966.2010.17255.x}
  {\bibfield  {journal} {\bibinfo  {journal} {"Mon. Not. Roy. Astron. Soc."}\
  }\textbf {\bibinfo {volume} {408}},\ \bibinfo {pages} {1897} (\bibinfo {year}
  {2010})},\ \Eprint {http://arxiv.org/abs/1003.5849} {arXiv:1003.5849
  [astro-ph.SR]} \BibitemShut {NoStop}%
\bibitem [{\citenamefont {Manuel}\ and\ \citenamefont
  {Tolos}(2013)}]{Manuel:2012rd}%
  \BibitemOpen
  \bibfield  {author} {\bibinfo {author} {\bibfnamefont {C.}~\bibnamefont
  {Manuel}}\ and\ \bibinfo {author} {\bibfnamefont {L.}~\bibnamefont {Tolos}},\
  }\href {\doibase 10.1103/PhysRevD.88.043001} {\bibfield  {journal} {\bibinfo
  {journal} {Phys. Rev.}\ }\textbf {\bibinfo {volume} {D88}},\ \bibinfo {pages}
  {043001} (\bibinfo {year} {2013})},\ \Eprint {http://arxiv.org/abs/1212.2075}
  {arXiv:1212.2075 [astro-ph.SR]} \BibitemShut {NoStop}%
\bibitem [{\citenamefont {Colucci}\ \emph {et~al.}(2013)\citenamefont
  {Colucci}, \citenamefont {Mannarelli},\ and\ \citenamefont
  {Manuel}}]{Colucci:2013sra}%
  \BibitemOpen
  \bibfield  {author} {\bibinfo {author} {\bibfnamefont {G.}~\bibnamefont
  {Colucci}}, \bibinfo {author} {\bibfnamefont {M.}~\bibnamefont {Mannarelli}},
  \ and\ \bibinfo {author} {\bibfnamefont {C.}~\bibnamefont {Manuel}},\ }\href
  {\doibase 10.1007/s10511-013-9271-z} {\bibfield  {journal} {\bibinfo
  {journal} {Astrophys.}\ }\textbf {\bibinfo {volume} {56}},\ \bibinfo {pages}
  {104} (\bibinfo {year} {2013})},\ \bibinfo {note}
  {[Astrofiz.56,117(2013)]}\BibitemShut {NoStop}%
\bibitem [{\citenamefont {Jaikumar}\ \emph {et~al.}(2006)\citenamefont
  {Jaikumar}, \citenamefont {Reddy},\ and\ \citenamefont
  {Steiner}}]{Jaikumar:2006rh}%
  \BibitemOpen
  \bibfield  {author} {\bibinfo {author} {\bibfnamefont {P.}~\bibnamefont
  {Jaikumar}}, \bibinfo {author} {\bibfnamefont {S.}~\bibnamefont {Reddy}}, \
  and\ \bibinfo {author} {\bibfnamefont {A.~W.}\ \bibnamefont {Steiner}},\
  }\href {\doibase 10.1142/S0217732306021396} {\bibfield  {journal} {\bibinfo
  {journal} {Mod. Phys. Lett.}\ }\textbf {\bibinfo {volume} {A21}},\ \bibinfo
  {pages} {1965} (\bibinfo {year} {2006})},\ \Eprint
  {http://arxiv.org/abs/astro-ph/0608345} {arXiv:astro-ph/0608345 [astro-ph]}
  \BibitemShut {NoStop}%
\bibitem [{\citenamefont {Madsen}(1992)}]{Madsen:1992}%
  \BibitemOpen
  \bibfield  {author} {\bibinfo {author} {\bibfnamefont {J.}~\bibnamefont
  {Madsen}},\ }\href {\doibase 10.1103/PhysRevD.46.3290} {\bibfield  {journal}
  {\bibinfo  {journal} {Phys. Rev. D}\ }\textbf {\bibinfo {volume} {46}},\
  \bibinfo {pages} {3290} (\bibinfo {year} {1992})}\BibitemShut {NoStop}%
\bibitem [{\citenamefont {Heiselberg}\ and\ \citenamefont
  {Pethick}(1993)}]{Heiselberg:1993}%
  \BibitemOpen
  \bibfield  {author} {\bibinfo {author} {\bibfnamefont {H.}~\bibnamefont
  {Heiselberg}}\ and\ \bibinfo {author} {\bibfnamefont {C.}~\bibnamefont
  {Pethick}},\ }\href@noop {} {\bibfield  {journal} {\bibinfo  {journal}
  {Physical Review D}\ }\textbf {\bibinfo {volume} {48}},\ \bibinfo {pages}
  {2916} (\bibinfo {year} {1993})}\BibitemShut {NoStop}%
\bibitem [{\citenamefont {Schwenzer}(2012)}]{Schwenzer:2012ga}%
  \BibitemOpen
  \bibfield  {author} {\bibinfo {author} {\bibfnamefont {K.}~\bibnamefont
  {Schwenzer}},\ }\href@noop {} {\  (\bibinfo {year} {2012})},\ \Eprint
  {http://arxiv.org/abs/1212.5242} {arXiv:1212.5242 [nucl-th]} \BibitemShut
  {NoStop}%
\bibitem [{\citenamefont {Iwamoto}(1980)}]{Iwamoto:1980}%
  \BibitemOpen
  \bibfield  {author} {\bibinfo {author} {\bibfnamefont {N.}~\bibnamefont
  {Iwamoto}},\ }\href {\doibase 10.1103/PhysRevLett.44.1637} {\bibfield
  {journal} {\bibinfo  {journal} {Phys. Rev. Lett.}\ }\textbf {\bibinfo
  {volume} {44}},\ \bibinfo {pages} {1637} (\bibinfo {year}
  {1980})}\BibitemShut {NoStop}%
\bibitem [{\citenamefont {{Iwamoto}}(1982)}]{Iwamoto:1982}%
  \BibitemOpen
  \bibfield  {author} {\bibinfo {author} {\bibfnamefont {N.}~\bibnamefont
  {{Iwamoto}}},\ }\href {\doibase 10.1016/0003-4916(82)90271-8} {\bibfield
  {journal} {\bibinfo  {journal} {Annals of Physics}\ }\textbf {\bibinfo
  {volume} {141}},\ \bibinfo {pages} {1} (\bibinfo {year} {1982})}\BibitemShut
  {NoStop}%
\bibitem [{\citenamefont {Rajagopal}\ and\ \citenamefont
  {Wilczek}(2000)}]{Rajagopal:2000wf}%
  \BibitemOpen
  \bibfield  {author} {\bibinfo {author} {\bibfnamefont {K.}~\bibnamefont
  {Rajagopal}}\ and\ \bibinfo {author} {\bibfnamefont {F.}~\bibnamefont
  {Wilczek}},\ }\href@noop {} {\  (\bibinfo {year} {2000})},\ \Eprint
  {http://arxiv.org/abs/hep-ph/0011333} {arXiv:hep-ph/0011333 [hep-ph]}
  \BibitemShut {NoStop}%
\bibitem [{\citenamefont {Alford}\ \emph
  {et~al.}(2001{\natexlab{a}})\citenamefont {Alford}, \citenamefont {Bowers},\
  and\ \citenamefont {Rajagopal}}]{Alford:2000sx}%
  \BibitemOpen
  \bibfield  {author} {\bibinfo {author} {\bibfnamefont {M.~G.}\ \bibnamefont
  {Alford}}, \bibinfo {author} {\bibfnamefont {J.~A.}\ \bibnamefont {Bowers}},
  \ and\ \bibinfo {author} {\bibfnamefont {K.}~\bibnamefont {Rajagopal}},\
  }\bibfield  {booktitle} {\emph {\bibinfo {booktitle} {{Strangeness in quark
  matter. Proceedings, 5th International Conference, Strangeness 2000,
  Berkeley, USA, July 20-25, 2000}}},\ }\href {\doibase
  10.1088/0954-3899/27/3/335} {\bibfield  {journal} {\bibinfo  {journal} {J.
  Phys.}\ }\textbf {\bibinfo {volume} {G27}},\ \bibinfo {pages} {541} (\bibinfo
  {year} {2001}{\natexlab{a}})},\ \bibinfo {note} {[Lect. Notes
  Phys.578,235(2001)]},\ \Eprint {http://arxiv.org/abs/hep-ph/0009357}
  {arXiv:hep-ph/0009357 [hep-ph]} \BibitemShut {NoStop}%
\bibitem [{\citenamefont {Alford}\ \emph
  {et~al.}(2008{\natexlab{a}})\citenamefont {Alford}, \citenamefont {Schmitt},
  \citenamefont {Rajagopal},\ and\ \citenamefont {Schäfer}}]{Alford:2007xm}%
  \BibitemOpen
  \bibfield  {author} {\bibinfo {author} {\bibfnamefont {M.~G.}\ \bibnamefont
  {Alford}}, \bibinfo {author} {\bibfnamefont {A.}~\bibnamefont {Schmitt}},
  \bibinfo {author} {\bibfnamefont {K.}~\bibnamefont {Rajagopal}}, \ and\
  \bibinfo {author} {\bibfnamefont {T.}~\bibnamefont {Schäfer}},\ }\href
  {\doibase 10.1103/RevModPhys.80.1455} {\bibfield  {journal} {\bibinfo
  {journal} {Rev. Mod. Phys.}\ }\textbf {\bibinfo {volume} {80}},\ \bibinfo
  {pages} {1455} (\bibinfo {year} {2008}{\natexlab{a}})},\ \Eprint
  {http://arxiv.org/abs/0709.4635} {arXiv:0709.4635 [hep-ph]} \BibitemShut
  {NoStop}%
\bibitem [{\citenamefont {Alford}\ \emph
  {et~al.}(1999{\natexlab{a}})\citenamefont {Alford}, \citenamefont
  {Rajagopal},\ and\ \citenamefont {Wilczek}}]{Alford:1998mk}%
  \BibitemOpen
  \bibfield  {author} {\bibinfo {author} {\bibfnamefont {M.~G.}\ \bibnamefont
  {Alford}}, \bibinfo {author} {\bibfnamefont {K.}~\bibnamefont {Rajagopal}}, \
  and\ \bibinfo {author} {\bibfnamefont {F.}~\bibnamefont {Wilczek}},\ }\href
  {\doibase 10.1016/S0550-3213(98)00668-3} {\bibfield  {journal} {\bibinfo
  {journal} {Nucl. Phys.}\ }\textbf {\bibinfo {volume} {B537}},\ \bibinfo
  {pages} {443} (\bibinfo {year} {1999}{\natexlab{a}})},\ \Eprint
  {http://arxiv.org/abs/hep-ph/9804403} {arXiv:hep-ph/9804403 [hep-ph]}
  \BibitemShut {NoStop}%
\bibitem [{\citenamefont {Manuel}\ \emph {et~al.}(2005)\citenamefont {Manuel},
  \citenamefont {Dobado},\ and\ \citenamefont
  {Llanes-Estrada}}]{Manuel:2004iv}%
  \BibitemOpen
  \bibfield  {author} {\bibinfo {author} {\bibfnamefont {C.}~\bibnamefont
  {Manuel}}, \bibinfo {author} {\bibfnamefont {A.}~\bibnamefont {Dobado}}, \
  and\ \bibinfo {author} {\bibfnamefont {F.~J.}\ \bibnamefont
  {Llanes-Estrada}},\ }\href {\doibase 10.1088/1126-6708/2005/09/076}
  {\bibfield  {journal} {\bibinfo  {journal} {JHEP}\ }\textbf {\bibinfo
  {volume} {09}},\ \bibinfo {pages} {076} (\bibinfo {year} {2005})},\ \Eprint
  {http://arxiv.org/abs/hep-ph/0406058} {arXiv:hep-ph/0406058 [hep-ph]}
  \BibitemShut {NoStop}%
\bibitem [{\citenamefont {Mannarelli}\ \emph {et~al.}(2008)\citenamefont
  {Mannarelli}, \citenamefont {Manuel},\ and\ \citenamefont
  {Sa'd}}]{Mannarelli:2008je}%
  \BibitemOpen
  \bibfield  {author} {\bibinfo {author} {\bibfnamefont {M.}~\bibnamefont
  {Mannarelli}}, \bibinfo {author} {\bibfnamefont {C.}~\bibnamefont {Manuel}},
  \ and\ \bibinfo {author} {\bibfnamefont {B.~A.}\ \bibnamefont {Sa'd}},\
  }\href {\doibase 10.1103/PhysRevLett.101.241101} {\bibfield  {journal}
  {\bibinfo  {journal} {Phys. Rev. Lett.}\ }\textbf {\bibinfo {volume} {101}},\
  \bibinfo {pages} {241101} (\bibinfo {year} {2008})},\ \Eprint
  {http://arxiv.org/abs/0807.3264} {arXiv:0807.3264 [hep-ph]} \BibitemShut
  {NoStop}%
\bibitem [{\citenamefont {Rupak}\ and\ \citenamefont
  {Jaikumar}(2010)}]{Rupak:2010qg}%
  \BibitemOpen
  \bibfield  {author} {\bibinfo {author} {\bibfnamefont {G.}~\bibnamefont
  {Rupak}}\ and\ \bibinfo {author} {\bibfnamefont {P.}~\bibnamefont
  {Jaikumar}},\ }\href {\doibase 10.1103/PhysRevC.82.055806} {\bibfield
  {journal} {\bibinfo  {journal} {Phys. Rev.}\ }\textbf {\bibinfo {volume}
  {C82}},\ \bibinfo {pages} {055806} (\bibinfo {year} {2010})},\ \Eprint
  {http://arxiv.org/abs/1005.4161} {arXiv:1005.4161 [nucl-th]} \BibitemShut
  {NoStop}%
\bibitem [{\citenamefont {Alford}\ and\ \citenamefont
  {Han}(2016)}]{Alford:2015gna}%
  \BibitemOpen
  \bibfield  {author} {\bibinfo {author} {\bibfnamefont {M.~G.}\ \bibnamefont
  {Alford}}\ and\ \bibinfo {author} {\bibfnamefont {S.}~\bibnamefont {Han}},\
  }\href {\doibase 10.1140/epja/i2016-16062-9} {\bibfield  {journal} {\bibinfo
  {journal} {Eur. Phys. J.}\ }\textbf {\bibinfo {volume} {A52}},\ \bibinfo
  {pages} {62} (\bibinfo {year} {2016})},\ \Eprint
  {http://arxiv.org/abs/1508.01261} {arXiv:1508.01261 [nucl-th]} \BibitemShut
  {NoStop}%
\bibitem [{\citenamefont {Alford}\ \emph
  {et~al.}(2001{\natexlab{b}})\citenamefont {Alford}, \citenamefont {Bowers},\
  and\ \citenamefont {Rajagopal}}]{Alford:2000ze}%
  \BibitemOpen
  \bibfield  {author} {\bibinfo {author} {\bibfnamefont {M.~G.}\ \bibnamefont
  {Alford}}, \bibinfo {author} {\bibfnamefont {J.~A.}\ \bibnamefont {Bowers}},
  \ and\ \bibinfo {author} {\bibfnamefont {K.}~\bibnamefont {Rajagopal}},\
  }\href {\doibase 10.1103/PhysRevD.63.074016} {\bibfield  {journal} {\bibinfo
  {journal} {Phys. Rev.}\ }\textbf {\bibinfo {volume} {D63}},\ \bibinfo {pages}
  {074016} (\bibinfo {year} {2001}{\natexlab{b}})},\ \Eprint
  {http://arxiv.org/abs/hep-ph/0008208} {arXiv:hep-ph/0008208 [hep-ph]}
  \BibitemShut {NoStop}%
\bibitem [{\citenamefont {Anglani}\ \emph {et~al.}(2014)\citenamefont
  {Anglani}, \citenamefont {Casalbuoni}, \citenamefont {Ciminale},
  \citenamefont {Ippolito}, \citenamefont {Gatto}, \citenamefont {Mannarelli},\
  and\ \citenamefont {Ruggieri}}]{Anglani:2013gfu}%
  \BibitemOpen
  \bibfield  {author} {\bibinfo {author} {\bibfnamefont {R.}~\bibnamefont
  {Anglani}}, \bibinfo {author} {\bibfnamefont {R.}~\bibnamefont {Casalbuoni}},
  \bibinfo {author} {\bibfnamefont {M.}~\bibnamefont {Ciminale}}, \bibinfo
  {author} {\bibfnamefont {N.}~\bibnamefont {Ippolito}}, \bibinfo {author}
  {\bibfnamefont {R.}~\bibnamefont {Gatto}}, \bibinfo {author} {\bibfnamefont
  {M.}~\bibnamefont {Mannarelli}}, \ and\ \bibinfo {author} {\bibfnamefont
  {M.}~\bibnamefont {Ruggieri}},\ }\href {\doibase 10.1103/RevModPhys.86.509}
  {\bibfield  {journal} {\bibinfo  {journal} {Rev. Mod. Phys.}\ }\textbf
  {\bibinfo {volume} {86}},\ \bibinfo {pages} {509} (\bibinfo {year} {2014})},\
  \Eprint {http://arxiv.org/abs/1302.4264} {arXiv:1302.4264 [hep-ph]}
  \BibitemShut {NoStop}%
\bibitem [{\citenamefont {Rajagopal}\ and\ \citenamefont
  {Sharma}(2006)}]{Rajagopal:2006ig}%
  \BibitemOpen
  \bibfield  {author} {\bibinfo {author} {\bibfnamefont {K.}~\bibnamefont
  {Rajagopal}}\ and\ \bibinfo {author} {\bibfnamefont {R.}~\bibnamefont
  {Sharma}},\ }\href {\doibase 10.1103/PhysRevD.74.094019} {\bibfield
  {journal} {\bibinfo  {journal} {Phys. Rev.}\ }\textbf {\bibinfo {volume}
  {D74}},\ \bibinfo {pages} {094019} (\bibinfo {year} {2006})},\ \Eprint
  {http://arxiv.org/abs/hep-ph/0605316} {arXiv:hep-ph/0605316 [hep-ph]}
  \BibitemShut {NoStop}%
\bibitem [{\citenamefont {Ippolito}\ \emph {et~al.}(2007)\citenamefont
  {Ippolito}, \citenamefont {Nardulli},\ and\ \citenamefont
  {Ruggieri}}]{Ippolito:2007uz}%
  \BibitemOpen
  \bibfield  {author} {\bibinfo {author} {\bibfnamefont {N.~D.}\ \bibnamefont
  {Ippolito}}, \bibinfo {author} {\bibfnamefont {G.}~\bibnamefont {Nardulli}},
  \ and\ \bibinfo {author} {\bibfnamefont {M.}~\bibnamefont {Ruggieri}},\
  }\href {\doibase 10.1088/1126-6708/2007/04/036} {\bibfield  {journal}
  {\bibinfo  {journal} {JHEP}\ }\textbf {\bibinfo {volume} {04}},\ \bibinfo
  {pages} {036} (\bibinfo {year} {2007})},\ \Eprint
  {http://arxiv.org/abs/hep-ph/0701113} {arXiv:hep-ph/0701113 [hep-ph]}
  \BibitemShut {NoStop}%
\bibitem [{\citenamefont {Cao}\ \emph {et~al.}(2015)\citenamefont {Cao},
  \citenamefont {He},\ and\ \citenamefont {Zhuang}}]{Cao:2015rea}%
  \BibitemOpen
  \bibfield  {author} {\bibinfo {author} {\bibfnamefont {G.}~\bibnamefont
  {Cao}}, \bibinfo {author} {\bibfnamefont {L.}~\bibnamefont {He}}, \ and\
  \bibinfo {author} {\bibfnamefont {P.}~\bibnamefont {Zhuang}},\ }\href
  {\doibase 10.1103/PhysRevD.91.114021} {\bibfield  {journal} {\bibinfo
  {journal} {Phys. Rev.}\ }\textbf {\bibinfo {volume} {D91}},\ \bibinfo {pages}
  {114021} (\bibinfo {year} {2015})},\ \Eprint
  {http://arxiv.org/abs/1502.03392} {arXiv:1502.03392 [nucl-th]} \BibitemShut
  {NoStop}%
\bibitem [{\citenamefont {Anglani}\ \emph {et~al.}(2006)\citenamefont
  {Anglani}, \citenamefont {Nardulli}, \citenamefont {Ruggieri},\ and\
  \citenamefont {Mannarelli}}]{Anglani:2006br}%
  \BibitemOpen
  \bibfield  {author} {\bibinfo {author} {\bibfnamefont {R.}~\bibnamefont
  {Anglani}}, \bibinfo {author} {\bibfnamefont {G.}~\bibnamefont {Nardulli}},
  \bibinfo {author} {\bibfnamefont {M.}~\bibnamefont {Ruggieri}}, \ and\
  \bibinfo {author} {\bibfnamefont {M.}~\bibnamefont {Mannarelli}},\ }\href
  {\doibase 10.1103/PhysRevD.74.074005} {\bibfield  {journal} {\bibinfo
  {journal} {Phys. Rev.}\ }\textbf {\bibinfo {volume} {D74}},\ \bibinfo {pages}
  {074005} (\bibinfo {year} {2006})},\ \Eprint
  {http://arxiv.org/abs/hep-ph/0607341} {arXiv:hep-ph/0607341 [hep-ph]}
  \BibitemShut {NoStop}%
\bibitem [{\citenamefont {Hess}\ and\ \citenamefont
  {Sedrakian}(2011)}]{Hess:2011}%
  \BibitemOpen
  \bibfield  {author} {\bibinfo {author} {\bibfnamefont {D.}~\bibnamefont
  {Hess}}\ and\ \bibinfo {author} {\bibfnamefont {A.}~\bibnamefont
  {Sedrakian}},\ }\href {\doibase 10.1103/PhysRevD.84.063015} {\bibfield
  {journal} {\bibinfo  {journal} {Phys. Rev. D}\ }\textbf {\bibinfo {volume}
  {84}},\ \bibinfo {pages} {063015} (\bibinfo {year} {2011})}\BibitemShut
  {NoStop}%
\bibitem [{\citenamefont {Fulde}\ and\ \citenamefont
  {Ferrell}(1964)}]{fulde1964superconductivity}%
  \BibitemOpen
  \bibfield  {author} {\bibinfo {author} {\bibfnamefont {P.}~\bibnamefont
  {Fulde}}\ and\ \bibinfo {author} {\bibfnamefont {R.~A.}\ \bibnamefont
  {Ferrell}},\ }\href@noop {} {\bibfield  {journal} {\bibinfo  {journal}
  {Physical Review}\ }\textbf {\bibinfo {volume} {135}},\ \bibinfo {pages}
  {A550} (\bibinfo {year} {1964})}\BibitemShut {NoStop}%
\bibitem [{\citenamefont {Alford}\ \emph {et~al.}(1998)\citenamefont {Alford},
  \citenamefont {Rajagopal},\ and\ \citenamefont {Wilczek}}]{Alford:1997zt}%
  \BibitemOpen
  \bibfield  {author} {\bibinfo {author} {\bibfnamefont {M.~G.}\ \bibnamefont
  {Alford}}, \bibinfo {author} {\bibfnamefont {K.}~\bibnamefont {Rajagopal}}, \
  and\ \bibinfo {author} {\bibfnamefont {F.}~\bibnamefont {Wilczek}},\ }\href
  {\doibase 10.1016/S0370-2693(98)00051-3} {\bibfield  {journal} {\bibinfo
  {journal} {Phys. Lett.}\ }\textbf {\bibinfo {volume} {B422}},\ \bibinfo
  {pages} {247} (\bibinfo {year} {1998})},\ \Eprint
  {http://arxiv.org/abs/hep-ph/9711395} {arXiv:hep-ph/9711395 [hep-ph]}
  \BibitemShut {NoStop}%
\bibitem [{\citenamefont {Rapp}\ \emph {et~al.}(1998)\citenamefont {Rapp},
  \citenamefont {Schäfer}, \citenamefont {Shuryak},\ and\ \citenamefont
  {Velkovsky}}]{Rapp:1997zu}%
  \BibitemOpen
  \bibfield  {author} {\bibinfo {author} {\bibfnamefont {R.}~\bibnamefont
  {Rapp}}, \bibinfo {author} {\bibfnamefont {T.}~\bibnamefont {Schäfer}},
  \bibinfo {author} {\bibfnamefont {E.~V.}\ \bibnamefont {Shuryak}}, \ and\
  \bibinfo {author} {\bibfnamefont {M.}~\bibnamefont {Velkovsky}},\ }\href
  {\doibase 10.1103/PhysRevLett.81.53} {\bibfield  {journal} {\bibinfo
  {journal} {Phys. Rev. Lett.}\ }\textbf {\bibinfo {volume} {81}},\ \bibinfo
  {pages} {53} (\bibinfo {year} {1998})},\ \Eprint
  {http://arxiv.org/abs/hep-ph/9711396} {arXiv:hep-ph/9711396 [hep-ph]}
  \BibitemShut {NoStop}%
\bibitem [{\citenamefont {Alford}\ \emph {et~al.}(2014)\citenamefont {Alford},
  \citenamefont {Nishimura},\ and\ \citenamefont {Sedrakian}}]{Alford:2014doa}%
  \BibitemOpen
  \bibfield  {author} {\bibinfo {author} {\bibfnamefont {M.~G.}\ \bibnamefont
  {Alford}}, \bibinfo {author} {\bibfnamefont {H.}~\bibnamefont {Nishimura}}, \
  and\ \bibinfo {author} {\bibfnamefont {A.}~\bibnamefont {Sedrakian}},\ }\href
  {\doibase 10.1103/PhysRevC.90.055205} {\bibfield  {journal} {\bibinfo
  {journal} {Phys.Rev.}\ }\textbf {\bibinfo {volume} {C90}},\ \bibinfo {pages}
  {055205} (\bibinfo {year} {2014})},\ \Eprint {http://arxiv.org/abs/1408.4999}
  {arXiv:1408.4999 [hep-ph]} \BibitemShut {NoStop}%
\bibitem [{\citenamefont {Alford}\ \emph
  {et~al.}(2000{\natexlab{a}})\citenamefont {Alford}, \citenamefont {Berges},\
  and\ \citenamefont {Rajagopal}}]{Alford:1999pb}%
  \BibitemOpen
  \bibfield  {author} {\bibinfo {author} {\bibfnamefont {M.~G.}\ \bibnamefont
  {Alford}}, \bibinfo {author} {\bibfnamefont {J.}~\bibnamefont {Berges}}, \
  and\ \bibinfo {author} {\bibfnamefont {K.}~\bibnamefont {Rajagopal}},\ }\href
  {\doibase 10.1016/S0550-3213(99)00830-5} {\bibfield  {journal} {\bibinfo
  {journal} {Nucl. Phys.}\ }\textbf {\bibinfo {volume} {B571}},\ \bibinfo
  {pages} {269} (\bibinfo {year} {2000}{\natexlab{a}})},\ \Eprint
  {http://arxiv.org/abs/hep-ph/9910254} {arXiv:hep-ph/9910254 [hep-ph]}
  \BibitemShut {NoStop}%
\bibitem [{\citenamefont {Rischke}(2000{\natexlab{a}})}]{Rischke:2000qz}%
  \BibitemOpen
  \bibfield  {author} {\bibinfo {author} {\bibfnamefont {D.~H.}\ \bibnamefont
  {Rischke}},\ }\href {\doibase 10.1103/PhysRevD.62.034007} {\bibfield
  {journal} {\bibinfo  {journal} {Phys. Rev.}\ }\textbf {\bibinfo {volume}
  {D62}},\ \bibinfo {pages} {034007} (\bibinfo {year} {2000}{\natexlab{a}})},\
  \Eprint {http://arxiv.org/abs/nucl-th/0001040} {arXiv:nucl-th/0001040
  [nucl-th]} \BibitemShut {NoStop}%
\bibitem [{\citenamefont {Rischke}\ \emph {et~al.}(2001)\citenamefont
  {Rischke}, \citenamefont {Son},\ and\ \citenamefont
  {Stephanov}}]{Rischke:2000cn}%
  \BibitemOpen
  \bibfield  {author} {\bibinfo {author} {\bibfnamefont {D.~H.}\ \bibnamefont
  {Rischke}}, \bibinfo {author} {\bibfnamefont {D.~T.}\ \bibnamefont {Son}}, \
  and\ \bibinfo {author} {\bibfnamefont {M.~A.}\ \bibnamefont {Stephanov}},\
  }\href {\doibase 10.1103/PhysRevLett.87.062001} {\bibfield  {journal}
  {\bibinfo  {journal} {Phys. Rev. Lett.}\ }\textbf {\bibinfo {volume} {87}},\
  \bibinfo {pages} {062001} (\bibinfo {year} {2001})},\ \Eprint
  {http://arxiv.org/abs/hep-ph/0011379} {arXiv:hep-ph/0011379 [hep-ph]}
  \BibitemShut {NoStop}%
\bibitem [{\citenamefont {Rischke}\ and\ \citenamefont
  {Shovkovy}(2002)}]{Rischke:2002rz}%
  \BibitemOpen
  \bibfield  {author} {\bibinfo {author} {\bibfnamefont {D.~H.}\ \bibnamefont
  {Rischke}}\ and\ \bibinfo {author} {\bibfnamefont {I.~A.}\ \bibnamefont
  {Shovkovy}},\ }\href {\doibase 10.1103/PhysRevD.66.054019} {\bibfield
  {journal} {\bibinfo  {journal} {Phys. Rev.}\ }\textbf {\bibinfo {volume}
  {D66}},\ \bibinfo {pages} {054019} (\bibinfo {year} {2002})},\ \Eprint
  {http://arxiv.org/abs/nucl-th/0205080} {arXiv:nucl-th/0205080 [nucl-th]}
  \BibitemShut {NoStop}%
\bibitem [{\citenamefont {Peskin}\ and\ \citenamefont
  {Schroeder}(1995)}]{Peskin:257493}%
  \BibitemOpen
  \bibfield  {author} {\bibinfo {author} {\bibfnamefont {M.~E.}\ \bibnamefont
  {Peskin}}\ and\ \bibinfo {author} {\bibfnamefont {D.~V.}\ \bibnamefont
  {Schroeder}},\ }\href {https://cds.cern.ch/record/257493} {\emph {\bibinfo
  {title} {{An Introduction to Quantum Field Theory; 1995 ed.}}}}\ (\bibinfo
  {publisher} {Westview},\ \bibinfo {address} {Boulder, CO},\ \bibinfo {year}
  {1995})\ \bibinfo {note} {includes exercises}\BibitemShut {NoStop}%
\bibitem [{\citenamefont {Nambu}\ and\ \citenamefont
  {Jona-Lasinio}(1961)}]{nambu1961dynamical}%
  \BibitemOpen
  \bibfield  {author} {\bibinfo {author} {\bibfnamefont {Y.}~\bibnamefont
  {Nambu}}\ and\ \bibinfo {author} {\bibfnamefont {G.}~\bibnamefont
  {Jona-Lasinio}},\ }\href@noop {} {\bibfield  {journal} {\bibinfo  {journal}
  {Physical Review}\ }\textbf {\bibinfo {volume} {122}},\ \bibinfo {pages}
  {345} (\bibinfo {year} {1961})}\BibitemShut {NoStop}%
\bibitem [{\citenamefont {Alford}\ \emph
  {et~al.}(1999{\natexlab{b}})\citenamefont {Alford}, \citenamefont {Berges},\
  and\ \citenamefont {Rajagopal}}]{Alford:1999pa}%
  \BibitemOpen
  \bibfield  {author} {\bibinfo {author} {\bibfnamefont {M.~G.}\ \bibnamefont
  {Alford}}, \bibinfo {author} {\bibfnamefont {J.}~\bibnamefont {Berges}}, \
  and\ \bibinfo {author} {\bibfnamefont {K.}~\bibnamefont {Rajagopal}},\ }\href
  {\doibase 10.1016/S0550-3213(99)00410-1} {\bibfield  {journal} {\bibinfo
  {journal} {Nucl. Phys.}\ }\textbf {\bibinfo {volume} {B558}},\ \bibinfo
  {pages} {219} (\bibinfo {year} {1999}{\natexlab{b}})},\ \Eprint
  {http://arxiv.org/abs/hep-ph/9903502} {arXiv:hep-ph/9903502 [hep-ph]}
  \BibitemShut {NoStop}%
\bibitem [{\citenamefont {Alford}\ \emph
  {et~al.}(2000{\natexlab{b}})\citenamefont {Alford}, \citenamefont {Berges},\
  and\ \citenamefont {Rajagopal}}]{Alford:1999xc}%
  \BibitemOpen
  \bibfield  {author} {\bibinfo {author} {\bibfnamefont {M.~G.}\ \bibnamefont
  {Alford}}, \bibinfo {author} {\bibfnamefont {J.}~\bibnamefont {Berges}}, \
  and\ \bibinfo {author} {\bibfnamefont {K.}~\bibnamefont {Rajagopal}},\ }\href
  {\doibase 10.1103/PhysRevLett.84.598} {\bibfield  {journal} {\bibinfo
  {journal} {Phys. Rev. Lett.}\ }\textbf {\bibinfo {volume} {84}},\ \bibinfo
  {pages} {598} (\bibinfo {year} {2000}{\natexlab{b}})},\ \Eprint
  {http://arxiv.org/abs/hep-ph/9908235} {arXiv:hep-ph/9908235 [hep-ph]}
  \BibitemShut {NoStop}%
\bibitem [{\citenamefont {Casalbuoni}\ \emph
  {et~al.}(2001{\natexlab{a}})\citenamefont {Casalbuoni}, \citenamefont
  {Gatto},\ and\ \citenamefont {Nardulli}}]{Casalbuoni:2000na}%
  \BibitemOpen
  \bibfield  {author} {\bibinfo {author} {\bibfnamefont {R.}~\bibnamefont
  {Casalbuoni}}, \bibinfo {author} {\bibfnamefont {R.}~\bibnamefont {Gatto}}, \
  and\ \bibinfo {author} {\bibfnamefont {G.}~\bibnamefont {Nardulli}},\ }\href
  {\doibase 10.1016/S0370-2693(01)01003-6, 10.1016/S0370-2693(00)01390-3}
  {\bibfield  {journal} {\bibinfo  {journal} {Phys. Lett.}\ }\textbf {\bibinfo
  {volume} {B498}},\ \bibinfo {pages} {179} (\bibinfo {year}
  {2001}{\natexlab{a}})},\ \bibinfo {note} {[Erratum: Phys.
  Lett.B517,483(2001)]},\ \Eprint {http://arxiv.org/abs/hep-ph/0010321}
  {arXiv:hep-ph/0010321 [hep-ph]} \BibitemShut {NoStop}%
\bibitem [{\citenamefont {Rischke}(2000{\natexlab{b}})}]{Rischke:2000ra}%
  \BibitemOpen
  \bibfield  {author} {\bibinfo {author} {\bibfnamefont {D.~H.}\ \bibnamefont
  {Rischke}},\ }\href {\doibase 10.1103/PhysRevD.62.054017} {\bibfield
  {journal} {\bibinfo  {journal} {Phys. Rev.}\ }\textbf {\bibinfo {volume}
  {D62}},\ \bibinfo {pages} {054017} (\bibinfo {year} {2000}{\natexlab{b}})},\
  \Eprint {http://arxiv.org/abs/nucl-th/0003063} {arXiv:nucl-th/0003063
  [nucl-th]} \BibitemShut {NoStop}%
\bibitem [{\citenamefont {Son}\ and\ \citenamefont
  {Stephanov}(2000{\natexlab{a}})}]{Son:1999cm}%
  \BibitemOpen
  \bibfield  {author} {\bibinfo {author} {\bibfnamefont {D.~T.}\ \bibnamefont
  {Son}}\ and\ \bibinfo {author} {\bibfnamefont {M.~A.}\ \bibnamefont
  {Stephanov}},\ }\href {\doibase 10.1103/PhysRevD.61.074012} {\bibfield
  {journal} {\bibinfo  {journal} {Phys. Rev.}\ }\textbf {\bibinfo {volume}
  {D61}},\ \bibinfo {pages} {074012} (\bibinfo {year} {2000}{\natexlab{a}})},\
  \Eprint {http://arxiv.org/abs/hep-ph/9910491} {arXiv:hep-ph/9910491 [hep-ph]}
  \BibitemShut {NoStop}%
\bibitem [{\citenamefont {Son}\ and\ \citenamefont
  {Stephanov}(2000{\natexlab{b}})}]{Son:2000tu}%
  \BibitemOpen
  \bibfield  {author} {\bibinfo {author} {\bibfnamefont {D.~T.}\ \bibnamefont
  {Son}}\ and\ \bibinfo {author} {\bibfnamefont {M.~A.}\ \bibnamefont
  {Stephanov}},\ }\href {\doibase 10.1103/PhysRevD.62.059902} {\bibfield
  {journal} {\bibinfo  {journal} {Phys. Rev.}\ }\textbf {\bibinfo {volume}
  {D62}},\ \bibinfo {pages} {059902} (\bibinfo {year} {2000}{\natexlab{b}})},\
  \Eprint {http://arxiv.org/abs/hep-ph/0004095} {arXiv:hep-ph/0004095 [hep-ph]}
  \BibitemShut {NoStop}%
\bibitem [{\citenamefont {Casalbuoni}\ and\ \citenamefont
  {Gatto}(1999)}]{Casalbuoni:1999zi}%
  \BibitemOpen
  \bibfield  {author} {\bibinfo {author} {\bibfnamefont {R.}~\bibnamefont
  {Casalbuoni}}\ and\ \bibinfo {author} {\bibfnamefont {R.}~\bibnamefont
  {Gatto}},\ }\href {\doibase 10.1016/S0370-2693(99)01274-5} {\bibfield
  {journal} {\bibinfo  {journal} {Phys. Lett.}\ }\textbf {\bibinfo {volume}
  {B469}},\ \bibinfo {pages} {213} (\bibinfo {year} {1999})},\ \Eprint
  {http://arxiv.org/abs/hep-ph/9909419} {arXiv:hep-ph/9909419 [hep-ph]}
  \BibitemShut {NoStop}%
\bibitem [{\citenamefont {Rho}\ \emph {et~al.}(2000{\natexlab{a}})\citenamefont
  {Rho}, \citenamefont {Wirzba},\ and\ \citenamefont {Zahed}}]{Rho:1999xf}%
  \BibitemOpen
  \bibfield  {author} {\bibinfo {author} {\bibfnamefont {M.}~\bibnamefont
  {Rho}}, \bibinfo {author} {\bibfnamefont {A.}~\bibnamefont {Wirzba}}, \ and\
  \bibinfo {author} {\bibfnamefont {I.}~\bibnamefont {Zahed}},\ }\href
  {\doibase 10.1016/S0370-2693(99)01420-3} {\bibfield  {journal} {\bibinfo
  {journal} {Phys. Lett.}\ }\textbf {\bibinfo {volume} {B473}},\ \bibinfo
  {pages} {126} (\bibinfo {year} {2000}{\natexlab{a}})},\ \Eprint
  {http://arxiv.org/abs/hep-ph/9910550} {arXiv:hep-ph/9910550 [hep-ph]}
  \BibitemShut {NoStop}%
\bibitem [{\citenamefont {Hong}\ \emph {et~al.}(2000)\citenamefont {Hong},
  \citenamefont {Lee},\ and\ \citenamefont {Min}}]{Hong:1999ei}%
  \BibitemOpen
  \bibfield  {author} {\bibinfo {author} {\bibfnamefont {D.~K.}\ \bibnamefont
  {Hong}}, \bibinfo {author} {\bibfnamefont {T.}~\bibnamefont {Lee}}, \ and\
  \bibinfo {author} {\bibfnamefont {D.-P.}\ \bibnamefont {Min}},\ }\href
  {\doibase 10.1016/S0370-2693(00)00188-X} {\bibfield  {journal} {\bibinfo
  {journal} {Phys. Lett.}\ }\textbf {\bibinfo {volume} {B477}},\ \bibinfo
  {pages} {137} (\bibinfo {year} {2000})},\ \Eprint
  {http://arxiv.org/abs/hep-ph/9912531} {arXiv:hep-ph/9912531 [hep-ph]}
  \BibitemShut {NoStop}%
\bibitem [{\citenamefont {Manuel}\ and\ \citenamefont
  {Tytgat}(2000)}]{Manuel:2000wm}%
  \BibitemOpen
  \bibfield  {author} {\bibinfo {author} {\bibfnamefont {C.}~\bibnamefont
  {Manuel}}\ and\ \bibinfo {author} {\bibfnamefont {M.~H.~G.}\ \bibnamefont
  {Tytgat}},\ }\href {\doibase 10.1016/S0370-2693(00)00331-2} {\bibfield
  {journal} {\bibinfo  {journal} {Phys. Lett.}\ }\textbf {\bibinfo {volume}
  {B479}},\ \bibinfo {pages} {190} (\bibinfo {year} {2000})},\ \Eprint
  {http://arxiv.org/abs/hep-ph/0001095} {arXiv:hep-ph/0001095 [hep-ph]}
  \BibitemShut {NoStop}%
\bibitem [{\citenamefont {Rho}\ \emph {et~al.}(2000{\natexlab{b}})\citenamefont
  {Rho}, \citenamefont {Shuryak}, \citenamefont {Wirzba},\ and\ \citenamefont
  {Zahed}}]{Rho:2000ww}%
  \BibitemOpen
  \bibfield  {author} {\bibinfo {author} {\bibfnamefont {M.}~\bibnamefont
  {Rho}}, \bibinfo {author} {\bibfnamefont {E.~V.}\ \bibnamefont {Shuryak}},
  \bibinfo {author} {\bibfnamefont {A.}~\bibnamefont {Wirzba}}, \ and\ \bibinfo
  {author} {\bibfnamefont {I.}~\bibnamefont {Zahed}},\ }\href {\doibase
  10.1016/S0375-9474(00)00190-1} {\bibfield  {journal} {\bibinfo  {journal}
  {Nucl. Phys.}\ }\textbf {\bibinfo {volume} {A676}},\ \bibinfo {pages} {273}
  (\bibinfo {year} {2000}{\natexlab{b}})},\ \Eprint
  {http://arxiv.org/abs/hep-ph/0001104} {arXiv:hep-ph/0001104 [hep-ph]}
  \BibitemShut {NoStop}%
\bibitem [{\citenamefont {Rajagopal}\ and\ \citenamefont
  {Schmitt}(2006)}]{Rajagopal:2005dg}%
  \BibitemOpen
  \bibfield  {author} {\bibinfo {author} {\bibfnamefont {K.}~\bibnamefont
  {Rajagopal}}\ and\ \bibinfo {author} {\bibfnamefont {A.}~\bibnamefont
  {Schmitt}},\ }\href {\doibase 10.1103/PhysRevD.73.045003} {\bibfield
  {journal} {\bibinfo  {journal} {Phys. Rev.}\ }\textbf {\bibinfo {volume}
  {D73}},\ \bibinfo {pages} {045003} (\bibinfo {year} {2006})},\ \Eprint
  {http://arxiv.org/abs/hep-ph/0512043} {arXiv:hep-ph/0512043 [hep-ph]}
  \BibitemShut {NoStop}%
\bibitem [{\citenamefont {Bowers}(2003)}]{Bowers:2003ye}%
  \BibitemOpen
  \bibfield  {author} {\bibinfo {author} {\bibfnamefont {J.~A.}\ \bibnamefont
  {Bowers}},\ }\emph {\bibinfo {title} {{Color superconducting phases of cold
  dense quark matter}}},\ \href
  {http://alice.cern.ch/format/showfull?sysnb=2377723} {Ph.D. thesis},\
  \bibinfo  {school} {MIT, LNS} (\bibinfo {year} {2003}),\ \Eprint
  {http://arxiv.org/abs/hep-ph/0305301} {arXiv:hep-ph/0305301 [hep-ph]}
  \BibitemShut {NoStop}%
\bibitem [{\citenamefont {Alford}\ \emph
  {et~al.}(2005{\natexlab{a}})\citenamefont {Alford}, \citenamefont
  {Kouvaris},\ and\ \citenamefont {Rajagopal}}]{Alford:2005}%
  \BibitemOpen
  \bibfield  {author} {\bibinfo {author} {\bibfnamefont {M.}~\bibnamefont
  {Alford}}, \bibinfo {author} {\bibfnamefont {C.}~\bibnamefont {Kouvaris}}, \
  and\ \bibinfo {author} {\bibfnamefont {K.}~\bibnamefont {Rajagopal}},\ }\href
  {\doibase 10.1103/PhysRevD.71.054009} {\bibfield  {journal} {\bibinfo
  {journal} {Phys. Rev. D}\ }\textbf {\bibinfo {volume} {71}},\ \bibinfo
  {pages} {054009} (\bibinfo {year} {2005}{\natexlab{a}})}\BibitemShut
  {NoStop}%
\bibitem [{\citenamefont {Bedaque}\ and\ \citenamefont
  {Schäfer}(2002)}]{Bedaque:2001je}%
  \BibitemOpen
  \bibfield  {author} {\bibinfo {author} {\bibfnamefont {P.~F.}\ \bibnamefont
  {Bedaque}}\ and\ \bibinfo {author} {\bibfnamefont {T.}~\bibnamefont
  {Schäfer}},\ }\href {\doibase 10.1016/S0375-9474(01)01272-6} {\bibfield
  {journal} {\bibinfo  {journal} {Nucl. Phys.}\ }\textbf {\bibinfo {volume}
  {A697}},\ \bibinfo {pages} {802} (\bibinfo {year} {2002})},\ \Eprint
  {http://arxiv.org/abs/hep-ph/0105150} {arXiv:hep-ph/0105150 [hep-ph]}
  \BibitemShut {NoStop}%
\bibitem [{\citenamefont {Schäfer}(2003)}]{Schafer:2002yy}%
  \BibitemOpen
  \bibfield  {author} {\bibinfo {author} {\bibfnamefont {T.}~\bibnamefont
  {Schäfer}},\ }\href {\doibase 10.1103/PhysRevD.67.074502} {\bibfield
  {journal} {\bibinfo  {journal} {Phys. Rev.}\ }\textbf {\bibinfo {volume}
  {D67}},\ \bibinfo {pages} {074502} (\bibinfo {year} {2003})},\ \Eprint
  {http://arxiv.org/abs/hep-lat/0211035} {arXiv:hep-lat/0211035 [hep-lat]}
  \BibitemShut {NoStop}%
\bibitem [{\citenamefont {Buballa}(2005)}]{Buballa:2004sx}%
  \BibitemOpen
  \bibfield  {author} {\bibinfo {author} {\bibfnamefont {M.}~\bibnamefont
  {Buballa}},\ }\href {\doibase 10.1016/j.physletb.2005.01.027} {\bibfield
  {journal} {\bibinfo  {journal} {Phys. Lett.}\ }\textbf {\bibinfo {volume}
  {B609}},\ \bibinfo {pages} {57} (\bibinfo {year} {2005})},\ \Eprint
  {http://arxiv.org/abs/hep-ph/0410397} {arXiv:hep-ph/0410397 [hep-ph]}
  \BibitemShut {NoStop}%
\bibitem [{\citenamefont {Forbes}(2005)}]{Forbes:2005jya}%
  \BibitemOpen
  \bibfield  {author} {\bibinfo {author} {\bibfnamefont {M.~M.}\ \bibnamefont
  {Forbes}},\ }\emph {\bibinfo {title} {{Fermionic Superfluids: From Cold Atoms
  To High Density Qcd Gapless (breached Pair) Superfluidity And Kaon
  Condensation}}},\ \href@noop {} {Ph.D. thesis},\ \bibinfo  {school}
  {Massachusetts Inst. Technology} (\bibinfo {year} {2005})\BibitemShut
  {NoStop}%
\bibitem [{\citenamefont {Warringa}(2006)}]{Warringa:2006dk}%
  \BibitemOpen
  \bibfield  {author} {\bibinfo {author} {\bibfnamefont {H.~J.}\ \bibnamefont
  {Warringa}},\ }\href@noop {} {\  (\bibinfo {year} {2006})},\ \Eprint
  {http://arxiv.org/abs/hep-ph/0606063} {arXiv:hep-ph/0606063 [hep-ph]}
  \BibitemShut {NoStop}%
\bibitem [{\citenamefont {Kryjevski}\ and\ \citenamefont
  {Yamada}(2005)}]{Kryjevski:2004kt}%
  \BibitemOpen
  \bibfield  {author} {\bibinfo {author} {\bibfnamefont {A.}~\bibnamefont
  {Kryjevski}}\ and\ \bibinfo {author} {\bibfnamefont {D.}~\bibnamefont
  {Yamada}},\ }\href {\doibase 10.1103/PhysRevD.71.014011} {\bibfield
  {journal} {\bibinfo  {journal} {Phys. Rev.}\ }\textbf {\bibinfo {volume}
  {D71}},\ \bibinfo {pages} {014011} (\bibinfo {year} {2005})},\ \Eprint
  {http://arxiv.org/abs/hep-ph/0407350} {arXiv:hep-ph/0407350 [hep-ph]}
  \BibitemShut {NoStop}%
\bibitem [{\citenamefont {Gerhold}\ \emph {et~al.}(2007)\citenamefont
  {Gerhold}, \citenamefont {Schäfer},\ and\ \citenamefont
  {Kryjevski}}]{Gerhold:2006np}%
  \BibitemOpen
  \bibfield  {author} {\bibinfo {author} {\bibfnamefont {A.}~\bibnamefont
  {Gerhold}}, \bibinfo {author} {\bibfnamefont {T.}~\bibnamefont {Schäfer}}, \
  and\ \bibinfo {author} {\bibfnamefont {A.}~\bibnamefont {Kryjevski}},\ }\href
  {\doibase 10.1103/PhysRevD.75.054012} {\bibfield  {journal} {\bibinfo
  {journal} {Phys. Rev.}\ }\textbf {\bibinfo {volume} {D75}},\ \bibinfo {pages}
  {054012} (\bibinfo {year} {2007})},\ \Eprint
  {http://arxiv.org/abs/hep-ph/0612181} {arXiv:hep-ph/0612181 [hep-ph]}
  \BibitemShut {NoStop}%
\bibitem [{\citenamefont {Kryjevski}(2008)}]{Kryjevski:2008zz}%
  \BibitemOpen
  \bibfield  {author} {\bibinfo {author} {\bibfnamefont {A.}~\bibnamefont
  {Kryjevski}},\ }\href {\doibase 10.1103/PhysRevD.77.014018} {\bibfield
  {journal} {\bibinfo  {journal} {Phys. Rev.}\ }\textbf {\bibinfo {volume}
  {D77}},\ \bibinfo {pages} {014018} (\bibinfo {year} {2008})},\ \Eprint
  {http://arxiv.org/abs/hep-ph/0508180} {arXiv:hep-ph/0508180 [hep-ph]}
  \BibitemShut {NoStop}%
\bibitem [{\citenamefont {Alford}\ \emph {et~al.}(2007)\citenamefont {Alford},
  \citenamefont {Braby}, \citenamefont {Reddy},\ and\ \citenamefont
  {Schäfer}}]{Alford:2007rw}%
  \BibitemOpen
  \bibfield  {author} {\bibinfo {author} {\bibfnamefont {M.~G.}\ \bibnamefont
  {Alford}}, \bibinfo {author} {\bibfnamefont {M.}~\bibnamefont {Braby}},
  \bibinfo {author} {\bibfnamefont {S.}~\bibnamefont {Reddy}}, \ and\ \bibinfo
  {author} {\bibfnamefont {T.}~\bibnamefont {Schäfer}},\ }\href {\doibase
  10.1103/PhysRevC.75.055209} {\bibfield  {journal} {\bibinfo  {journal} {Phys.
  Rev.}\ }\textbf {\bibinfo {volume} {C75}},\ \bibinfo {pages} {055209}
  (\bibinfo {year} {2007})},\ \Eprint {http://arxiv.org/abs/nucl-th/0701067}
  {arXiv:nucl-th/0701067 [nucl-th]} \BibitemShut {NoStop}%
\bibitem [{\citenamefont {Alford}\ \emph
  {et~al.}(2008{\natexlab{b}})\citenamefont {Alford}, \citenamefont {Braby},\
  and\ \citenamefont {Schmitt}}]{Alford:2008pb}%
  \BibitemOpen
  \bibfield  {author} {\bibinfo {author} {\bibfnamefont {M.~G.}\ \bibnamefont
  {Alford}}, \bibinfo {author} {\bibfnamefont {M.}~\bibnamefont {Braby}}, \
  and\ \bibinfo {author} {\bibfnamefont {A.}~\bibnamefont {Schmitt}},\ }\href
  {\doibase 10.1088/0954-3899/35/11/115007} {\bibfield  {journal} {\bibinfo
  {journal} {J. Phys.}\ }\textbf {\bibinfo {volume} {G35}},\ \bibinfo {pages}
  {115007} (\bibinfo {year} {2008}{\natexlab{b}})},\ \Eprint
  {http://arxiv.org/abs/0806.0285} {arXiv:0806.0285 [nucl-th]} \BibitemShut
  {NoStop}%
\bibitem [{\citenamefont {Casalbuoni}\ \emph
  {et~al.}(2005{\natexlab{a}})\citenamefont {Casalbuoni}, \citenamefont
  {Gatto}, \citenamefont {Mannarelli}, \citenamefont {Nardulli},\ and\
  \citenamefont {Ruggieri}}]{Casalbuoni:2004tb}%
  \BibitemOpen
  \bibfield  {author} {\bibinfo {author} {\bibfnamefont {R.}~\bibnamefont
  {Casalbuoni}}, \bibinfo {author} {\bibfnamefont {R.}~\bibnamefont {Gatto}},
  \bibinfo {author} {\bibfnamefont {M.}~\bibnamefont {Mannarelli}}, \bibinfo
  {author} {\bibfnamefont {G.}~\bibnamefont {Nardulli}}, \ and\ \bibinfo
  {author} {\bibfnamefont {M.}~\bibnamefont {Ruggieri}},\ }\href {\doibase
  10.1016/j.physletb.2004.11.045, 10.1016/j.physletb.2005.04.025} {\bibfield
  {journal} {\bibinfo  {journal} {Phys. Lett.}\ }\textbf {\bibinfo {volume}
  {B605}},\ \bibinfo {pages} {362} (\bibinfo {year} {2005}{\natexlab{a}})},\
  \bibinfo {note} {[Erratum: Phys. Lett.B615,297(2005)]},\ \Eprint
  {http://arxiv.org/abs/hep-ph/0410401} {arXiv:hep-ph/0410401 [hep-ph]}
  \BibitemShut {NoStop}%
\bibitem [{\citenamefont {Fukushima}(2005)}]{Fukushima:2005cm}%
  \BibitemOpen
  \bibfield  {author} {\bibinfo {author} {\bibfnamefont {K.}~\bibnamefont
  {Fukushima}},\ }\href {\doibase 10.1103/PhysRevD.72.074002} {\bibfield
  {journal} {\bibinfo  {journal} {Phys. Rev.}\ }\textbf {\bibinfo {volume}
  {D72}},\ \bibinfo {pages} {074002} (\bibinfo {year} {2005})},\ \Eprint
  {http://arxiv.org/abs/hep-ph/0506080} {arXiv:hep-ph/0506080 [hep-ph]}
  \BibitemShut {NoStop}%
\bibitem [{\citenamefont {Alford}\ and\ \citenamefont
  {Wang}(2005)}]{Alford:2005qw}%
  \BibitemOpen
  \bibfield  {author} {\bibinfo {author} {\bibfnamefont {M.}~\bibnamefont
  {Alford}}\ and\ \bibinfo {author} {\bibfnamefont {Q.-h.}\ \bibnamefont
  {Wang}},\ }\href {\doibase 10.1088/0954-3899/31/7/017} {\bibfield  {journal}
  {\bibinfo  {journal} {J. Phys.}\ }\textbf {\bibinfo {volume} {G31}},\
  \bibinfo {pages} {719} (\bibinfo {year} {2005})},\ \Eprint
  {http://arxiv.org/abs/hep-ph/0501078} {arXiv:hep-ph/0501078 [hep-ph]}
  \BibitemShut {NoStop}%
\bibitem [{\citenamefont {Huang}\ and\ \citenamefont
  {Shovkovy}(2004{\natexlab{a}})}]{Huang:2004bg}%
  \BibitemOpen
  \bibfield  {author} {\bibinfo {author} {\bibfnamefont {M.}~\bibnamefont
  {Huang}}\ and\ \bibinfo {author} {\bibfnamefont {I.~A.}\ \bibnamefont
  {Shovkovy}},\ }\href {\doibase 10.1103/PhysRevD.70.051501} {\bibfield
  {journal} {\bibinfo  {journal} {Phys. Rev.}\ }\textbf {\bibinfo {volume}
  {D70}},\ \bibinfo {pages} {051501} (\bibinfo {year} {2004}{\natexlab{a}})},\
  \Eprint {http://arxiv.org/abs/hep-ph/0407049} {arXiv:hep-ph/0407049 [hep-ph]}
  \BibitemShut {NoStop}%
\bibitem [{\citenamefont {Huang}\ and\ \citenamefont
  {Shovkovy}(2004{\natexlab{b}})}]{Huang:2004am}%
  \BibitemOpen
  \bibfield  {author} {\bibinfo {author} {\bibfnamefont {M.}~\bibnamefont
  {Huang}}\ and\ \bibinfo {author} {\bibfnamefont {I.~A.}\ \bibnamefont
  {Shovkovy}},\ }\href {\doibase 10.1103/PhysRevD.70.094030} {\bibfield
  {journal} {\bibinfo  {journal} {Phys. Rev.}\ }\textbf {\bibinfo {volume}
  {D70}},\ \bibinfo {pages} {094030} (\bibinfo {year} {2004}{\natexlab{b}})},\
  \Eprint {http://arxiv.org/abs/hep-ph/0408268} {arXiv:hep-ph/0408268 [hep-ph]}
  \BibitemShut {NoStop}%
\bibitem [{\citenamefont {Giannakis}\ and\ \citenamefont
  {Ren}(2005{\natexlab{a}})}]{Giannakis:2004pf}%
  \BibitemOpen
  \bibfield  {author} {\bibinfo {author} {\bibfnamefont {I.}~\bibnamefont
  {Giannakis}}\ and\ \bibinfo {author} {\bibfnamefont {H.-C.}\ \bibnamefont
  {Ren}},\ }\href {\doibase 10.1016/j.physletb.2005.02.020} {\bibfield
  {journal} {\bibinfo  {journal} {Phys. Lett.}\ }\textbf {\bibinfo {volume}
  {B611}},\ \bibinfo {pages} {137} (\bibinfo {year} {2005}{\natexlab{a}})},\
  \Eprint {http://arxiv.org/abs/hep-ph/0412015} {arXiv:hep-ph/0412015 [hep-ph]}
  \BibitemShut {NoStop}%
\bibitem [{\citenamefont {Fukushima}(2006)}]{Fukushima:2006su}%
  \BibitemOpen
  \bibfield  {author} {\bibinfo {author} {\bibfnamefont {K.}~\bibnamefont
  {Fukushima}},\ }\href {\doibase 10.1103/PhysRevD.73.094016} {\bibfield
  {journal} {\bibinfo  {journal} {Phys. Rev.}\ }\textbf {\bibinfo {volume}
  {D73}},\ \bibinfo {pages} {094016} (\bibinfo {year} {2006})},\ \Eprint
  {http://arxiv.org/abs/hep-ph/0603216} {arXiv:hep-ph/0603216 [hep-ph]}
  \BibitemShut {NoStop}%
\bibitem [{\citenamefont {Larkin}\ and\ \citenamefont
  {Ovchinnikov}(1964)}]{larkin1964nonuniform}%
  \BibitemOpen
  \bibfield  {author} {\bibinfo {author} {\bibfnamefont {A.}~\bibnamefont
  {Larkin}}\ and\ \bibinfo {author} {\bibfnamefont {Y.~N.}\ \bibnamefont
  {Ovchinnikov}},\ }\href@noop {} {\bibfield  {journal} {\bibinfo  {journal}
  {Zh. Eksperim. i Teor. Fiz.}\ }\textbf {\bibinfo {volume} {47}} (\bibinfo
  {year} {1964})}\BibitemShut {NoStop}%
\bibitem [{\citenamefont {Gorbar}\ \emph {et~al.}(2006)\citenamefont {Gorbar},
  \citenamefont {Hashimoto},\ and\ \citenamefont {Miransky}}]{Gorbar:2005rx}%
  \BibitemOpen
  \bibfield  {author} {\bibinfo {author} {\bibfnamefont {E.~V.}\ \bibnamefont
  {Gorbar}}, \bibinfo {author} {\bibfnamefont {M.}~\bibnamefont {Hashimoto}}, \
  and\ \bibinfo {author} {\bibfnamefont {V.~A.}\ \bibnamefont {Miransky}},\
  }\href {\doibase 10.1016/j.physletb.2005.10.063} {\bibfield  {journal}
  {\bibinfo  {journal} {Phys. Lett.}\ }\textbf {\bibinfo {volume} {B632}},\
  \bibinfo {pages} {305} (\bibinfo {year} {2006})},\ \Eprint
  {http://arxiv.org/abs/hep-ph/0507303} {arXiv:hep-ph/0507303 [hep-ph]}
  \BibitemShut {NoStop}%
\bibitem [{\citenamefont {Kiriyama}\ \emph {et~al.}(2006)\citenamefont
  {Kiriyama}, \citenamefont {Rischke},\ and\ \citenamefont
  {Shovkovy}}]{Kiriyama:2006ui}%
  \BibitemOpen
  \bibfield  {author} {\bibinfo {author} {\bibfnamefont {O.}~\bibnamefont
  {Kiriyama}}, \bibinfo {author} {\bibfnamefont {D.~H.}\ \bibnamefont
  {Rischke}}, \ and\ \bibinfo {author} {\bibfnamefont {I.~A.}\ \bibnamefont
  {Shovkovy}},\ }\href {\doibase 10.1016/j.physletb.2006.10.067} {\bibfield
  {journal} {\bibinfo  {journal} {Phys. Lett.}\ }\textbf {\bibinfo {volume}
  {B643}},\ \bibinfo {pages} {331} (\bibinfo {year} {2006})},\ \Eprint
  {http://arxiv.org/abs/hep-ph/0606030} {arXiv:hep-ph/0606030 [hep-ph]}
  \BibitemShut {NoStop}%
\bibitem [{\citenamefont {Alford}\ and\ \citenamefont
  {Schmitt}(2007)}]{Alford:2006gy}%
  \BibitemOpen
  \bibfield  {author} {\bibinfo {author} {\bibfnamefont {M.~G.}\ \bibnamefont
  {Alford}}\ and\ \bibinfo {author} {\bibfnamefont {A.}~\bibnamefont
  {Schmitt}},\ }\href {\doibase 10.1088/0954-3899/34/1/005} {\bibfield
  {journal} {\bibinfo  {journal} {J. Phys.}\ }\textbf {\bibinfo {volume}
  {G34}},\ \bibinfo {pages} {67} (\bibinfo {year} {2007})},\ \Eprint
  {http://arxiv.org/abs/nucl-th/0608019} {arXiv:nucl-th/0608019 [nucl-th]}
  \BibitemShut {NoStop}%
\bibitem [{\citenamefont {Abuki}\ and\ \citenamefont
  {Kunihiro}(2006)}]{Abuki:2005ms}%
  \BibitemOpen
  \bibfield  {author} {\bibinfo {author} {\bibfnamefont {H.}~\bibnamefont
  {Abuki}}\ and\ \bibinfo {author} {\bibfnamefont {T.}~\bibnamefont
  {Kunihiro}},\ }\href {\doibase 10.1016/j.nuclphysa.2005.12.019} {\bibfield
  {journal} {\bibinfo  {journal} {Nucl. Phys.}\ }\textbf {\bibinfo {volume}
  {A768}},\ \bibinfo {pages} {118} (\bibinfo {year} {2006})},\ \Eprint
  {http://arxiv.org/abs/hep-ph/0509172} {arXiv:hep-ph/0509172 [hep-ph]}
  \BibitemShut {NoStop}%
\bibitem [{\citenamefont {Ruester}\ \emph {et~al.}(2005)\citenamefont
  {Ruester}, \citenamefont {Werth}, \citenamefont {Buballa}, \citenamefont
  {Shovkovy},\ and\ \citenamefont {Rischke}}]{Ruester:2005jc}%
  \BibitemOpen
  \bibfield  {author} {\bibinfo {author} {\bibfnamefont {S.~B.}\ \bibnamefont
  {Ruester}}, \bibinfo {author} {\bibfnamefont {V.}~\bibnamefont {Werth}},
  \bibinfo {author} {\bibfnamefont {M.}~\bibnamefont {Buballa}}, \bibinfo
  {author} {\bibfnamefont {I.~A.}\ \bibnamefont {Shovkovy}}, \ and\ \bibinfo
  {author} {\bibfnamefont {D.~H.}\ \bibnamefont {Rischke}},\ }\href {\doibase
  10.1103/PhysRevD.72.034004} {\bibfield  {journal} {\bibinfo  {journal} {Phys.
  Rev.}\ }\textbf {\bibinfo {volume} {D72}},\ \bibinfo {pages} {034004}
  (\bibinfo {year} {2005})},\ \Eprint {http://arxiv.org/abs/hep-ph/0503184}
  {arXiv:hep-ph/0503184 [hep-ph]} \BibitemShut {NoStop}%
\bibitem [{\citenamefont {Alford}\ and\ \citenamefont
  {Rajagopal}(2002)}]{Alford:2002kj}%
  \BibitemOpen
  \bibfield  {author} {\bibinfo {author} {\bibfnamefont {M.}~\bibnamefont
  {Alford}}\ and\ \bibinfo {author} {\bibfnamefont {K.}~\bibnamefont
  {Rajagopal}},\ }\href {\doibase 10.1088/1126-6708/2002/06/031} {\bibfield
  {journal} {\bibinfo  {journal} {JHEP}\ }\textbf {\bibinfo {volume} {06}},\
  \bibinfo {pages} {031} (\bibinfo {year} {2002})},\ \Eprint
  {http://arxiv.org/abs/hep-ph/0204001} {arXiv:hep-ph/0204001 [hep-ph]}
  \BibitemShut {NoStop}%
\bibitem [{\citenamefont {Steiner}\ \emph {et~al.}(2002)\citenamefont
  {Steiner}, \citenamefont {Reddy},\ and\ \citenamefont
  {Prakash}}]{Steiner:2002gx}%
  \BibitemOpen
  \bibfield  {author} {\bibinfo {author} {\bibfnamefont {A.~W.}\ \bibnamefont
  {Steiner}}, \bibinfo {author} {\bibfnamefont {S.}~\bibnamefont {Reddy}}, \
  and\ \bibinfo {author} {\bibfnamefont {M.}~\bibnamefont {Prakash}},\ }\href
  {\doibase 10.1103/PhysRevD.66.094007} {\bibfield  {journal} {\bibinfo
  {journal} {Phys. Rev.}\ }\textbf {\bibinfo {volume} {D66}},\ \bibinfo {pages}
  {094007} (\bibinfo {year} {2002})},\ \Eprint
  {http://arxiv.org/abs/hep-ph/0205201} {arXiv:hep-ph/0205201 [hep-ph]}
  \BibitemShut {NoStop}%
\bibitem [{\citenamefont {Shovkovy}\ and\ \citenamefont
  {Huang}(2003)}]{Shovkovy:2003uu}%
  \BibitemOpen
  \bibfield  {author} {\bibinfo {author} {\bibfnamefont {I.}~\bibnamefont
  {Shovkovy}}\ and\ \bibinfo {author} {\bibfnamefont {M.}~\bibnamefont
  {Huang}},\ }\href {\doibase 10.1016/S0370-2693(03)00748-2} {\bibfield
  {journal} {\bibinfo  {journal} {Phys. Lett.}\ }\textbf {\bibinfo {volume}
  {B564}},\ \bibinfo {pages} {205} (\bibinfo {year} {2003})},\ \Eprint
  {http://arxiv.org/abs/hep-ph/0302142} {arXiv:hep-ph/0302142 [hep-ph]}
  \BibitemShut {NoStop}%
\bibitem [{\citenamefont {Huang}\ and\ \citenamefont
  {Shovkovy}(2003)}]{Huang:2003xd}%
  \BibitemOpen
  \bibfield  {author} {\bibinfo {author} {\bibfnamefont {M.}~\bibnamefont
  {Huang}}\ and\ \bibinfo {author} {\bibfnamefont {I.}~\bibnamefont
  {Shovkovy}},\ }\href {\doibase 10.1016/j.nuclphysa.2003.10.005} {\bibfield
  {journal} {\bibinfo  {journal} {Nucl. Phys.}\ }\textbf {\bibinfo {volume}
  {A729}},\ \bibinfo {pages} {835} (\bibinfo {year} {2003})},\ \Eprint
  {http://arxiv.org/abs/hep-ph/0307273} {arXiv:hep-ph/0307273 [hep-ph]}
  \BibitemShut {NoStop}%
\bibitem [{\citenamefont {Alford}\ \emph {et~al.}(2003)\citenamefont {Alford},
  \citenamefont {Bowers}, \citenamefont {Cheyne},\ and\ \citenamefont
  {Cowan}}]{Alford:2002rz}%
  \BibitemOpen
  \bibfield  {author} {\bibinfo {author} {\bibfnamefont {M.~G.}\ \bibnamefont
  {Alford}}, \bibinfo {author} {\bibfnamefont {J.~A.}\ \bibnamefont {Bowers}},
  \bibinfo {author} {\bibfnamefont {J.~M.}\ \bibnamefont {Cheyne}}, \ and\
  \bibinfo {author} {\bibfnamefont {G.~A.}\ \bibnamefont {Cowan}},\ }\href
  {\doibase 10.1103/PhysRevD.67.054018} {\bibfield  {journal} {\bibinfo
  {journal} {Phys. Rev.}\ }\textbf {\bibinfo {volume} {D67}},\ \bibinfo {pages}
  {054018} (\bibinfo {year} {2003})},\ \Eprint
  {http://arxiv.org/abs/hep-ph/0210106} {arXiv:hep-ph/0210106 [hep-ph]}
  \BibitemShut {NoStop}%
\bibitem [{\citenamefont {Schäfer}(2000)}]{Schafer:2000cj}%
  \BibitemOpen
  \bibfield  {author} {\bibinfo {author} {\bibfnamefont {T.}~\bibnamefont
  {Schäfer}},\ }\href {\doibase 10.1103/PhysRevD.62.035013} {\bibfield
  {journal} {\bibinfo  {journal} {Phys. Rev.}\ }\textbf {\bibinfo {volume}
  {D62}},\ \bibinfo {pages} {035013} (\bibinfo {year} {2000})},\ \Eprint
  {http://arxiv.org/abs/hep-ph/0003290} {arXiv:hep-ph/0003290 [hep-ph]}
  \BibitemShut {NoStop}%
\bibitem [{\citenamefont {Buballa}\ \emph {et~al.}(2003)\citenamefont
  {Buballa}, \citenamefont {Hosek},\ and\ \citenamefont
  {Oertel}}]{Buballa:2002wy}%
  \BibitemOpen
  \bibfield  {author} {\bibinfo {author} {\bibfnamefont {M.}~\bibnamefont
  {Buballa}}, \bibinfo {author} {\bibfnamefont {J.}~\bibnamefont {Hosek}}, \
  and\ \bibinfo {author} {\bibfnamefont {M.}~\bibnamefont {Oertel}},\ }\href
  {\doibase 10.1103/PhysRevLett.90.182002} {\bibfield  {journal} {\bibinfo
  {journal} {Phys. Rev. Lett.}\ }\textbf {\bibinfo {volume} {90}},\ \bibinfo
  {pages} {182002} (\bibinfo {year} {2003})},\ \Eprint
  {http://arxiv.org/abs/hep-ph/0204275} {arXiv:hep-ph/0204275 [hep-ph]}
  \BibitemShut {NoStop}%
\bibitem [{\citenamefont {Schmitt}(2005)}]{Schmitt:2004et}%
  \BibitemOpen
  \bibfield  {author} {\bibinfo {author} {\bibfnamefont {A.}~\bibnamefont
  {Schmitt}},\ }\href {\doibase 10.1103/PhysRevD.71.054016} {\bibfield
  {journal} {\bibinfo  {journal} {Phys. Rev.}\ }\textbf {\bibinfo {volume}
  {D71}},\ \bibinfo {pages} {054016} (\bibinfo {year} {2005})},\ \Eprint
  {http://arxiv.org/abs/nucl-th/0412033} {arXiv:nucl-th/0412033 [nucl-th]}
  \BibitemShut {NoStop}%
\bibitem [{\citenamefont {Schmitt}\ \emph {et~al.}(2006)\citenamefont
  {Schmitt}, \citenamefont {Shovkovy},\ and\ \citenamefont
  {Wang}}]{Schmitt:2005wg}%
  \BibitemOpen
  \bibfield  {author} {\bibinfo {author} {\bibfnamefont {A.}~\bibnamefont
  {Schmitt}}, \bibinfo {author} {\bibfnamefont {I.~A.}\ \bibnamefont
  {Shovkovy}}, \ and\ \bibinfo {author} {\bibfnamefont {Q.}~\bibnamefont
  {Wang}},\ }\href {\doibase 10.1103/PhysRevD.73.034012} {\bibfield  {journal}
  {\bibinfo  {journal} {Phys. Rev.}\ }\textbf {\bibinfo {volume} {D73}},\
  \bibinfo {pages} {034012} (\bibinfo {year} {2006})},\ \Eprint
  {http://arxiv.org/abs/hep-ph/0510347} {arXiv:hep-ph/0510347 [hep-ph]}
  \BibitemShut {NoStop}%
\bibitem [{\citenamefont {Schmitt}\ \emph {et~al.}(2003)\citenamefont
  {Schmitt}, \citenamefont {Wang},\ and\ \citenamefont
  {Rischke}}]{Schmitt:2003xq}%
  \BibitemOpen
  \bibfield  {author} {\bibinfo {author} {\bibfnamefont {A.}~\bibnamefont
  {Schmitt}}, \bibinfo {author} {\bibfnamefont {Q.}~\bibnamefont {Wang}}, \
  and\ \bibinfo {author} {\bibfnamefont {D.~H.}\ \bibnamefont {Rischke}},\
  }\href {\doibase 10.1103/PhysRevLett.91.242301} {\bibfield  {journal}
  {\bibinfo  {journal} {Phys. Rev. Lett.}\ }\textbf {\bibinfo {volume} {91}},\
  \bibinfo {pages} {242301} (\bibinfo {year} {2003})},\ \Eprint
  {http://arxiv.org/abs/nucl-th/0301090} {arXiv:nucl-th/0301090 [nucl-th]}
  \BibitemShut {NoStop}%
\bibitem [{\citenamefont {Kundu}\ and\ \citenamefont
  {Rajagopal}(2002)}]{Kundu:2001tt}%
  \BibitemOpen
  \bibfield  {author} {\bibinfo {author} {\bibfnamefont {J.}~\bibnamefont
  {Kundu}}\ and\ \bibinfo {author} {\bibfnamefont {K.}~\bibnamefont
  {Rajagopal}},\ }\href {\doibase 10.1103/PhysRevD.65.094022} {\bibfield
  {journal} {\bibinfo  {journal} {Phys. Rev.}\ }\textbf {\bibinfo {volume}
  {D65}},\ \bibinfo {pages} {094022} (\bibinfo {year} {2002})},\ \Eprint
  {http://arxiv.org/abs/hep-ph/0112206} {arXiv:hep-ph/0112206 [hep-ph]}
  \BibitemShut {NoStop}%
\bibitem [{\citenamefont {Bowers}\ and\ \citenamefont
  {Rajagopal}(2002)}]{Bowers:2002xr}%
  \BibitemOpen
  \bibfield  {author} {\bibinfo {author} {\bibfnamefont {J.~A.}\ \bibnamefont
  {Bowers}}\ and\ \bibinfo {author} {\bibfnamefont {K.}~\bibnamefont
  {Rajagopal}},\ }\href {\doibase 10.1103/PhysRevD.66.065002} {\bibfield
  {journal} {\bibinfo  {journal} {Phys. Rev.}\ }\textbf {\bibinfo {volume}
  {D66}},\ \bibinfo {pages} {065002} (\bibinfo {year} {2002})},\ \Eprint
  {http://arxiv.org/abs/hep-ph/0204079} {arXiv:hep-ph/0204079 [hep-ph]}
  \BibitemShut {NoStop}%
\bibitem [{\citenamefont {Mannarelli}\ \emph {et~al.}(2006)\citenamefont
  {Mannarelli}, \citenamefont {Rajagopal},\ and\ \citenamefont
  {Sharma}}]{Mannarelli:2006}%
  \BibitemOpen
  \bibfield  {author} {\bibinfo {author} {\bibfnamefont {M.}~\bibnamefont
  {Mannarelli}}, \bibinfo {author} {\bibfnamefont {K.}~\bibnamefont
  {Rajagopal}}, \ and\ \bibinfo {author} {\bibfnamefont {R.}~\bibnamefont
  {Sharma}},\ }\href {\doibase 10.1103/PhysRevD.73.114012} {\bibfield
  {journal} {\bibinfo  {journal} {Phys. Rev. D}\ }\textbf {\bibinfo {volume}
  {73}},\ \bibinfo {pages} {114012} (\bibinfo {year} {2006})}\BibitemShut
  {NoStop}%
\bibitem [{\citenamefont {Leibovich}\ \emph {et~al.}(2001)\citenamefont
  {Leibovich}, \citenamefont {Rajagopal},\ and\ \citenamefont
  {Shuster}}]{Leibovich:2001xr}%
  \BibitemOpen
  \bibfield  {author} {\bibinfo {author} {\bibfnamefont {A.~K.}\ \bibnamefont
  {Leibovich}}, \bibinfo {author} {\bibfnamefont {K.}~\bibnamefont
  {Rajagopal}}, \ and\ \bibinfo {author} {\bibfnamefont {E.}~\bibnamefont
  {Shuster}},\ }\href {\doibase 10.1103/PhysRevD.64.094005} {\bibfield
  {journal} {\bibinfo  {journal} {Phys. Rev.}\ }\textbf {\bibinfo {volume}
  {D64}},\ \bibinfo {pages} {094005} (\bibinfo {year} {2001})},\ \Eprint
  {http://arxiv.org/abs/hep-ph/0104073} {arXiv:hep-ph/0104073 [hep-ph]}
  \BibitemShut {NoStop}%
\bibitem [{\citenamefont {Casalbuoni}\ \emph
  {et~al.}(2005{\natexlab{b}})\citenamefont {Casalbuoni}, \citenamefont
  {Gatto}, \citenamefont {Ippolito}, \citenamefont {Nardulli},\ and\
  \citenamefont {Ruggieri}}]{Casalbuoni:2005zp}%
  \BibitemOpen
  \bibfield  {author} {\bibinfo {author} {\bibfnamefont {R.}~\bibnamefont
  {Casalbuoni}}, \bibinfo {author} {\bibfnamefont {R.}~\bibnamefont {Gatto}},
  \bibinfo {author} {\bibfnamefont {N.}~\bibnamefont {Ippolito}}, \bibinfo
  {author} {\bibfnamefont {G.}~\bibnamefont {Nardulli}}, \ and\ \bibinfo
  {author} {\bibfnamefont {M.}~\bibnamefont {Ruggieri}},\ }\href {\doibase
  10.1016/j.physletb.2005.08.123, 10.1016/j.physletb.2006.01.057} {\bibfield
  {journal} {\bibinfo  {journal} {Phys. Lett.}\ }\textbf {\bibinfo {volume}
  {B627}},\ \bibinfo {pages} {89} (\bibinfo {year} {2005}{\natexlab{b}})},\
  \bibinfo {note} {[Erratum: Phys. Lett.B634,565(2006)]},\ \Eprint
  {http://arxiv.org/abs/hep-ph/0507247} {arXiv:hep-ph/0507247 [hep-ph]}
  \BibitemShut {NoStop}%
\bibitem [{\citenamefont {Baym}\ \emph {et~al.}(1990)\citenamefont {Baym},
  \citenamefont {Monien}, \citenamefont {Pethick},\ and\ \citenamefont
  {Ravenhall}}]{Baym:1990}%
  \BibitemOpen
  \bibfield  {author} {\bibinfo {author} {\bibfnamefont {G.}~\bibnamefont
  {Baym}}, \bibinfo {author} {\bibfnamefont {H.}~\bibnamefont {Monien}},
  \bibinfo {author} {\bibfnamefont {C.~J.}\ \bibnamefont {Pethick}}, \ and\
  \bibinfo {author} {\bibfnamefont {D.~G.}\ \bibnamefont {Ravenhall}},\ }\href
  {\doibase 10.1103/PhysRevLett.64.1867} {\bibfield  {journal} {\bibinfo
  {journal} {Phys. Rev. Lett.}\ }\textbf {\bibinfo {volume} {64}},\ \bibinfo
  {pages} {1867} (\bibinfo {year} {1990})}\BibitemShut {NoStop}%
\bibitem [{\citenamefont {Alford}\ \emph
  {et~al.}(2005{\natexlab{b}})\citenamefont {Alford}, \citenamefont {Jotwani},
  \citenamefont {Kouvaris}, \citenamefont {Kundu},\ and\ \citenamefont
  {Rajagopal}}]{Alford:2005pooja}%
  \BibitemOpen
  \bibfield  {author} {\bibinfo {author} {\bibfnamefont {M.}~\bibnamefont
  {Alford}}, \bibinfo {author} {\bibfnamefont {P.}~\bibnamefont {Jotwani}},
  \bibinfo {author} {\bibfnamefont {C.}~\bibnamefont {Kouvaris}}, \bibinfo
  {author} {\bibfnamefont {J.}~\bibnamefont {Kundu}}, \ and\ \bibinfo {author}
  {\bibfnamefont {K.}~\bibnamefont {Rajagopal}},\ }\href@noop {} {\bibfield
  {journal} {\bibinfo  {journal} {Physical Review D}\ }\textbf {\bibinfo
  {volume} {71}},\ \bibinfo {pages} {114011} (\bibinfo {year}
  {2005}{\natexlab{b}})}\BibitemShut {NoStop}%
\bibitem [{\citenamefont {Casalbuoni}\ \emph {et~al.}(2002)\citenamefont
  {Casalbuoni}, \citenamefont {Fabiano}, \citenamefont {Gatto}, \citenamefont
  {Mannarelli},\ and\ \citenamefont {Nardulli}}]{Casalbuoni:2002my}%
  \BibitemOpen
  \bibfield  {author} {\bibinfo {author} {\bibfnamefont {R.}~\bibnamefont
  {Casalbuoni}}, \bibinfo {author} {\bibfnamefont {E.}~\bibnamefont {Fabiano}},
  \bibinfo {author} {\bibfnamefont {R.}~\bibnamefont {Gatto}}, \bibinfo
  {author} {\bibfnamefont {M.}~\bibnamefont {Mannarelli}}, \ and\ \bibinfo
  {author} {\bibfnamefont {G.}~\bibnamefont {Nardulli}},\ }\href {\doibase
  10.1103/PhysRevD.66.094006} {\bibfield  {journal} {\bibinfo  {journal} {Phys.
  Rev.}\ }\textbf {\bibinfo {volume} {D66}},\ \bibinfo {pages} {094006}
  (\bibinfo {year} {2002})},\ \Eprint {http://arxiv.org/abs/hep-ph/0208121}
  {arXiv:hep-ph/0208121 [hep-ph]} \BibitemShut {NoStop}%
\bibitem [{\citenamefont {Giannakis}\ and\ \citenamefont
  {Ren}(2005{\natexlab{b}})}]{Giannakis:2005vw}%
  \BibitemOpen
  \bibfield  {author} {\bibinfo {author} {\bibfnamefont {I.}~\bibnamefont
  {Giannakis}}\ and\ \bibinfo {author} {\bibfnamefont {H.-C.}\ \bibnamefont
  {Ren}},\ }\href {\doibase 10.1016/j.nuclphysb.2005.06.008} {\bibfield
  {journal} {\bibinfo  {journal} {Nucl. Phys.}\ }\textbf {\bibinfo {volume}
  {B723}},\ \bibinfo {pages} {255} (\bibinfo {year} {2005}{\natexlab{b}})},\
  \Eprint {http://arxiv.org/abs/hep-th/0504053} {arXiv:hep-th/0504053 [hep-th]}
  \BibitemShut {NoStop}%
\bibitem [{\citenamefont {Ciminale}\ \emph {et~al.}(2006)\citenamefont
  {Ciminale}, \citenamefont {Nardulli}, \citenamefont {Ruggieri},\ and\
  \citenamefont {Gatto}}]{Ciminale:2006sm}%
  \BibitemOpen
  \bibfield  {author} {\bibinfo {author} {\bibfnamefont {M.}~\bibnamefont
  {Ciminale}}, \bibinfo {author} {\bibfnamefont {G.}~\bibnamefont {Nardulli}},
  \bibinfo {author} {\bibfnamefont {M.}~\bibnamefont {Ruggieri}}, \ and\
  \bibinfo {author} {\bibfnamefont {R.}~\bibnamefont {Gatto}},\ }\href
  {\doibase 10.1016/j.physletb.2006.03.075} {\bibfield  {journal} {\bibinfo
  {journal} {Phys. Lett.}\ }\textbf {\bibinfo {volume} {B636}},\ \bibinfo
  {pages} {317} (\bibinfo {year} {2006})},\ \Eprint
  {http://arxiv.org/abs/hep-ph/0602180} {arXiv:hep-ph/0602180 [hep-ph]}
  \BibitemShut {NoStop}%
\bibitem [{InP()}]{InPreparation}%
  \BibitemOpen
  \href@noop {} {\bibinfo  {journal} {In preparation}\ }\BibitemShut {NoStop}%
\bibitem [{\citenamefont {Casalbuoni}\ \emph
  {et~al.}(2001{\natexlab{b}})\citenamefont {Casalbuoni}, \citenamefont
  {Gatto}, \citenamefont {Mannarelli},\ and\ \citenamefont
  {Nardulli}}]{Casalbuoni:2001gt}%
  \BibitemOpen
\bibfield  {journal} {  }\bibfield  {author} {\bibinfo {author} {\bibfnamefont
  {R.}~\bibnamefont {Casalbuoni}}, \bibinfo {author} {\bibfnamefont
  {R.}~\bibnamefont {Gatto}}, \bibinfo {author} {\bibfnamefont
  {M.}~\bibnamefont {Mannarelli}}, \ and\ \bibinfo {author} {\bibfnamefont
  {G.}~\bibnamefont {Nardulli}},\ }\href {\doibase
  10.1016/S0370-2693(01)00645-1} {\bibfield  {journal} {\bibinfo  {journal}
  {Phys. Lett.}\ }\textbf {\bibinfo {volume} {B511}},\ \bibinfo {pages} {218}
  (\bibinfo {year} {2001}{\natexlab{b}})},\ \Eprint
  {http://arxiv.org/abs/hep-ph/0101326} {arXiv:hep-ph/0101326 [hep-ph]}
  \BibitemShut {NoStop}%
\bibitem [{\citenamefont {Mannarelli}\ \emph {et~al.}(2007)\citenamefont
  {Mannarelli}, \citenamefont {Rajagopal},\ and\ \citenamefont
  {Sharma}}]{Mannarelli:2007bs}%
  \BibitemOpen
  \bibfield  {author} {\bibinfo {author} {\bibfnamefont {M.}~\bibnamefont
  {Mannarelli}}, \bibinfo {author} {\bibfnamefont {K.}~\bibnamefont
  {Rajagopal}}, \ and\ \bibinfo {author} {\bibfnamefont {R.}~\bibnamefont
  {Sharma}},\ }\href {\doibase 10.1103/PhysRevD.76.074026} {\bibfield
  {journal} {\bibinfo  {journal} {Phys. Rev.}\ }\textbf {\bibinfo {volume}
  {D76}},\ \bibinfo {pages} {074026} (\bibinfo {year} {2007})},\ \Eprint
  {http://arxiv.org/abs/hep-ph/0702021} {arXiv:hep-ph/0702021 [hep-ph]}
  \BibitemShut {NoStop}%
\bibitem [{\citenamefont {{Radzihovsky}}\ and\ \citenamefont
  {{Vishwanath}}(2009)}]{Radzihovsky:2009}%
  \BibitemOpen
  \bibfield  {author} {\bibinfo {author} {\bibfnamefont {L.}~\bibnamefont
  {{Radzihovsky}}}\ and\ \bibinfo {author} {\bibfnamefont {A.}~\bibnamefont
  {{Vishwanath}}},\ }\href {\doibase 10.1103/PhysRevLett.103.010404} {\bibfield
   {journal} {\bibinfo  {journal} {Physical Review Letters}\ }\textbf {\bibinfo
  {volume} {103}},\ \bibinfo {eid} {010404} (\bibinfo {year} {2009})},\ \Eprint
  {http://arxiv.org/abs/0812.3945} {arXiv:0812.3945 [cond-mat.supr-con]}
  \BibitemShut {NoStop}%
\bibitem [{\citenamefont {{Adams}}(1960)}]{Ziman:1960}%
  \BibitemOpen
  \bibfield  {author} {\bibinfo {author} {\bibfnamefont {E.}~\bibnamefont
  {{Adams}}},\ }\href {\doibase 10.1016/0022-3697(60)90260-2} {\bibfield
  {journal} {\bibinfo  {journal} {Journal of Physics and Chemistry of Solids}\
  }\textbf {\bibinfo {volume} {15}},\ \bibinfo {pages} {359} (\bibinfo {year}
  {1960})}\BibitemShut {NoStop}%
\bibitem [{\citenamefont {Landau}\ and\ \citenamefont
  {Khalatnikov}(1949)}]{landau1949theory}%
  \BibitemOpen
  \bibfield  {author} {\bibinfo {author} {\bibfnamefont {L.}~\bibnamefont
  {Landau}}\ and\ \bibinfo {author} {\bibfnamefont {I.}~\bibnamefont
  {Khalatnikov}},\ }\href@noop {} {\bibfield  {journal} {\bibinfo  {journal}
  {Zh. Eksp. Teor. Fiz.}\ }\textbf {\bibinfo {volume} {19}} (\bibinfo {year}
  {1949})}\BibitemShut {NoStop}%
\bibitem [{\citenamefont {Maris}(1973)}]{maris1973hydrodynamics}%
  \BibitemOpen
  \bibfield  {author} {\bibinfo {author} {\bibfnamefont {H.~J.}\ \bibnamefont
  {Maris}},\ }\href@noop {} {\bibfield  {journal} {\bibinfo  {journal}
  {Physical Review A}\ }\textbf {\bibinfo {volume} {8}},\ \bibinfo {pages}
  {1980} (\bibinfo {year} {1973})}\BibitemShut {NoStop}%
\bibitem [{\citenamefont {Rupak}\ and\ \citenamefont
  {Schäfer}(2007)}]{Rupak:2007vp}%
  \BibitemOpen
  \bibfield  {author} {\bibinfo {author} {\bibfnamefont {G.}~\bibnamefont
  {Rupak}}\ and\ \bibinfo {author} {\bibfnamefont {T.}~\bibnamefont
  {Schäfer}},\ }\href {\doibase 10.1103/PhysRevA.76.053607} {\bibfield
  {journal} {\bibinfo  {journal} {Phys. Rev.}\ }\textbf {\bibinfo {volume}
  {A76}},\ \bibinfo {pages} {053607} (\bibinfo {year} {2007})},\ \Eprint
  {http://arxiv.org/abs/0707.1520} {arXiv:0707.1520 [cond-mat.other]}
  \BibitemShut {NoStop}%
\bibitem [{\citenamefont {Manuel}\ and\ \citenamefont
  {Tolos}(2011)}]{Manuel:2011ed}%
  \BibitemOpen
  \bibfield  {author} {\bibinfo {author} {\bibfnamefont {C.}~\bibnamefont
  {Manuel}}\ and\ \bibinfo {author} {\bibfnamefont {L.}~\bibnamefont {Tolos}},\
  }\href {\doibase 10.1103/PhysRevD.84.123007} {\bibfield  {journal} {\bibinfo
  {journal} {Phys. Rev.}\ }\textbf {\bibinfo {volume} {D84}},\ \bibinfo {pages}
  {123007} (\bibinfo {year} {2011})},\ \Eprint {http://arxiv.org/abs/1110.0669}
  {arXiv:1110.0669 [astro-ph.SR]} \BibitemShut {NoStop}%
\bibitem [{\citenamefont {Aguilera}\ \emph {et~al.}(2009)\citenamefont
  {Aguilera}, \citenamefont {Cirigliano}, \citenamefont {Pons}, \citenamefont
  {Reddy},\ and\ \citenamefont {Sharma}}]{Aguilera:2008ed}%
  \BibitemOpen
  \bibfield  {author} {\bibinfo {author} {\bibfnamefont {D.~N.}\ \bibnamefont
  {Aguilera}}, \bibinfo {author} {\bibfnamefont {V.}~\bibnamefont
  {Cirigliano}}, \bibinfo {author} {\bibfnamefont {J.~A.}\ \bibnamefont
  {Pons}}, \bibinfo {author} {\bibfnamefont {S.}~\bibnamefont {Reddy}}, \ and\
  \bibinfo {author} {\bibfnamefont {R.}~\bibnamefont {Sharma}},\ }\href
  {\doibase 10.1103/PhysRevLett.102.091101} {\bibfield  {journal} {\bibinfo
  {journal} {Phys. Rev. Lett.}\ }\textbf {\bibinfo {volume} {102}},\ \bibinfo
  {pages} {091101} (\bibinfo {year} {2009})},\ \Eprint
  {http://arxiv.org/abs/0807.4754} {arXiv:0807.4754 [nucl-th]} \BibitemShut
  {NoStop}%
\bibitem [{\citenamefont {Schrieffer}(1983)}]{schrieffer1983theory}%
  \BibitemOpen
  \bibfield  {author} {\bibinfo {author} {\bibfnamefont {J.}~\bibnamefont
  {Schrieffer}},\ }\href {https://books.google.co.in/books?id=TzGAgVD0p38C}
  {\emph {\bibinfo {title} {Theory of Superconductivity}}},\ Advanced Book
  Program Series\ (\bibinfo  {publisher} {Advanced Book Program, Perseus
  Books},\ \bibinfo {year} {1983})\BibitemShut {NoStop}%
\bibitem [{\citenamefont {Levin}(1999)}]{Levin:1998wa}%
  \BibitemOpen
  \bibfield  {author} {\bibinfo {author} {\bibfnamefont {Y.}~\bibnamefont
  {Levin}},\ }\href {\doibase 10.1086/307196} {\bibfield  {journal} {\bibinfo
  {journal} {Astrophys. J.}\ }\textbf {\bibinfo {volume} {517}},\ \bibinfo
  {pages} {328} (\bibinfo {year} {1999})},\ \Eprint
  {http://arxiv.org/abs/astro-ph/9810471} {arXiv:astro-ph/9810471 [astro-ph]}
  \BibitemShut {NoStop}%
\end{thebibliography}%

\end{document}